\documentclass[floats,floatfix,amssymb,prd,twocolumn,superscriptaddress,nofootinbib]{revtex4-2}

\usepackage[english]{babel}
\usepackage[utf8]{inputenc}
\usepackage{amsmath}
\usepackage{mathbbol}
\usepackage{amssymb}
\usepackage{bbold}
\usepackage{graphicx,amsfonts}
\usepackage{epsfig}
\usepackage[svgnames]{xcolor}
\definecolor{crimsonglory}{rgb}{0.75,0.0,0.2}
\usepackage[colorlinks=true,
linkcolor=crimsonglory,
urlcolor=crimsonglory,
citecolor=crimsonglory]{hyperref}
\usepackage{bm}
\usepackage{mathrsfs}
\usepackage{mathtools}
\usepackage{enumerate}
\usepackage{amsthm}
\usepackage{bbm}
\usepackage{comment}
\usepackage{physics}
\usepackage{url}
\usepackage{upgreek}
\usepackage[title]{appendix}
\usepackage{float} 
\usepackage{rotating}
\usepackage{booktabs}
\usepackage{underscore} 

\begin{document}

\title{Cumulative effect of orbital resonances in extreme-mass-ratio inspirals}

\author{Edoardo Levati}
\email{levae25@wfu.edu}
\affiliation{Department of Physics, Wake Forest University, Winston-Salem, North Carolina 27109, USA}
\affiliation{Dipartimento di Fisica, Sapienza Università di Roma, Italy}
\affiliation{INFN, Sezione di Roma, Piazzale Aldo Moro 5, 00185, Roma, Italy}
\author{Alejandro C\'ardenas-Avenda\~no}
\affiliation{Computational Physics and Methods (CCS-2), Center for Nonlinear Studies (CNLS) \& Center for Theoretical Astrophysics (CTA), Los Alamos National Laboratory, Los Alamos NM 87545, USA}
\author{Kyriakos Destounis}
\affiliation{Dipartimento di Fisica, Sapienza Università di Roma, Italy}
\affiliation{INFN, Sezione di Roma, Piazzale Aldo Moro 5, 00185, Roma, Italy}
\affiliation{CENTRA, Departamento de Física, Instituto Superior Técnico – IST, Universidade
de Lisboa – UL, Avenida Rovisco Pais 1, 1049-001 Lisboa, Portugal}
\author{Paolo Pani}
\affiliation{Dipartimento di Fisica, Sapienza Università di Roma, Italy}
\affiliation{INFN, Sezione di Roma, Piazzale Aldo Moro 5, 00185, Roma, Italy}


\begin{abstract}
Orbital resonances in extreme-mass-ratio inspirals (EMRIs) have been proven to be a key feature for accurate gravitational-wave template modeling. Decades of research have led to schemes that can not only model the adiabatic inspiral of such a binary system, but also account for the effects of resonances on their evolution. In this work, we use an effective resonance model that includes analytically derived corrections to the radiation reaction fluxes, to study the combined effects of both dominant (low-order) and subdominant (high-order) orbital resonances in EMRIs. We show that using single, universal shifts for all fluxes overestimates the resonance impact, and therefore individualized shifts for each resonance crossing are needed for accurate modeling. Our analysis reveals that the cumulative effects from multiple resonance crossings can significantly impact the orbital evolution of EMRIs, especially for highly eccentric orbits. Our results provide further evidence that resonance effects have to be included in template production to extract detailed astrophysical parameters from EMRI signals. 
\end{abstract}

\maketitle

\section{Introduction}

The detection of gravitational waves (GWs)~\cite{LIGO,LIGO:2021ppb,KAGRA:2021vkt}, predicted by Einstein in 1916~\cite{Einstein:1916cc}, has provided a new powerful tool to probe gravity in its most extreme regime, and to search for signatures of \emph{new physics}~\cite{Barack:2018yly}. GWs have the potential to examine largely unexplored regions of the universe that are otherwise electromagnetically obscure, thus extending the knowledge obtained from electromagnetic observations~\cite{EventHorizonTelescope:2019dse,EventHorizonTelescope:2022wkp,Bambi:2024hhi}. These advancements pave the way to the new field of multimessenger astrophysics~\cite{Schutz:1999xj,Bishop:2021rye,Bailes:2021tot}. Space-based GW detectors, such as the Laser Interferometer Space Antenna (LISA)~\cite{LISA,LISA:2024hlh}, will open a new frequency window centered in the mHz band that is inaccessible by ground-based GW interferometers. These detectors will be sensitive to new astrophysical sources and augment the information we have so far from ground-based interferometers. One of the most promising targets is stellar-mass compact objects inspiralling into supermassive black holes (BHs) in the centers of galaxies, known as~\emph{extreme-mass-ratio inspirals} (EMRIs).

EMRIs are extraordinary natural laboratories for strong-gravity physics~\cite{Barack:2018yly,CardenasSopuertaReview} that consist of a (\emph{primary}) supermassive BH and a (\emph{secondary}) stellar-mass compact object. Due to their slow (adiabatic) evolution, EMRIs are expected to execute around $10^4 - 10^5$ orbits before merging~\cite{Hughes}, thus potentially staying in LISA's sensitivity band for years. Estimates suggest that LISA could detect up to $10^3$ EMRI events during its lifetime~\cite{Kludge,Babak:2017tow,LISA:2024hlh}, which, if possible, can reshape the way we understand gravitation. The signal produced by this slow inspiral can be used to map the spacetime geometry of the central object, providing unprecedentedly precise measurements of BH masses and spins~\cite{LISA,LISA:2024hlh}, and even test if the received signal is consistent with general relativity (GR) predictions~\cite{CardenasSopuertaReview, Speri:2024qak}. Observing several EMRI events will also probe the population of massive BHs and stellar-mass compact objects in the galactic nuclei. 

EMRI signals are expected to be found and characterized by comparing the data against a large number of waveform templates that have yet to be produced. The templates are required to remain phase coherent with the true ones to an accuracy of about $1$ cycle over a complete inspiral~\cite{FlanaganHinderer1,Kludge}. Therefore, accurate waveform modeling that includes all phenomenology across the entire parameter space of EMRIs is essential to maximize the scientific yield of space-based GW observations~\cite{LISAConsortiumWaveformWorkingGroup:2023arg}. 

In this work, we first investigate three methods to build adiabatic inspirals of EMRIs in the Kerr spacetime~\cite{Kerr}. One is the \textit{numerical kludge}~(NK) model~\cite{Kludge}, where the integrals of motion are used to decouple the geodesic equations and derive a set of first-order differential equations for the orbital motion. The other two methods directly solve the second-order system, i.e., the coupled radial and polar geodesic equations. In particular, the \emph{second technique} integrates the geodesic equations and re-initializes the velocities at each step of the inspiral. The radiation reaction is included as the new initial velocities are computed by updating the integrals of motion with the adiabatic fluxes. The \emph{third technique} adds forcing terms to the geodesic equation in order to include the leading order self-acceleration, and integrates the so-called MiSaTaQuWa equations~\cite{Mino:1996nk,Quinn:1996am,Gralla:2008fg}. These forcing terms depend on the derivative of the integrals of motion with respect to time, which we approximate with the adiabatic fluxes. We consider an adiabatic inspiral of two generic EMRI systems and provide the first \textit{systematic comparison} between these three techniques. Our findings indicate that the second technique is extremely robust with respect to the NK scheme, while the third technique is less robust to error for long-lasting inspirals. 

After understanding the nature of errors arising from the methods used, we transition to the primary focus of our study: analyzing the impact of transient orbital resonances in Kerr EMRIs evolved using an \emph{augmented NK technique}. During its evolution, an EMRI system experiences a slow, adiabatic change in the orbital parameters. This gradual evolution implies that the fundamental frequencies—radial ($\omega_r$), polar ($\omega_\theta$), and azimuthal ($\omega_\phi$)—vary over time, inevitably leading the system to cross multiple resonances (specific periodic orbits). These resonance crossings are significant because they induce drastic alterations in the system's dynamics, potentially causing dephasing and affecting GW parameter estimation. 

This particular challenge was first highlighted in the seminal work by Hinderer and Flanagan~\cite{FlanaganHinderer1}, who demonstrated that \emph{orbital resonances} (a purely conservative effect) can lead to order-unity changes in the fluxes (associated with the energy, angular momentum, and Carter constant). For clarity, these (``self-force'') orbital resonances are different from \emph{tidal resonances}~\cite{Yang:2017aht,Bonga:2019ycj,Gupta:2022fbe,Bronicki:2022eqa}, which are caused by the tidal force of a tertiary body. In particular, if an orbit is (nearly) circular, the orbital resonances are minimized (as their strength is proportional to eccentricity for low-order resonances) but there are still tidal resonances between $\omega_\theta$ and $\omega_\phi$. Nevertheless, their effects are qualitatively similar: they induce changes in the fluxes that can be physically visualized as a ``kick'' in the trajectory. These effects must be considered for accurate waveform modeling.

Resonance modeling is configuration-dependent and analytically demanding, even after the realization of a toy-model scalar self-force \cite{Nasipak:2021qfu,Nasipak:2022xjh}. To address this, a phenomenological approach—the \textit{effective resonance model} (ERM)—was introduced in Ref.~\cite{SperiGair}. The ERM phenomenologically modifies the fluxes at resonance crossings by introducing an impulse function parameterized by \emph{resonance coefficients}, which encapsulate the resonance's impact on the orbital parameters. While the precise value of these coefficients can, in principle, be derived analytically, as shown in Ref.~\cite{FlanaganHinderer1}, their computation remains theoretically expensive. The ERM enables the study of various EMRI systems without the need for full gravitational self-force (GSF) calculations, thus broadening its applicability.

In this study, we focus on resonances with low multiplicity number $m$, adhering to the standard definition of multiplicity in dynamical systems: the number of stable periodic points in the phase space~\cite{Contopoulos,Contopoulos:2011dz,Gerakopoulos,Destounis,Lukes-Gerakopoulos:2021ybx}. We refer to the $m/n=3:2$ and $2:1$ orbital resonances, or equivalently $\omega_r/\omega_\theta=2:3$ and $1:2$, as \emph{low-order, dominant,} resonances, due to their low multiplicity, while resonances with higher multiplicity are referred to as \emph{higher-order, subdominant} resonances. Low-order resonances correspond to broader, more stable surfaces in the 3-$D$ space of orbital elements $(p,e,\iota)$ where the resonance condition $\nu\equiv m/n=\omega_\theta/\omega_r$ is satisfied~\cite{Brink:2013nna, Brink:2015roa}. These surfaces are wider and more accessible because the derivatives of the frequency ratios with respect to orbital elements are relatively small near these points and the volume in parameter space near a low-order commensurability (rational frequency ratio) is larger \cite{Warburton:2011fk,Ruangsri:2013hra,Osburn:2015duj}, due to the way irrational frequency ratios are distributed.

More formally, the resonant surface for a given ratio (such as $3:2$) is the set of points in $(p,e,\iota)$ space where the resonant ratio holds: $\nu(e,p,\iota)=3:2$. The ``thickness" or ``volume" around that surface that corresponds to near-resonant conditions depends on how flat the function $\nu(e,p,\iota)$ is near $3:2$, i.e., how close $d\nu/dt=\delta\nu$ is to zero: 
\begin{equation*}
    \delta\nu=\left|\frac{m}{n}-\nu\right|<\epsilon.
\end{equation*}
Thus, low-order $m/n$ ratios like $3:2$ or $2:1$ will define larger regions of parameter space, that satisfy the resonance condition within a small tolerance $\epsilon$, since $\delta\nu$ is satisfied for a longer period of time. Ultimately, from a dynamical system point of view, Kolmogorov--Arnold--Moser (KAM) theory states that low-order commensurabilities are more dynamically significant because they correspond to resonant tori that are more robust and occupy larger phase space volume \cite{Contopoulos,Arnold:1989who,Lichtenberg:1992}. The rest of the higher-order, subdominant resonances typically occur in much smaller volumes where the resonant condition is met. This leads to less time spent around, and in, these resonances.

According to this classification, the strongest resonance is expected to be the $1:1$. However, the ratio $1:1$ is only satisfied at infinity in vacuum EMRIs~\cite{vandeMeent:2013sza} (for non-vacuum EMRIs the ratio $1:1$ can be satisfied at a finite distance from the primary supermassive BH \cite{Destounis:2022obl}). Since the radial frequency is always the smallest frequency in the Kerr spacetime~\cite{Schmidt}, meaning that $m\geq n$, the $3:2$ resonance emerges as the lowest order integer ratio and plays a significant role. This low order resonance is known to produce significant phase shifts~\cite{Berry}, it is common in several EMRI systems~\cite{Ruangsri:2013hra}, and is capable of producing sustained resonances~\cite{vandeMeent:2013sza}.

In this work, after re-visiting the $3:2$ resonance analysis conducted in Ref.~\cite{SperiGair} and relaxing the assumption that all the resonance coefficients are equal during the crossing, we extend our analysis beyond this single resonance. We investigate other individual, subdominant resonances, such as $4:3$, $3:1$, and $2:1$, as well as scenarios that involve multiple resonance crossings within a single waveform. By varying the resonance coefficients that ``kick'' the orbital parameters, we quantify the cumulative effects of these crossings. Our results show that the dephasing between waveforms accounting only for the dominant $3:2$ resonance and those incorporating multiple resonances is substantial. This underscores the necessity of understanding multiple resonance effects for accurate EMRI waveform modeling. Our results indicate that for future detections by LISA to accurately represent GWs from Kerr EMRIs, it is crucial to consider multiple resonance crossings of the secondary. 

In Sec.~\ref{sec::EMRI} we give an overview of EMRIs and how to properly model their orbital evolution within the assumption of the adiabatic approximation. We then provide a consistent comparison between the three methods utilized for the evolution of Kerr EMRIs. In Sec.~\ref{sec::ERM} we discuss the effective resonance model, first proposed in Ref.~\cite{SperiGair}, that we use to augment the adiabatic evolution of EMRIs, in order for orbital resonances to be taken into account in radiation reaction.  We then revisit some of the results of Ref.~\cite{SperiGair} regarding the effect of the $ 3:2 $ resonance on the EMRI fluxes and GWs. In Sec.~\ref{sec::cumulative_effect} we extend the work of Ref.~\cite{SperiGair} to other subdominant resonances and calculate their contribution to flux augmentation. To understand if the accumulation of more than one orbital resonance is important in EMRI phenomenology, we provide the dephasing and GW mismatch analysis between EMRIs that only account for the dominant $ 3:2 $ resonance, and EMRIs that allow for further subdominant resonances to be included in the inspiral fluxes, with and without equal resonant coefficients. Finally, in Sec.~\ref{sec::discussion} we conclude with the discussion of our results and present future outlook. We relegate more technical details in a series of appendices.

Throughout this work, we use geometrized units setting $G = 1 = c$, unless otherwise specified. We use the Einstein summation convention for repeated indexes. Greek letters are used to indicate a sum over all spacetime indexes, whereas Roman letters are used to indicate a sum over the spatial indexes only. Only in Sec.~\ref{sec::EMRI}, we follow the notation adopted in Ref.~\cite{FlanaganHinderer1}, where lowercase Greek letters from the start of the alphabet ($\alpha$, $\beta$, $\cdots$) label position or momentum coordinates on the phase space, that are not associated with coordinates on spacetime and do not transform under spacetime coordinate transformations.

\section{An overview of EMRI modeling}\label{sec::EMRI}

To model the trajectory of the secondary around the primary, we can utilize perturbation theory because of the significant mass disparity between the two components of the EMRI. The most astrophysically relevant, theoretically and observationally, compact objects known to date are well-described by the Kerr metric~\cite{Kerr,Teukolsky:2014vca}. As a result, the geodesic motion of massive test particles around Kerr BHs serves as a zeroth-order approximation to EMRI evolution for (short) timescales of a couple of orbital revolutions around the primary. At the geodesic level, stationary and axisymmetric spacetimes, like the Kerr metric, possess at least two integrals of motion, namely the orbital energy $\mathcal{E}$ and the azimuthal angular momentum $\mathcal{L}_z$ of the test particle. Together with the conservation of the four-velocity $u^\mu$ of the particle and the Carter constant $\mathcal{Q}$~\cite{Carter}, the dynamical system of geodesics in the Kerr spacetime is integrable, i.e., there are as many degrees of freedom as constants of the motion and one can reduce the system to first-order differential equations. However, integrability is a fragile feature and it is broken for almost any spacetime other than Kerr, including BHs in modified gravity, exotic compact objects, and environmental effects. In these cases the radial and polar geodesics are second-order, coupled differential equations and the system is nonintegrable. Such behavior goes beyond the inability to write close-formed solutions to the geodesic motion or even to reduce it to first quadratures; it can lead to very interesting chaotic phenomenology (see Refs.~\cite{Destounis:2021mqv,Destounis:2021rko,Cardenas,Destounis:2023khj,Destounis:2023gpw,Destounis:2023cim,Eleni:2024fgs} for several examples).

Modeling the long-term evolution of EMRIs is feasible only when the back-reaction of the secondary compact object on the background metric of the primary massive BH is accounted for. This back-reaction arises from the GSF~\cite{GSF}, which can be split into conservative and dissipative parts. The former is responsible for shifting the orbital frequencies away from their geodesic values, while the latter drives the small body's inspiral~\cite{Ruangsri:2013hra}. The overall effects of the GSF can be more readily understood from the relativistic generalization of the \textit{action-angle} variables for bound orbits in the Kerr spacetime~\cite{Schmidt, Glampedakis:2005cf, FlanaganHinderer1}. The self-acceleration experienced by the smaller object can then be expanded as a series in powers of the system's mass ratio, providing a systematic framework for studying the impact of the GSF, as
\begin{equation}
    a_{\mu} = \eta a^{(1)}_{\mu} + \eta^2 a^{(2)}_{\mu} + \order{\eta^3},
\end{equation}
where $a^{(1)}_{\mu}$ is the leading-order self-acceleration derived in Refs.~\cite{Mino:1996nk,Quinn:1996am}, whereas $\eta\equiv\mu/M$ is the mass ratio of the EMRI, with $\mu$ the mass of the secondary and $M$ the mass of the primary. Using the action-angle formalism, we can express Kerr geodesics in Boyer--Lindquist coordinates to include the effect of the GSF as~\cite{FlanaganHinderer1}
\begin{align}
    & \dv{q_{\alpha}}{\tau} = \Tilde{\omega}_{\alpha} (J_k) + \eta g_{\alpha}^{(1)}(q_A, J_k) + \eta^2 g_{\alpha}^{(2)}(q_A, J_k)+ \order{\eta^3}, \label{angle} \\
    & \dv{J_{i}}{\tau} = \eta G_{i}^{(1)}(q_A, J_k) + \eta^2 G_{i}^{(2)}(q_A, J_k) + \order{\eta^3}, \label{action}
\end{align}
where $\alpha$ runs over $0, 1, 2, 3$, $(i, k)$ run over $1, 2, 3$, $q_A = (q_r, q_{\theta})$ and $\tau$ is the proper time. As a manifestation of the weak equivalence principle, geodesic motion in Kerr spacetime has only three physical degrees of freedom since orbits are the same regardless of the mass of the particle~\cite{Schmidt}. The functions $\Tilde{\omega}_{i} (J_k)$ represent the fundamental frequencies associated with the radial, polar, and azimuthal components of Kerr geodesics. The action variables $J_{i} = (\mathcal{E}, \mathcal{L}_z, \mathcal{Q})$ describe the ``shape'' of the orbit and determine its physical attributes, as these can be mapped to familiar Keplerian-like orbital elements ($p, e, \iota$), where $p$ is the semilatus rectum, $e$ is the eccentricity and $\iota$ is the inclination angle. Their explicit definition can be found in App.~\ref{sec::NK}. The angle variables $q_{\alpha}$ represent the orbit's positional elements and give information about the phase, the direction of the periapsis and the orientation of the orbital plane~\cite{GSF}. The forcing terms $g_{\alpha}^{(1)}$, $G_{i}^{(1)}$ and $g_{\alpha}^{(2)}$, $G_{i}^{(2)}$ include first-order and second-order corrections due to the GSF, respectively, and only depend on $q_r, q_{\theta}$ and $J_k$. At zeroth order in $\eta$, Eqs.~(\ref{angle})-(\ref{action}) reduce to Kerr geodesics, and the integrals of motion and the fundamental frequencies are conserved quantities. At higher orders in $\eta$, the small body’s interaction with the background spacetime changes the trajectory, thus deviating from the geodesic worldline~\cite{FlanaganHughes}. The average over the $2$-torus parameterized by $(q_r, q_{\theta})$ of the conservative part of $G_{i}^{(1)}$ vanishes~\cite{FlanaganHinderer1}. For phase-space filling orbits, this torus-average is equivalent to a time average, thus the time average of the conservative part of $G_{i}^{(1)}$ vanishes~\cite{Mino}. Similarly, the time average of the dissipative part of $g_{\alpha}^{(1)}$ vanishes~\cite{FlanaganHinderer1}.

\subsection{The adiabatic approximation}
\label{sec::adiab_approx}

The previous discussion highlights a method to simplify Eqs.~(\ref{angle})-(\ref{action}) using the adiabatic approximation. In this approach, we treat the motion as nearly geodesic under the assumption that the time scale of the radiation reaction, $\mathcal{T}_{rad} \equiv J / \dot J \sim \order{1/\eta}$ (where the overdot symbol denotes differentiation with respect to $\tau$), is much longer than the dynamical time scales of the three components of the orbital motion, $\mathcal{T}_{i} \equiv q_i / \Tilde{\omega}_{i} \sim \order{1}$~\cite{PhysRevLett.94.221101, FlanaganHinderer1}. Over long time scales, we can expand the orbital phases $\Phi_i$ as~\cite{FlanaganHinderer1}
\begin{equation}\label{adiabaticPhase}
    \Phi_i(\tau, \eta) = \frac{1}{\eta} \Phi_i^{(0)} (\tau, \eta) + \frac{1}{\sqrt{\eta}} \Phi_i^{\textrm{res}} (\tau, \eta) + \Phi_i^{(1)} (\tau, \eta) + \mathcal{O}({\eta}),
\end{equation}
where $\Phi_i^{(0)}$ depends on the dissipative first-order self-force,  $\Phi_i^{\textrm{res}}$ represents the corrections due to orbital resonances, and $\Phi_i^{(1)}$ denotes post-adiabatic effects, i.e., the conservative first-order and the dissipative second-order self-force. The leading order correction to the orbital motion contains only $\Phi_i^{(0)}$~\cite{PhysRevLett.94.221101}, and neglects both $\Phi_i^{\textrm{res}}$ and $\Phi_i^{(1)}$. This assumption is equivalent to~\cite{FlanaganHinderer1}
\begin{itemize}
    \item truncating Eqs.~(\ref{angle})-(\ref{action}) to leading order in $\eta$, 
    \item dropping the forcing term $g_{\alpha}^{(1)}$ in Eq.~(\ref{angle}),
    \item and replacing the term $G_{i}^{(1)}$ in Eq.~(\ref{action}) with its time average $\langle \dot J_i \rangle$ (see, for instance, Eq. (3) in Ref.~\cite{SperiGair}).
\end{itemize}

We use second-order post-Newtonian (PN) formulae augmented by additional corrections from fits of Teukolsky-based inspirals to compute the adiabatic fluxes, as provided in Ref.~\cite{Gair}. In the adiabatic approximation, the inspiral is approximated as a flow through a sequence of geodesic orbits~\cite{PhysRevLett.94.221101}, each one with different values of the integrals of motion, and only the dissipative parts of the self-force are included~\cite{FlanaganHughes}. We neglect the backreaction of the inspiral on the mass $M$ and the spin $a$ of the central massive BH, which gradually evolve due to absorption of gravitational radiation crossing the event horizon (see Eq. (2.47c) in Ref.~\cite{FlanaganHinderer1}). The time scale for these parameters to change by a factor of $\order{1}$ is $\simeq M/\eta^2$, thus introducing corrections to the inspiral at the first post-adiabatic order~\cite{FlanaganHinderer1}.

\subsection{Comparing first-order NK integration with second-order methods}\label{sec:method_comparison}

Throughout the paper, we will employ the standard methodology introduced in Ref.~\cite{Gair,Glampedakis:2005cf} to evolve EMRIs with the first-order NK kludge scheme, augmented as proposed in Ref.~\cite{SperiGair} to include the effects of orbital resonances. In what follows we systematically compare it with the two second-order integration methods discussed in depth in Apps.~\ref{sec::NK} and ~\ref{sec::coupled}, and assess their robustness as alternative methods for EMRI evolutions.

\begin{figure}[t]
\centering
\includegraphics[scale=0.40]{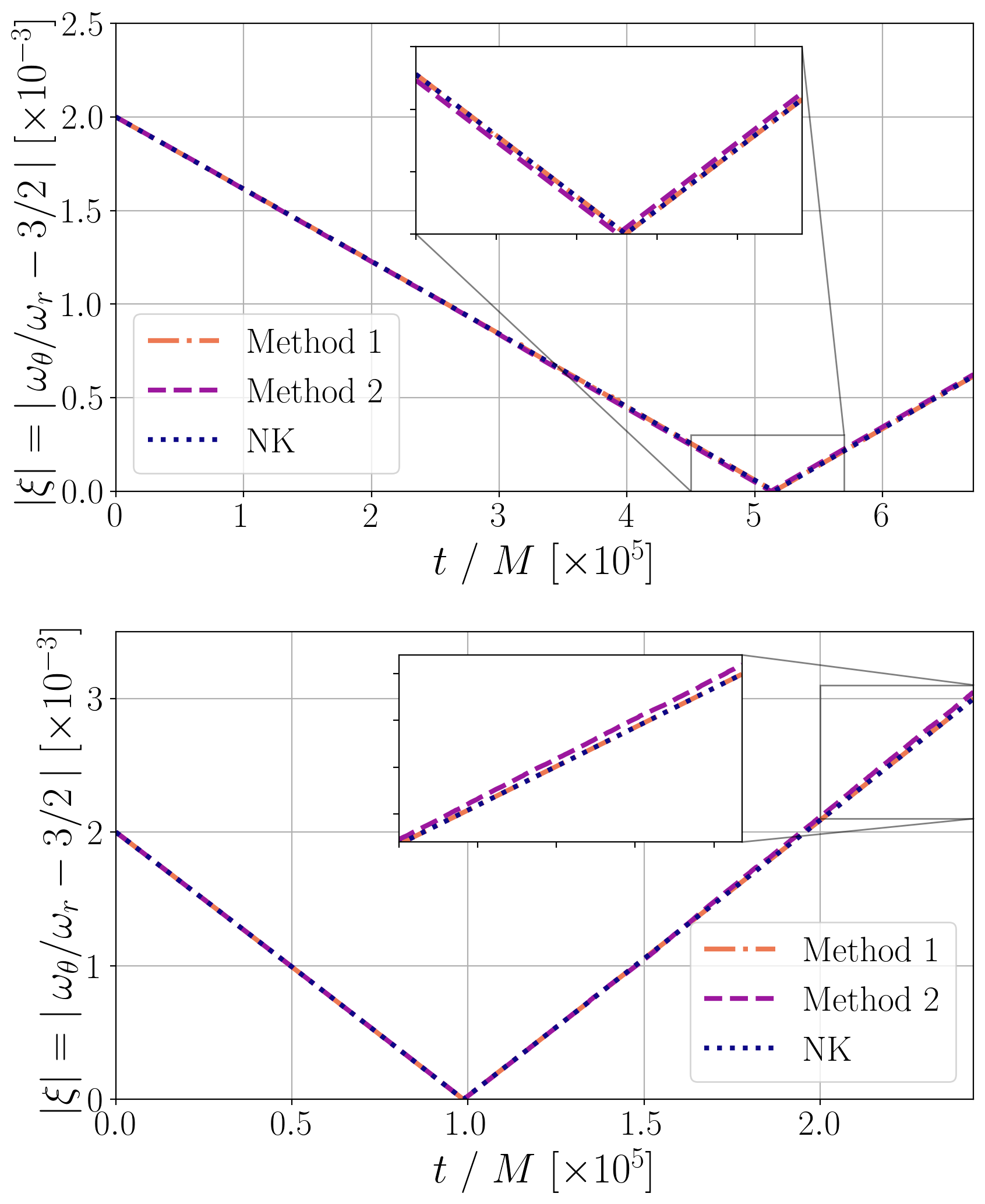} 
\caption{Example of an adiabatic inspiral of two EMRI systems with initial parameters ($\eta$ = $3 \times 10^{-6}$, $M$ = $10^6 M_{\odot}$, $a$ = $0.95$, $p/M$ = $8.49$, $e$ = $0.7$, $\iota$ = $1.2$) and ($\eta$ = $10^{-5}$, $M$ = $10^6 M_{\odot}$, $a$ = $0.95$, $p/M$ = $8.91$, $e$ = $0.1$, $\iota$ = $1.3$). The total integration times are T$_{1\:{\rm max}} \approx 0.7 \times 10^6 M$ and T$_{2\:{\rm max}} \approx 0.25 \times 10^6 M$, respectively. We plot the evolution of $|\xi| = | \omega_{\theta} / {\omega_r} - 1.5 |$, which tracks how far the system is from the 3:2 resonance, and compare the first-order NK model with the two second-order integration methods discussed in Apps.~\ref{sec::NK} and ~\ref{sec::coupled}. With ``Method 2,'' the numerical error grows with time and the solution deviates from the NK model (see insets). With ``Method 1,'' the error does not grow with time: this method is more robust with respect to ``Method 2'' over the whole evolution.}
\label{coupled_vs_carter_dissipation}
\end{figure}

We consider two EMRI systems with initial conditions $r(0) = r_p \sim 5M$ and $8M$, respectively, while $\theta(0) = \pi / 2$, $\varphi(0) = 0$. The total integration times are T$_{1 \: \textnormal{max}} \approx 0.7 \times 10^6 M$ and T$_{2 \: \textnormal{max}} \approx 0.25 \times 10^6 M$, respectively. In Fig.~\ref{coupled_vs_carter_dissipation} we plot the evolution of $|\xi|$, which, in this case, tracks how far the system is from the $3:2$ resonance. 

With ``Method 2,'' the numerical error grows with time and it is larger for $\mathcal Q$, compared to the ones in ($\mathcal E$, $\mathcal L_z$, $\lVert u \rVert$), by at least an order of magnitude. When the error in $\mathcal Q$ is $\gtrsim 10^{-2}$, we find a significant deviation from the NK solution, at least for the EMRI systems considered here. We keep the drift in the integrals of motion within around $[10^{-8}, \: 10^{-5}]$. ``Method 1,'' on the other side, has an error that does not grow with time and the maximum values are smaller than $10^{-14}$. This occurs because we re-initialize the velocities at the beginning of each step of the inspiral, and then use the invariants to constraint the trajectory during each integration time step. Therefore, the second-order ``Method 1'' solver is more robust with respect to ``Method 2'' over the whole evolution. To the best of our knowledge, this is the first time a systematic comparison between the three aforementioned techniques is investigated. 

\section{An Effective Resonance Model}
\label{sec::ERM}

Since EMRIs are expected to complete hundreds of thousands of orbits in the strong gravitational field regime, and their orbital evolution occurs at a much slower pace compared to symmetric binaries, new phenomena—such as transient orbital resonances—are expected to influence GW generation and could therefore be detectable. Resonances are expected to impact the detection, characterization, and parameter estimation of EMRI systems.

A transient orbital resonance occurs when the ratio of the polar frequency $(\omega_\theta)$ and the radial frequency $(\omega_r)$ is commensurate~\cite{Contopoulos}, i.e., $l^*\omega_{r_0}+m^*\omega_{\theta_0} = 0$ at $t=t_0$, with $l^*$, $m^*$ $\in \mathbb{Z}$. One of the first analyses regarding the effect of orbital resonances in EMRIs was performed in Ref.~\cite{FlanaganHinderer2}. When the system enters the resonance, the evolution of the integrals of motion gains a sudden contribution, a ``kick,'' due to additional contributions to the adiabatic fluxes, and the trajectory deviates from the adiabatic one. This leads to an overall dephasing in the emitted GWs. Since the leading-order dissipative component of the self-force alone is insufficient to model resonant orbits, one must either conduct theoretically intensive GSF computations or find a way to enhance the NK scheme. The latter approach was pursued recently in Ref.~\cite{SperiGair}, where a phenomenological impulse function was added to the PN fluxes of the NK scheme. Near resonances, an impulse function is activated, adding extra terms to the PN fluxes from the conservative part of the self-force, and turned off away from resonances.

To activate the resonance effects when the orbit is close to resonance, we expand the threshold function 
\begin{equation}
\label{Eq:ThresholdFunction}
\xi = (\omega_{\theta}/\omega_r) - \mathscr{R}
\end{equation}
around $t_0$, where $\mathscr{R}$ denotes the resonance value (e.g., $\mathscr{R}=1.5$ for the $3:2$ resonance). This expansion leads to $\xi^* = - \pi/(t_\textnormal{res}\, \omega_{r 0})$ when $t_\textrm{start}=t_0-t_\textrm{res}/2$, where $t_\textrm{res}$ is the total duration of the resonance in coordinate time $t$~\cite{SperiGair}. Assuming that $t_\textrm{res}$ and $\omega_{r0}$ are approximately constant around $t_0$, we can calculate $\xi$ and $\xi^*$ and trigger the start of the resonance when the starting condition $|\xi^*| > |\xi|$ is satisfied. Therefore, if $|\xi^*| > |\xi|$ and the orbit is non-circular and non-equatorial, we modify the adiabatic fluxes as~\cite{SperiGair}
\begin{equation}
    \label{Eq:FluxMod}
    \frac{dJ_i}{dt}=f_\mathrm{NK}\left[1+\mathcal{C}_i\,w(t)\right],
\end{equation}
where $f_\mathrm{NK}$ are the standard PN fluxes from Ref.~\cite{Gair}, $w(t)$ is the impulse function~\cite{SperiGair}\footnote{We find that the normalization factor of Eq.~(14) in Ref.~\cite{SperiGair} requires to compute the integral in $[-1/2, 1/2]$, i.e., where the auxiliary variable $x$ is defined.}
\begin{equation}
    w(t) = 
\begin{cases}
  \frac{1 +\cos \left[4 \pi \left(\frac{t - t_0}{t_\textnormal{res}}\right)^2\right]}{\int _{-1/2} ^{+1/2} 1 + \cos \left[4 \pi x^2\right] \, dx}
  &\quad t \in[t_\textrm{start}, t_\textrm{start} +t_\textrm{res}],\\
  0 &\quad \text{elsewhere},
\end{cases}
\end{equation}
and the resonance strength is parameterized by the resonance coefficients, $\mathcal{C}_i = (\mathcal{C}_{\mathcal{E}}, \mathcal{C}_{\mathcal{L}_z}, \mathcal{C}_{\mathcal{Q}})$. These coefficients approximate the amplitude of the augmented fluxes (the ``kick'') for the entire resonance duration, which can be roughly defined as~\cite{Ruangsri:2013hra,SperiGair}
\begin{equation}
    \label{tRes}
    t_\textrm{res}=\sqrt{\frac{4\pi}{|l^*\dot{\omega}_{r_0}+m^*\dot{\omega}_{\theta_0}|}}.
\end{equation}
Given the chosen functional form of the impulse function, the transition between the resonance regime and the adiabatic regime is smooth rather than sharp. Hence, $t_\textrm{res}$ is an arbitrary definition up to overall factors, in accordance with Ref.~\cite{SperiGair}. For specific orbits passing through different resonances, the values for $\mathcal{C}_i$ were computed in Ref.~\cite{FlanaganHughes}, with the largest values being of $\order{10^{-2}}$. 

\begin{table*}[t]
\scriptsize
\renewcommand\arraystretch{1.2}
\setlength{\tabcolsep}{6pt}
\begin{tabular}{l|c|c|c|c|c|c|c|c|c}
Resonance & $e$ & $\iota$ & $p / \textnormal{M}$ & t$_{\textnormal{res}}$ [d] & T$_{\textnormal{max}}$ [d] & $|\mathcal{C}_{\mathcal{E}}|$ & $|\mathcal{C}_{\mathcal{L}_z}|$ & $|\mathcal{C}_{\mathcal{Q}}|$ & $|\mathcal{C}_{\textnormal{tot}}|$ \\ 
\hline
\hline
Case (i) \\
\hline
4:3 & 0.30 & 0.35 & 7.46 & 5.7 & 362 & 0.00001 & 0.00001 & 0.00006 & 0.00008 \\
3:2 & 0.30 & 0.35 & 5.36 & 2.3 & 84 & 0.00102 & 0.00067 & 0.00208 & 0.00377 \\
2:1 & 0.30 & 0.35 & 3.59 & 0.7 & 10 & 0.00131 & 0.00179 & 0.00046 & 0.00356 \\
3:1 & 0.30 & 0.35 & 2.92 & 0.2 & 1 & 0.00059 & 0.00070 & $\textbf{0.00310}$ & 0.00439 \\
\hline
Case (ii) \\
\hline
4:3 & 0.30 & 1.22 & 11.36 & 14.6 & 365 &  0.00003 & 0.00004 & 0.00002 & 0.00009  \\
3:2 & 0.30 & 1.22 & 8.67 & 6.5 & 365 & 0.00303 & 0.00123 & 0.00123 & 0.00549 \\
2:1 & 0.30 & 1.22 & 6.18 & 2.0 & 55 & 0.00004 & 0.00080 & 0.00002 & 0.00086 \\
3:1 & 0.30 & 1.22 & 5.07 & 0.5 & 5 & 0.00008 & 0.00024 & 0.00033 & 0.00065 \\
\hline
Case (iii) \\
\hline
4:3 & 0.70 & 0.35 & 7.58 & 11.4 & 365 & 0.00001 & 0.00002 & 0.00023 & 0.00026 \\
3:2 & 0.70 & 0.35 & 5.50 & 4.7 & 144 & 0.00127 & 0.00078 & 0.00210 & 0.00415  \\
2:1 & 0.70 & 0.35 & 3.82 & 1.5 & 30 & 0.00167 & 0.00067 & 0.00357 & 0.00591 \\
3:1 & 0.70 & 0.35 & 3.28 & 0.5 & 17 & 0.00026 & 0.00009 & 0.00035 & 0.00070 \\
\hline
Case (iv) \\
\hline
4:3 & 0.70 & 1.22 & 11.47 & 28.2 & 365 & 0.00047 & $\textbf{0.00060}$ & 0.00002 & 0.00109  \\
3:2 & 0.70 & 1.22 & 8.80 & 12.4 & 365 & $\textbf{0.01030}$ & 0.00489 & 0.00261 & 0.01780  \\
2:1 & 0.70 & 1.22 & 6.36 & 3.7 & 88 & $\textbf{0.00662}$ & 0.00270 & 0.00494 & 0.01426  \\
3:1 & 0.70 & 1.22 & 5.41 & 0.8 & 5 & 0.00125 & 0.00027 & 0.00126 & 0.00278 \\
\bottomrule
\end{tabular}
\caption{Parameter-space location and resonance coefficients for the investigated orbital resonances, with $a = 0.9$, $\eta = 10^{-5}$ and $M = 10^6 M_{\odot}$. We consider four orbital configurations: case (i) $e=0.3$ and $\iota=0.35$, case (ii) $e=0.3$ and $\iota=1.22$, case (iii) $e=0.7$ and $\iota=0.35$, and case (iv) $e=0.7$ and $\iota=1.22$. The EMRI evolution begins close to the resonance, by setting the initial value of $p/M$ such that the threshold function (Eq.~\ref{Eq:ThresholdFunction}) is $\xi_0 = - (0.002, 0.002, 0.02, 0.05)$ for the $4:3$, $3:2$, $2:1$ and $3:1$ resonance, respectively. We fix the observation time to $T_{max} = 1$ year, and terminate the evolution earlier if the secondary BH reaches the separatrix. The resonance durations, $t_{\textnormal{res}}$, are computed using Eq.~(\ref{tRes}). For the resonance coefficients, we use the values provided in Ref.~\cite{FlanaganHughes}. For each resonance, the maximum $\mathcal{C}$ is highlighted in bold. In the last column, $|\mathcal{C}_{\textnormal{tot}}|$ is the sum of $|\mathcal{C}_{\mathcal{E}}|$, $|\mathcal{C}_{\mathcal{L}_z}|$ and $|\mathcal{C}_{\mathcal{Q}}|$.}
\label{Table_1}
\end{table*}

For ease of notation, from now how we will write $J\equiv J_i$ and $\mathcal{C}=\mathcal{C}_i$ whenever dropping the index is not ambiguous. In what follows we will briefly revisit the dominant $3:2$ resonance, analyzed in Ref.~\cite{SperiGair}, and study the orbital dephasing as well as the waveform mismatch between two EMRIs: one that does not take into account the contribution of the $3:2$ resonance, and one that does it through Eq. \eqref{Eq:FluxMod}. After revisiting the results of Ref.~\cite{SperiGair}, in Sec.~\ref{sec::cumulative_effect} we will extend the ERM to subdominant resonances and assess if their cumulative effect leads to further dephasing with respect to EMRIs where only the dominant resonance is taken into account, as well as allowing the resonance coefficients to vary for each integral of motion. 

The resonant coefficients for all considered resonances in this work are presented in Tab.~\ref{Table_1}. These coefficients represent the peak-to-trough variation in the fluxes, as provided in Ref.~\cite{FlanaganHughes}. The exact magnitude of these fluxes at resonance depends on the phase $\chi_0$, which defines the value of the orbit’s $\theta$ angle when it reaches the periapsis. Thus, for a given inspiral, the fluxes will take a value between $-|\mathcal{C}_i|$ and $|\mathcal{C}_i|$.

\subsection{Revisiting the $3:2$ resonance}

Let us apply the ERM to the $3:2$ resonance. Following Ref.~\cite{SperiGair}, we will begin by setting $\mathcal{C} = -(0.01, 0.01,$ $0.01)$ as the reference case to benchmark our methods of integrating EMRIs, and study the effects of orbital resonances across various EMRI systems. To compute the adiabatic fluxes, we use Eqs. (44), (45), and (60) provided in Ref.~\cite{Gair}, respectively for ($\mathcal E, \mathcal L_z, \mathcal Q$).

\begin{figure}[]
\centering
\includegraphics[scale = 0.51]{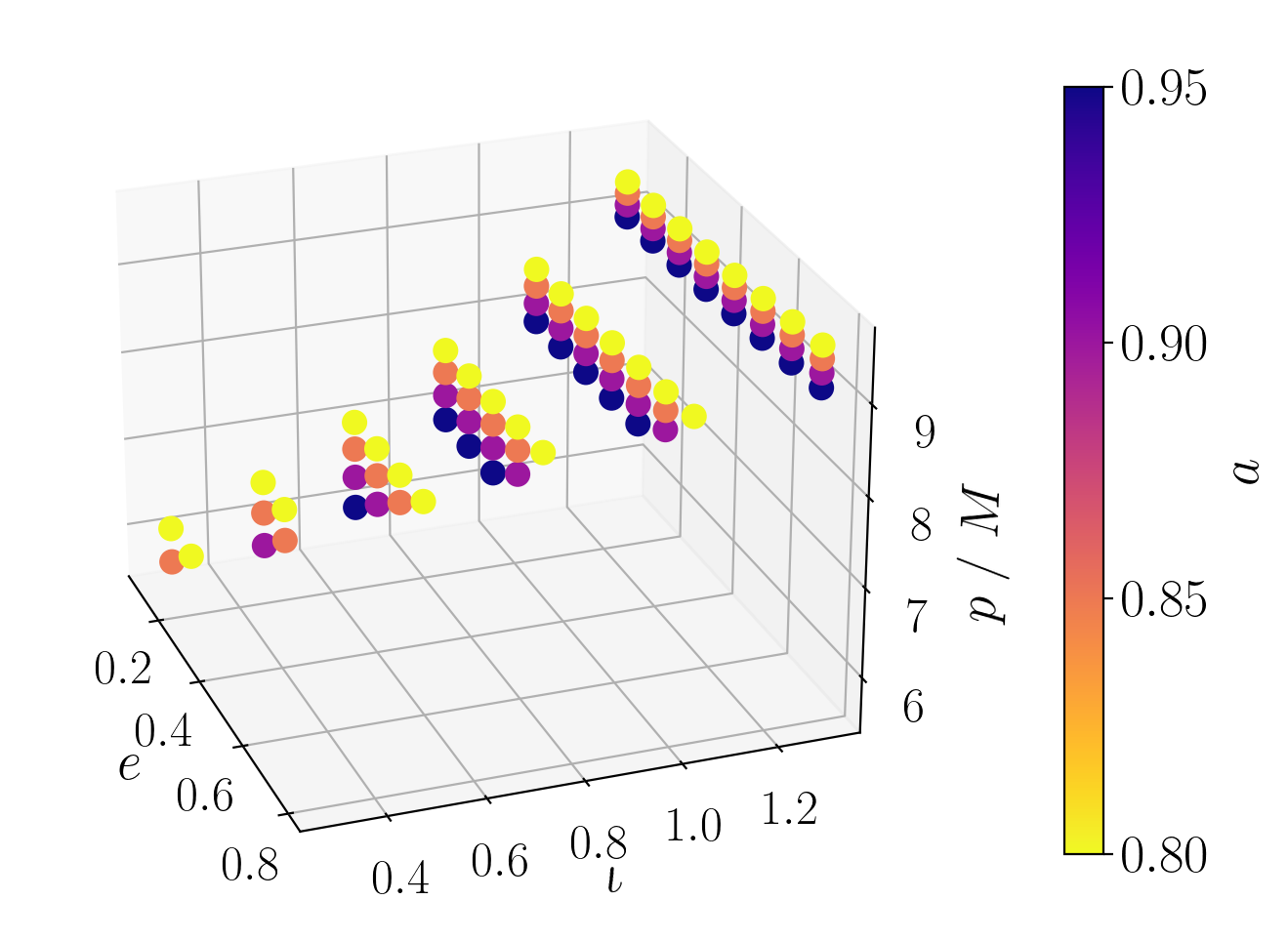}
\caption{The location of the $3:2$ resonance in the orbital parameter space. For all the cases, we start the EMRI evolution close to resonance by setting the initial value of $p/M$ such that $\omega_{\theta} / \omega_r = 1.498$, that is  $\xi = -0.002$. We consider only systems with an initial periapsis $r_p > 5M$, per the limitations of the NK model imposed by the PN approximation~\cite{Kludge}.}
\label{param_space}
\end{figure}

Following Ref.~\cite{SperiGair} we consider several EMRI systems with varying BH spin parameter $a \in [0.8, 0.95]$, eccentricity $e \in [0.1, 0.8]$, and inclination angle $\iota \in [0.3, 1.3]$ with steps $\Delta a = 0.05$, $\Delta e = 0.1$, $\Delta \iota = 0.2$. We focus on prograde orbits, as we expect the largest dephasing effects around $\iota = 1.2$~\cite{SperiGair}. Since, a priori, we do not know the exact location of the resonances, we investigate a wide range of initial conditions, with $p \in [5.5 - 10.0]M$ and a resolution $\Delta p = 10^{-4}$. The location of the $3:2$ orbital resonances in the phase-space ($p/M$, $e$, $\iota$, $a$) is shown in Fig.~\ref{param_space}. 

Once the resonance is found, the EMRI system is evolved by setting the initial value of the semilatus rectum such that $\omega_{\theta} / \omega_r = 1.498$~\cite{SperiGair}, so that we start the evolution near the resonance.  We consider only EMRIs with an initial periapsis radius $r_p \gtrsim 5M$, per the limitations of the NK model, which includes some PN approximations~\cite{Kludge}.

The crossing through the resonance causes the systems to deviate from the adiabatic trajectory, as modeled in Eq.~(\ref{Eq:FluxMod}). Since the change in the integrals of motion accumulated over the resonance is 
\begin{equation}
    \Delta J = \mathcal{C} \: \dot{J} \: t_{res},
\end{equation}
and $J / \dot{J} \sim 1 / \eta$, the flux changes scale as 

\begin{equation}    
    \frac{\Delta J}{J_r \: \eta} \sim \mathcal{C} \: t_{res}.
\end{equation}

We compute the relative changes in ($\mathcal E, \mathcal L_z, \mathcal Q$) from the start to the end of the $3:2$ resonance for the EMRI system shown in Fig.~1 of Ref.~\cite{SperiGair}: we find agreement within a few percent.

\subsubsection{Relationship between orbital dephasing, observation time, resonance duration and resonance coefficients}

Since the GW phase of an EMRI system consists of a complex combination of the evolution of the phases ($\psi$, $\chi$, $\varphi$)\footnote{See Eqs. ~\eqref{parameterizationR} and ~\eqref{parameterizationTheta} for the definitions of the phases $\psi$ and $\chi$, respectively. The phase $\varphi$ is the azimuthal angle.}, it is illustrative to study the change of each phase independently over time. That is, we are interested in studying the orbital dephasing
\begin{equation}
    \Delta \Phi_i = \frac{(|\Delta \psi|, |\Delta  \chi|, |\Delta  \varphi|)}{2 \pi}.
\end{equation}
This dephasing refers to the number of cycles by which the resonant evolution diverges from the adiabatic evolution at the end of the observation time. In Ref.~\cite{SperiGair} it was argued that $\Delta \Phi_i$ scales as
\begin{equation}\label{dephaseScaling}
    \Delta \Phi_i \sim \Delta \dot{\omega}(\mathcal{C}, t_{\textnormal{res}}) \: T^2,
\end{equation}
where $\Delta \dot{\omega}$ is the change in the frequency derivative that is accumulated over the resonance, and $T$ is the time between the start of the resonance and the end of the observation. We will now, for the single $3:2$ resonance, verify that scaling and obtain a relationship between orbital dephasing, observation time, resonance duration and resonance coefficients ($\mathcal{C}$). 

For this analysis, we fix the observation time to $T_{\textnormal{max}} = 1$ year, and terminate the evolution earlier if the secondary reaches the periastron limit of $r_p = 5M$. We set the mass ratio to $\eta = 10^{-5}$ and the mass of the central BH to $M = 10^6 M_{\odot}$. The initial phases are $\psi(0) = 0$, $\chi(0) = \pi / 2$, $\varphi(0) = 0$. 

\begin{figure}[]
\centering
\includegraphics[scale=0.349]{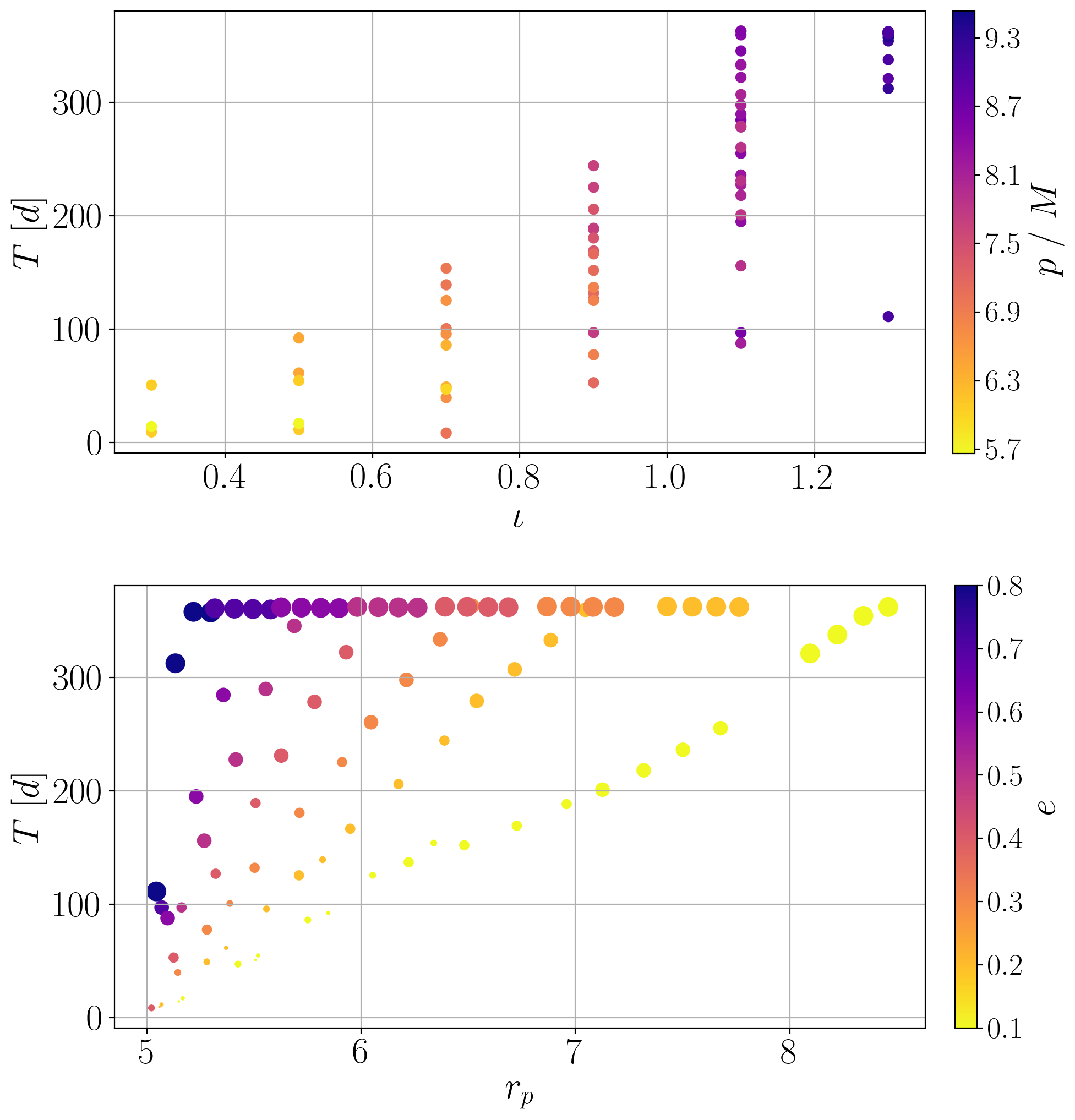}
\caption{Observation times for the EMRI systems of Fig.~\ref{param_space}. We set $T_\textrm{max} = 1$ year and terminate the evolution earlier if the secondary BH reaches $r_p = 5M$, per the limitations of the NK model, which includes some PN approximations~\cite{Kludge}. We set the mass ratio to $\eta = 10^{-5}$ and the mass of the primary BH to $M = 10^6 M_{\odot}$. Here, $T$ is the time, in days, between the start of the resonance and the end of the observation. \textit{Top panel}: $T$ as a function of the orbital inclination angle and the initial semi-latus rectum. EMRI systems with higher $\iota$ obtain higher $T$. \textit{Bottom panel}: The time between the start of the resonance and the end of the observation, $T$, as a function of the initial periapsis and the orbital eccentricity. The marker size increases with the orbital inclination angle. $T$ increases linearly with $r_p$; systems with higher eccentricity $e$ show a higher slope.}
\label{dephasing_tmax}
\end{figure}

As shown in Fig.~\ref{dephasing_tmax}, in the case of the $3:2$ resonance, for most of the orbital configurations, especially those with $\iota \lesssim 1$, the location of the resonance in the parameter space is such that $T$ is much less than one year. Overall, we find that:
\begin{itemize}
    \item systems with higher $\iota$ have longer $T$,
    \item $T$ increases linearly with the initial periapsis radius, and systems with higher orbital eccentricity show higher slopes.
\end{itemize}

\begin{figure}
\centering
\includegraphics[scale=0.5]{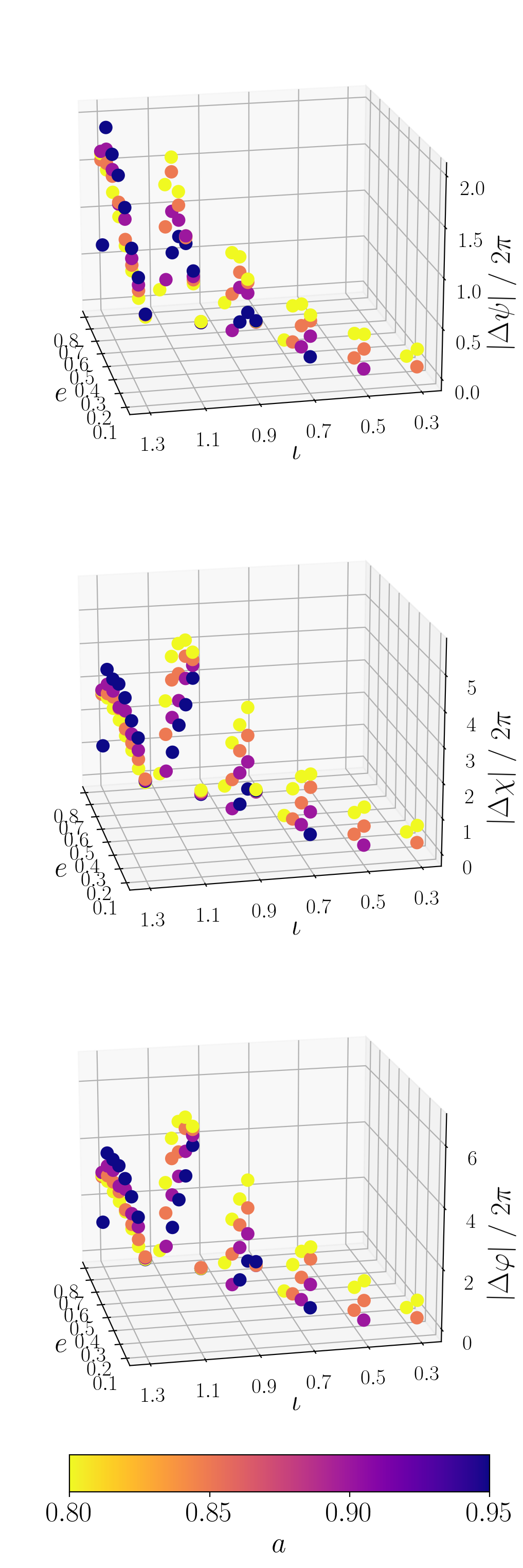}
\caption{Orbital dephasing between ERM ($3 : 2$ resonance) and NK orbits for the EMRI systems of Fig.~\ref{param_space}, with parameters $\eta = 10^{-5}$, $M = 10^6 M_{\odot}$ and resonance coefficients $\mathcal{C} = - (0.01, 0.01, 0.01)$. Since the phases oscillate quickly, we take the average approximately over the last $1000$ s~\cite{SperiGair}. This figure should be compared with Fig. 2 in Ref.~\cite{SperiGair}. However, we noticed that the values for the $\psi$ and $\varphi$ coordinates appear to be exchanged and that the larger phase shifts reported in Ref.~\cite{SperiGair} are due to the fact that they evolved until plunge, whereas we have, conservatively, stopped the evolutions when the systems arrives to $r_{p}=5\,M$.}
\label{dephasing_1}
\end{figure}

In Fig.~\ref{dephasing_1} we show the orbital dephasing between ERM and NK orbits, where we find that:
\begin{itemize}
    \item the maximum values amount to $|\Delta \Phi_i| = (2.0, 5.6, 6.6)$ cycles for ($\psi$, $\chi$, $\varphi$), respectively,
    \item the largest dephasings occur around $\iota = 1.1$ for $\chi$ and $\varphi$, and around $\iota = 1.3$ for $\psi$,
    \item there is a slight dependence on BH spin of the primary. When $\iota = 1.3$, systems with higher $a$ show higher $\Delta \Phi_i$. When $\iota \leq 1.1$, larger phase shifts occur for lower spin configurations.
\end{itemize}

\begin{figure*}[t]
\centering
\includegraphics[scale=0.39]{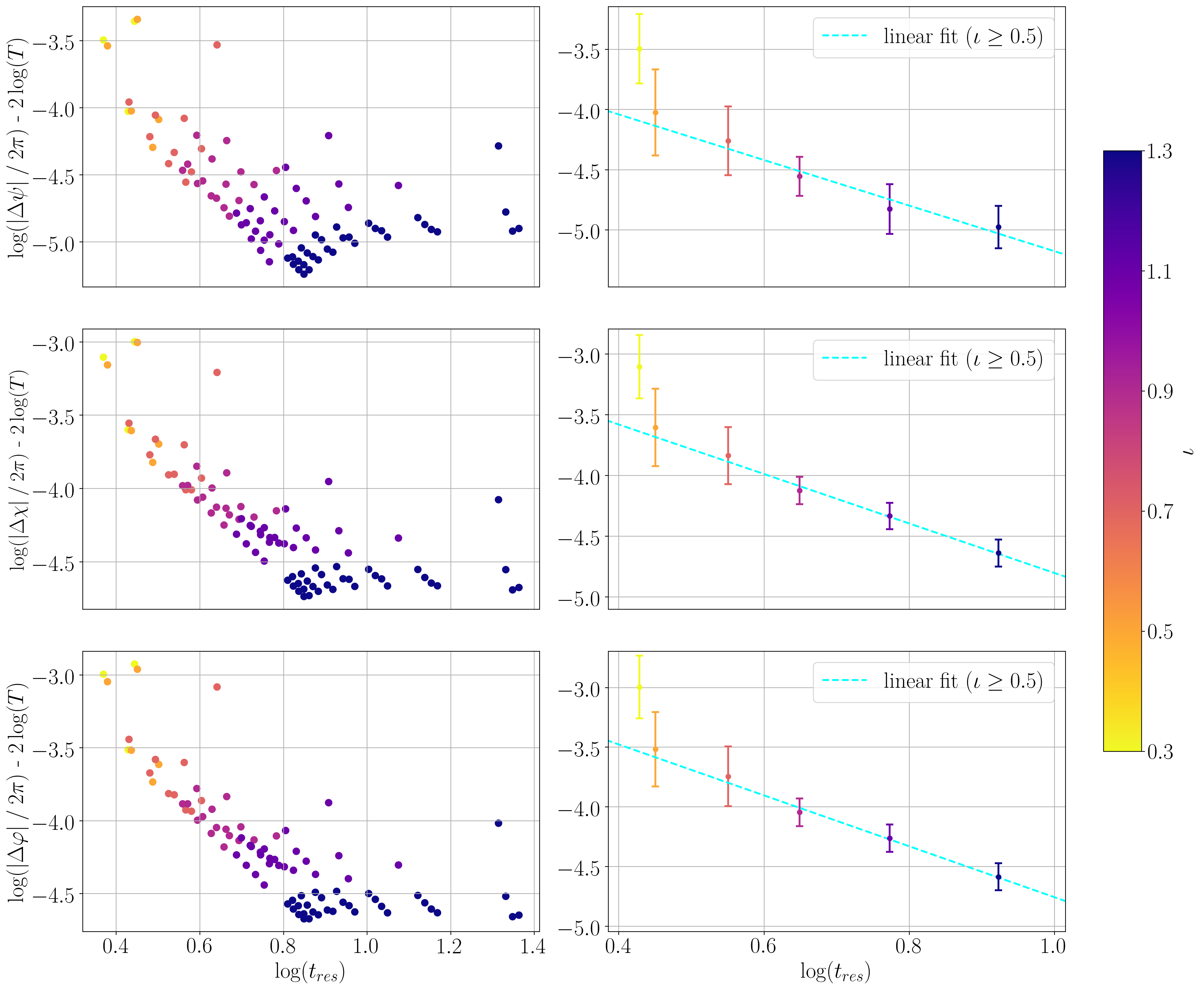}
\caption{Relation between orbital dephasing, observation time and resonance duration in the case of the $3 : 2$ resonance, for the EMRI systems of Fig.~\ref{param_space} with parameters $\eta = 10^{-5}$, $M = 10^6 M_{\odot}$ and resonance coefficients $\mathcal{C} = -(0.01, 0.01, 0.01)$. We group the systems with the same $\iota$ and take the median and the standard deviation of $\Delta \Phi_i$. We exclude EMRI evolutions with $\iota = 0.3$, as they exhibit small dephasing effects, and fit the data using Eq.~(\ref{fit1Eq}).}
\label{dephasing_2}
\end{figure*}

Now, let us quantify the relation between orbital dephasing, observation time, resonance duration and resonance coefficients resulting from Eq.~(\ref{dephaseScaling}). We assume the functional form of $\Delta \dot{\omega}\,(\mathcal{C}, t_{\textnormal{res}})$ to be a power law, i.e.,
\begin{equation}\label{ansatz}
   \Delta \dot{\omega}\,(\mathcal{C}, t_{\textnormal{res}}) \sim \mathcal{C}^m \: t_{\textnormal{res}}^n,
\end{equation}
where we expect $m = 1$ ~\cite{SperiGair} and $n = -2$, with the latter value resulting from a dimensional analysis. Starting from the results in Fig.~\ref{dephasing_1}, we group the systems with same $\iota$ and take the median and the standard deviation of $\Delta \Phi_i$. We fit these values to verify
\begin{equation}\label{fit1Eq}
   \log{\Delta \Phi_i} - 2 \log{T} = n_i \log{t_{\textnormal{res}}} + c_{1, \:i},
\end{equation}
where 
\begin{equation}\label{fit2Eq}
   c_{1, \:i} = m_i \log{|\mathcal{C}|} + c_{2, \:i},
\end{equation}
and the index $i$ labels the coefficient for the phases $\psi$, $\chi$, and $\varphi$. The results are illustrated in Fig.~\ref{dephasing_2}, where we find that EMRI systems with $\iota = 0.3$ do not follow the linear relation~(\ref{fit1Eq}). This is due to low-inclined orbits exhibiting small dephasing effects, as shown in Fig.~\ref{dephasing_1}. We then fit the data for EMRIs with $0.5 \leq \iota \leq 1.3$ obtaining
\begin{align*}
    & n_i|_{\textnormal{reference} \: \textnormal{case} \:} = - (1.89, 2.04, 2.13) \pm (0.40, 0.21, 0.21), \\
    & c_{1, \:i}|_{\textnormal{reference} \: \textnormal{case} \:} = - (3.28, 2.77, 2.62) \pm (0.22, 0.12, 0.13).
\end{align*}

So far, we have only considered $|\mathcal{C}| = 0.01$, but since the orbital dephasing scales with the resonance coefficients, we will now evolve the same EMRI systems and vary $|\mathcal{C}| \in [1.0 \times 10^{-4}, 5.0 \times 10^{-1}]$\footnote{The lower bound is consistent with Fig. 3 of Ref.~\cite{SperiGair}. To explore larger effects on the fluxes, we choose as upper bound the maximum value of $|\mathcal{C}|$ provided in Ref.~\cite{FlanaganHughes}, increased by one order of magnitude.} and repeat the above fit procedure. In Tab.~\ref{fit1Table} we list the results of these fits.

\begin{figure}[b]
\centering
\includegraphics[scale=0.32]{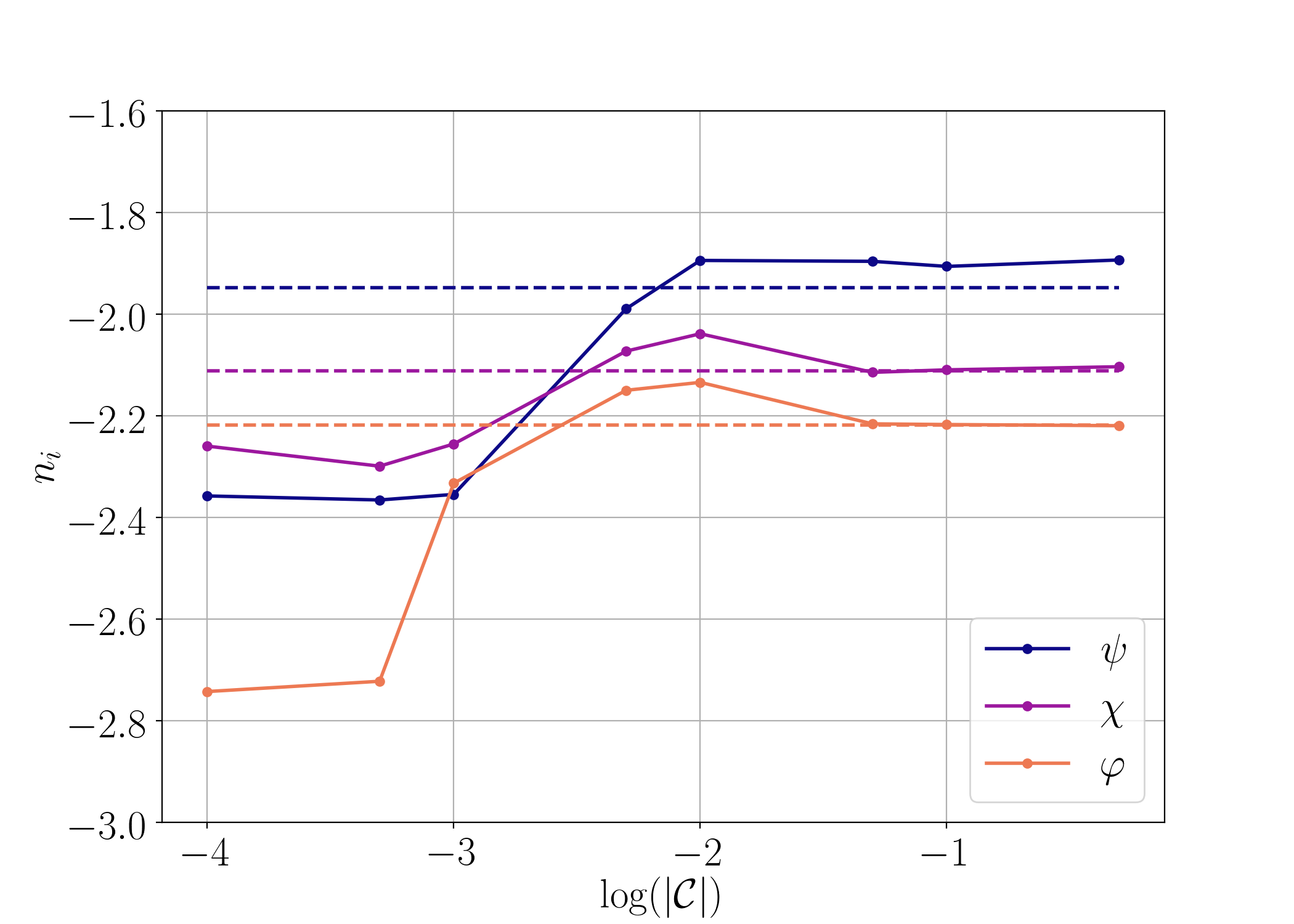}
\caption[Coefficients $n_i$ as a function of $|\mathcal{C}|$]{The coefficients $n_i$, the power for the growth of the orbital dephasings, in Table~\ref{fit1Table} increase for stronger resonances and, only when $|\mathcal{C}| \gtrsim 0.01$ they converge to the median values. The lowest values for $\varphi$ are also the ones with the highest standard deviations. These results support the scaling presented in Eq.~\eqref{dephaseScalingFinal}.}
\label{fit_median}
\end{figure}

\begin{figure}[t]
\centering
\includegraphics[scale=0.4]{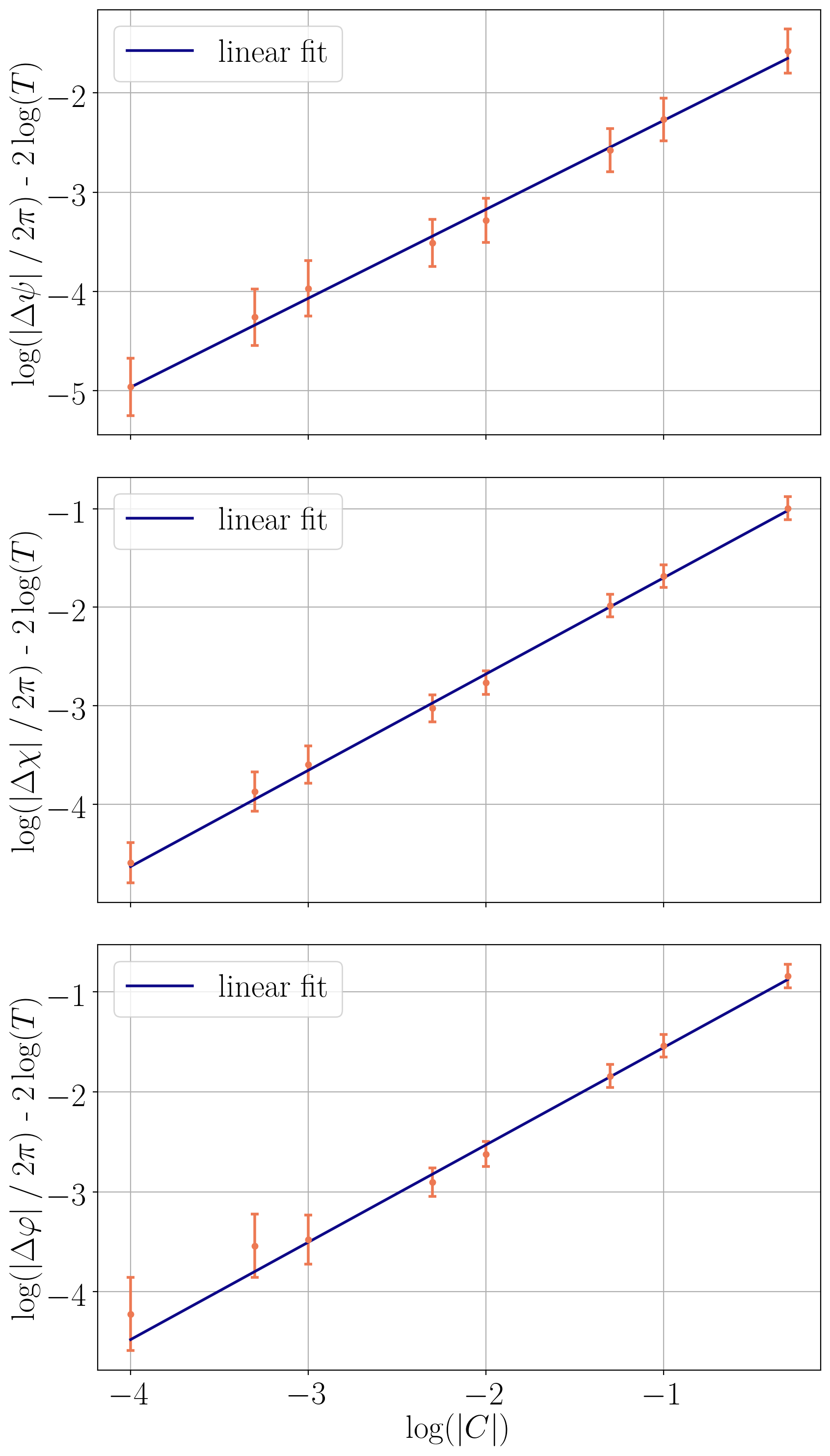}
\caption{Relation between orbital dephasing and resonance coefficients for the EMRI systems of Fig.~\ref{param_space}, with parameters $\eta = 10^{-5}$, $M = 10^6 M_{\odot}$. We fit $m_i$ and $c_{2, \:i}$ from Eq.~(\ref{fit1Eq}) using the data in Table~\ref{fit1Table}. The error bars indicate one standard deviation. The results are consistent with $m \sim 1$ in Eq.~\eqref{ansatz}.}
\label{dephasing_3}
\end{figure}

As shown in Fig. ~\ref{fit_median}, we found that coefficients $n_i$ increase with stronger resonances and, for $|\mathcal{C}| \gtrsim 0.01$, converge to the median values 
\begin{equation*}
    \overline{n}_i = - (1.95, 2.11, 2.22) \pm (0.42, 0.22, 0.23), 
\end{equation*}
i.e., these results are consistent with $n \approx -2$ in Eq.~\eqref{ansatz}. We use the values of $c_{1, \:i}$ in Tab.~\ref{fit1Table} to fit $m_i$ and $c_{2, \:i}$ from Eq.~(\ref{fit2Eq}), as illustrated in Fig.~\ref{dephasing_3}. We find that:
\begin{align*}
    & m_i = (0.895, 0.977, 0.974) \pm (0.006, 0.002, 0.003), \\
    & c_{2, \:i} = - (1.383, 0.725, 0.582) \pm (0.028, 0.009, 0.010),
\end{align*}
i.e., these results are consistent with $m \approx 1$ in Eq.~\eqref{ansatz}. Hence, we verify that, for systems with $0.5 \leq \iota \leq 1.3$, the dephasing scales approximately as
\begin{equation}\label{dephaseScalingFinal}
    \Delta \Phi_i \sim \mathcal{C} \: \Biggl ( \frac{T}{t_{\textnormal{res}}} \Biggr )^2. 
\end{equation}

Going back to Fig.~\ref{dephasing_1}, as the orbital inclination angle increases, $\Delta \Phi_i$ grows because of the longer observation time. The latter value converges to $T_{\textnormal{max}} = 1$ year when $\iota \geq 1.1$, as shown in Fig.~\ref{dephasing_tmax}. We then expect to see lower phase shifts when $\iota$ increases from $1.1$ to $1.3$, because in Eq.~(\ref{dephaseScalingFinal}) the resonance duration grows, with $T\approx const$. These systems, indeed, enter the $3:2$ resonance farther away from the central BH, where the magnitude of the GW fluxes is smaller ~\cite{SperiGair}.

However, the dephasing in $\psi$ does not decrease when $\iota$ increases from $1.1$ to $1.3$, as it happens for $\chi$ and $\varphi$. We notice that the phase shifts in $\psi$ are about three times lower than the ones $\chi$ and $\varphi$, and a degeneracy with the spin parameter seems to take place. 

\begin{table}[]
\tiny
\renewcommand\arraystretch{1.3}
\setlength{\tabcolsep}{2pt}
\begin{tabular}{c|c|c}
\toprule
$|\mathcal{C}|$ & $n_i$ & $c_{1, \:i}$ \\ 
\hline
\hline
0.0001 & - (2.4, 2.3, 2.7) $\pm$ (0.6, 0.4, 0.9) & -(5.0, 4.6, 4.2) $\pm$ (0.3, 0.2, 0.4) \\
0.0005 & - (2.4, 2.3, 2.7) $\pm$ (0.6, 0.4, 0.7) & -(4.3, 3.9, 3.5) $\pm$ (0.3, 0.2, 0.3) \\
0.001 & - (2.3, 2.3, 2.3) $\pm$ (0.5, 0.4, 0.5) & -(4.0, 3.6, 3.5) $\pm$ (0.3, 0.2, 0.3) \\
0.005 & - (2.0, 2.1, 2.2) $\pm$ (0.4, 0.2, 0.3) & -(3.5, 3.0, 2.9) $\pm$ (0.2, 0.1, 0.1) \\
0.01 & - (1.9, 2.0, 2.1) $\pm$ (0.4, 0.2, 0.2) & -(3.3, 2.8, 2.6) $\pm$ (0.2, 0.1, 0.1) \\
0.05 & - (1.9, 2.1, 2.2) $\pm$ (0.4, 0.2, 0.2) & -(2.6, 2.0, 1.8) $\pm$ (0.2, 0.1, 0.1) \\
0.1 & - (1.9, 2.1, 2.2) $\pm$ (0.4, 0.2, 0.2) & -(2.3, 1.7, 1.5) $\pm$ (0.2, 0.1, 0.1) \\
0.5 & - (1.9, 2.1, 2.2) $\pm$ (0.4, 0.2, 0.2) & -(1.6, 1.0, 0.8) $\pm$ (0.2, 0.1, 0.1) \\
\hline
\textbf{Median} & \textbf{- (2.0, 2.1, 2.2)} $\pm$ \textbf{(0.4, 0.2, 0.2)} & \textbf{- (3.4, 2.9, 2.8)} $\pm$ \textbf{(0.2, 0.1, 0.1)} \\
\bottomrule
\end{tabular}
\caption[Fit of the coefficients $n_i$, $c_{1, \:i}$ with different values of $|\mathcal{C}|$]{We evolve the investigated configurations for the $3:2$ resonance varying $|\mathcal{C}|$ in the range $\in [0.0001, 0.5]$. The reference case is $|\mathcal{C}| = 0.01$. For each scenario, we compute $\Delta \Phi_i$, where $i$ ranges over ($\psi$, $\chi$, $\varphi$), following the procedure described in Fig.~\ref{dephasing_2}. We fit $n_i$ and $c_{1, \:i}$ using Eq.~\eqref{fit1Eq}. The results, reported with one standard deviation error, are consistent with $n \sim -2$ in Eq.~\eqref{ansatz}.}
\label{fit1Table}
\end{table}

\subsubsection{Mismatch between waveforms}
\label{sec::mismatch}

The orbital dephasing we studied in the previous sections will manifest as an overall dephasing in the GW signals. Following Ref.~\cite{Kludge}, in this section we will compute the produced gravitational waveforms by applying the quadrupole formula in the transverse-traceless (TT) gauge to the Boyer--Lindquist coordinates of the inspiralling object as a function of coordinate time $t$. This mapping is achieved by projecting the Boyer--Lindquist coordinates onto a fictitious spherical-polar coordinate grid, where we define the corresponding Cartesian coordinate system, and assume that these new coordinates are true flat-space Cartesian coordinates. We obtain the two GW polarizations, $h_+$ and $h_{\times}$, for every EMRI system, and then use the low-frequency approximation to the LISA response function in order to approximate the waveforms as~\cite{Barack:2003fp}
\begin{equation}
    h_{\alpha}(t) = \frac{\sqrt{3}}{2} [F_{\alpha}^{+}(t) h_{+}(t) + F_{\alpha}^{\times}(t) h_{\times}(t)],
\end{equation}
where $F_{\alpha}^{(+, \times)}$ is the detector's antennae patterns, the subindex $\alpha = {I,II}$ denotes one of the two LISA channels, and the luminosity distance of the source has been included in $h_+$ and $h_{\times}$. We assume that the two data streams are uncorrelated and that the detector's noise power spectral density, $S_n(f)$, is the same for both channels. In this work, we use the LISA's sensitivity curve available within the PyCBC software package~\cite{pycbc}.

To compare resonant-crossing and non-resonant-crossing waveforms, we use the overlap $\mathcal{O}(a, b)$~\cite{Cutler:1994ys, Owen:1995tm, Moore:2014lga} which gives an estimate of how indistinguishable the signals $a$ and $b$ are. From the overlap, we compute the mismatch as $\mathcal{M} = 1 - |\mathcal{O}|$. The mismatch caused by orbital resonances should affect the detection and parameter estimation of EMRI systems~\cite{Berry}. A criterion used to assess if two waveforms are indistinguishable is $||\delta h || < 1$, where $|| \delta h ||$ is the norm of the waveform difference. The mismatch is approximately $\mathcal{M} \approx || \delta h || / (2 \rho^2)$, where $\rho$ is the signal-to-noise ratio. As in Ref.~\cite{SperiGair}, we fix $\rho = 20$ and the above criterion is satisfied when $\mathcal{M} \lesssim 10^{-3}$.

\begin{figure*}
\centering
\includegraphics[scale=0.09]{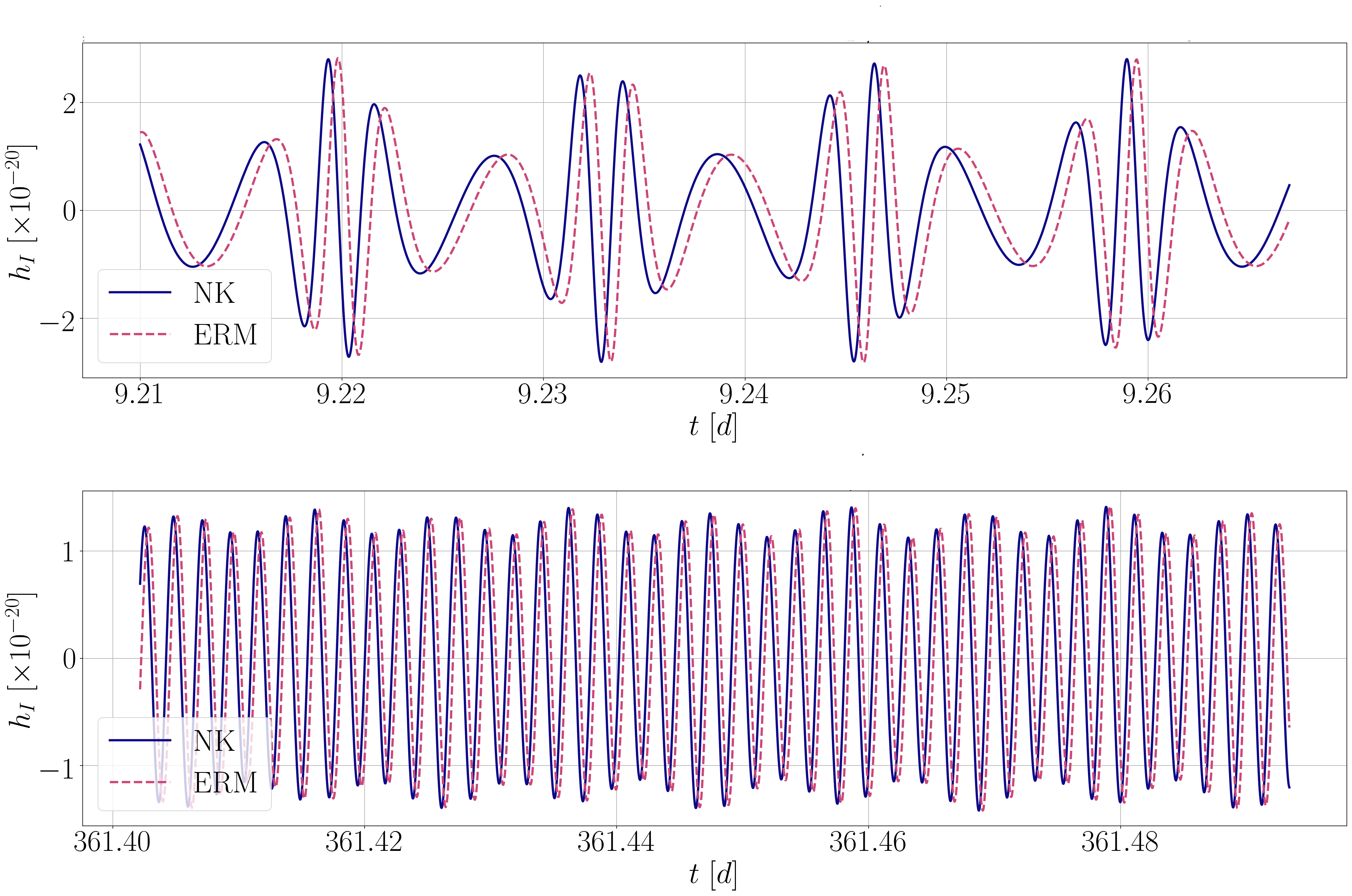} 
\caption{Gravitational waveforms at the end of the observation time for the two EMRI systems of Fig.~\ref{dephasing_1} with the lowest (EMRI$_1$) and highest (EMRI$_2$) orbital dephasing, i.e., ($a$ = $0.80$, $p/M$ = $7.03$, $e$ = $0.40$, $\iota$ = $0.70$) and ($a$ = $0.80$, $p/M$ = $8.46$, $e$ = $0.20$, $\iota$ = $1.10$), respectively. The blue solid line corresponds to an EMRI evolution without accounting for the changes to the fluxes around the resonance (NK), while the purple dashed line corresponds to an evolution with the effective resonance model (ERM). We compute the mismatch between the two waveforms finding $\mathcal{M}_1 = 2.4 \times 10^{-1}$, $\mathcal{M}_2 = 9.9 \times 10^{-1}$ at $t ={T}_{\textnormal{max}}$. While we are only presenting $h_I$, the results of the other LISA channel are similar.}
\label{GW_3}
\end{figure*}

We compute the gravitational waveforms for the lowest (EMRI$_1$) and highest (EMRI$_2$) orbital dephasing found in the EMRI systems of Fig.~\ref{dephasing_1}. By the time the systems exit the resonance, a phase difference is accumulated between the resonant and non-resonant cases. The dephasing further increases over time, as we expect from the scaling relation of Eq.~(\ref{dephaseScalingFinal}). In Fig.~\ref{GW_3}, we plot the gravitational waveforms at the end of the observation time.

We compute the mismatch both when the system exits the resonance and at the end of the observation time. We consider the time $T$ between the start of the resonance and the end of the observation, i.e., $T_1 = 9$ days for EMRI$_1$, and $T_2 = 360$ days for EMRI$_2$. For the reference case with equal resonance coefficients, $\mathcal{C} =$ $-(0.01,$ $0.01,$ $0.01)$, we find the following mismatches:
\begin{align*}
    & \mathcal{M}_1 = 2.2 \times 10^{-2}, \hspace{0.3cm} \mathcal{M}_2 = 5.2 \times 10^{-3}, \hspace{5pt} t = t_{\textnormal{start}} + t_{\textnormal{res}}, \\ 
    & \mathcal{M}_1 = 2.4 \times 10^{-1}, \hspace{0.3cm} \mathcal{M}_2 = 9.9 \times 10^{-1}, \hspace{5pt} t = {T}_{{max}}.
\end{align*}
Both at $t = t_{\textnormal{start}} + t_{\textnormal{res}}$ and $t = {T}_{\textnormal{max}}$, the mismatch is higher than the indistinguishability threshold $\mathcal{M}\lesssim 10^{-3}$. Note that naively one can expect that a system with a higher orbital dephasing will immediately imply a larger mismatch. However, at the end of the resonance, $\mathcal{M}_1 > \mathcal{M}_2$. For EMRI$_1$, despite the resonance being shorter ($4$ vs. $6$ days), it is located closer to the separatrix, where the magnitude of the GW fluxes is higher. At the end of the observation time, on the other hand, $\mathcal{M}_1 < \mathcal{M}_2$. The dephasing increases with $T$, that is much longer for EMRI$_2$ and compensates for the lower flux amplitudes. 

As reported in Ref.~\cite{SperiGair}, we can conclude that the mismatch caused by the $3:2$ orbital resonance is likely to affect the detection of the EMRIs of Fig.~\ref{param_space}. Systems with longer observation times have longer resonances located farther away from the central BH, while for lower $T_{\textnormal{max}}$ resonances are shorter and located in stronger field regions. The overall effect is a mismatch of $\order{10^{-1}}$ between resonant-crossing and non-resonant-crossing waveforms. 

\section{Subdominant resonances and multiple resonant-crossings}
\label{sec::cumulative_effect}

Let us now study the phenomenology of transient orbital resonances other than the dominant $3:2$, and further extend the previous analysis to the cumulative effects of crossing \emph{multiple} resonances in a single inspiral. If the order of the resonance is high, i.e., the number of cycles in $r$ with respect to the cycles in $\theta$ is high enough, an average over the $2$-torus effectively occurs~\cite{SperiGair}. On the other hand, for low-order resonances, the phase-space trajectories do not fill out the phase-space $(r, \theta)$, as generic orbits do, and they do not effectively average over their domain~\cite{FlanaganHughes}. Thus, as we will see, other low-order resonances, while subdominant, can also play a significant role. 

For concreteness, we will focus on these resonances: $4:3$, $2:1$, and $3:1$. The ERM is implemented around each resonance exactly as done for the $3:2$ resonance. So, for example, the starting condition is computing according to Eq.~\eqref{tRes}; such that for the $4:3$ resonance, one gets $ t_{\textnormal{res}}^{4:3} = \sqrt{4 \pi/|-4 \dot{\omega}_{r0} + 3 \dot{\omega}_{\theta 0}|}$ and $\xi^*_{4:3} = - (2 \pi)/(3 t_{\textnormal{res}}) \omega_{r 0}$. 

\subsection{The investigated parameter space}

As we explore the strongest low-order resonances, we also want to test the ERM when we allow the resonance coefficients $\mathcal{C}$ to vary for each integral of motion. For this purpose, we will only consider four orbital configurations, for which the resonance coefficients have already been computed~\cite{FlanaganHughes}: case (i) $e=0.3$ and $\iota=0.35$, case (ii) $e=0.3$ and $\iota=1.22$, case (iii) $e=0.7$ and $\iota=0.35$, and case (iv) $e=0.7$ and $\iota=1.22$. We set $a = 0.9$, $\eta = 10^{-5}$ and $M = 10^6 M_{\odot}$ for all these cases. 

In Tab.~\ref{Table_1} we show the initial location, the resonance duration, the observation time and the resonance coefficients for the investigated orbital resonances. For each configuration, we start the evolution close to the resonance, by setting the initial value of $p/M$ such that the threshold function (Eq.~\ref{Eq:ThresholdFunction}) is $\xi_0 = - (0.002, 0.002, 0.02, 0.05)$ for the $4:3$, $3:2$, $2:1$ and $3:1$ resonances, respectively. We fix the observation time to T$_{\textnormal{max}} = 1$ year, and terminate the evolution earlier if the secondary reaches the separatrix. For this analysis, since the resonances $2:1$ and $3:1$ can be located beyond $r_p = 5M$, we relax the condition $r_p\gtrsim5M$, since our goal is to investigate the phenomenology of different type of resonances, and at which order of magnitude they affect the orbital evolution. 

Regarding the resonance coefficients, we consider the following scenarios:
\begin{itemize}
    \item \textit{Equal} $\mathcal{C}$: we set all the coefficients equal to their maximum values found in Table~\ref{Table_1}, rounded to the first significant digit, i.e.,
    \begin{align*}
        & |\mathcal{C}|_{4:3} = 0.0006, \:\: |\mathcal{C}|_{3:2} = 0.01, \\
        & |\mathcal{C}|_{2:1} = 0.007, \:\:\,\,\, |\mathcal{C}|_{3:1} = 0.003.
    \end{align*}
    \item \textit{Varying} $\mathcal{C}$: for each orbital configuration, we use the coefficients computed in Ref.~\cite{FlanaganHughes} as presented in Table~\ref{Table_1}.
\end{itemize} 

We stress that the magnitude of the fluxes at resonance varies considerably with the phase $\chi_0$, which defines the value of the orbit’s $\theta$ angle at the moment it reaches the periapsis. For instance, two orbits with the same values of ($\mathcal{E}$, $\mathcal{L}_z$, $\mathcal{E}$) will evolve differently if they enter the resonance with different phases, meaning that resonances enhance the dependence of the orbital evolution on initial conditions \cite{FlanaganHughes}. This suggests that, in principle, one should introduce an additional parameter in the model to track the phase at resonance. As we are interested in studying the phenomenology of different types of resonances, we choose not to explicitly model the dependence on the phase $\chi_0$, but just its sign. Hence, the coefficients of Table~\ref{Table_1} represent the peak-to-trough variation in the fluxes, as provided in Ref.~\cite{FlanaganHughes}, and can be set either positive (negative ``kicks'') or negative (positive ``kicks''): in what follows, we choose the coefficients to be negative for all the investigated resonance.

We now have all the necessary ingredients to systematically study these EMRI systems using the ERM, considering different resonance crosses and various values for the resonance coefficients.

\subsection{Resonant Flux Augmentation}

\begin{figure*}[t]
\centering
\includegraphics[scale=0.4]{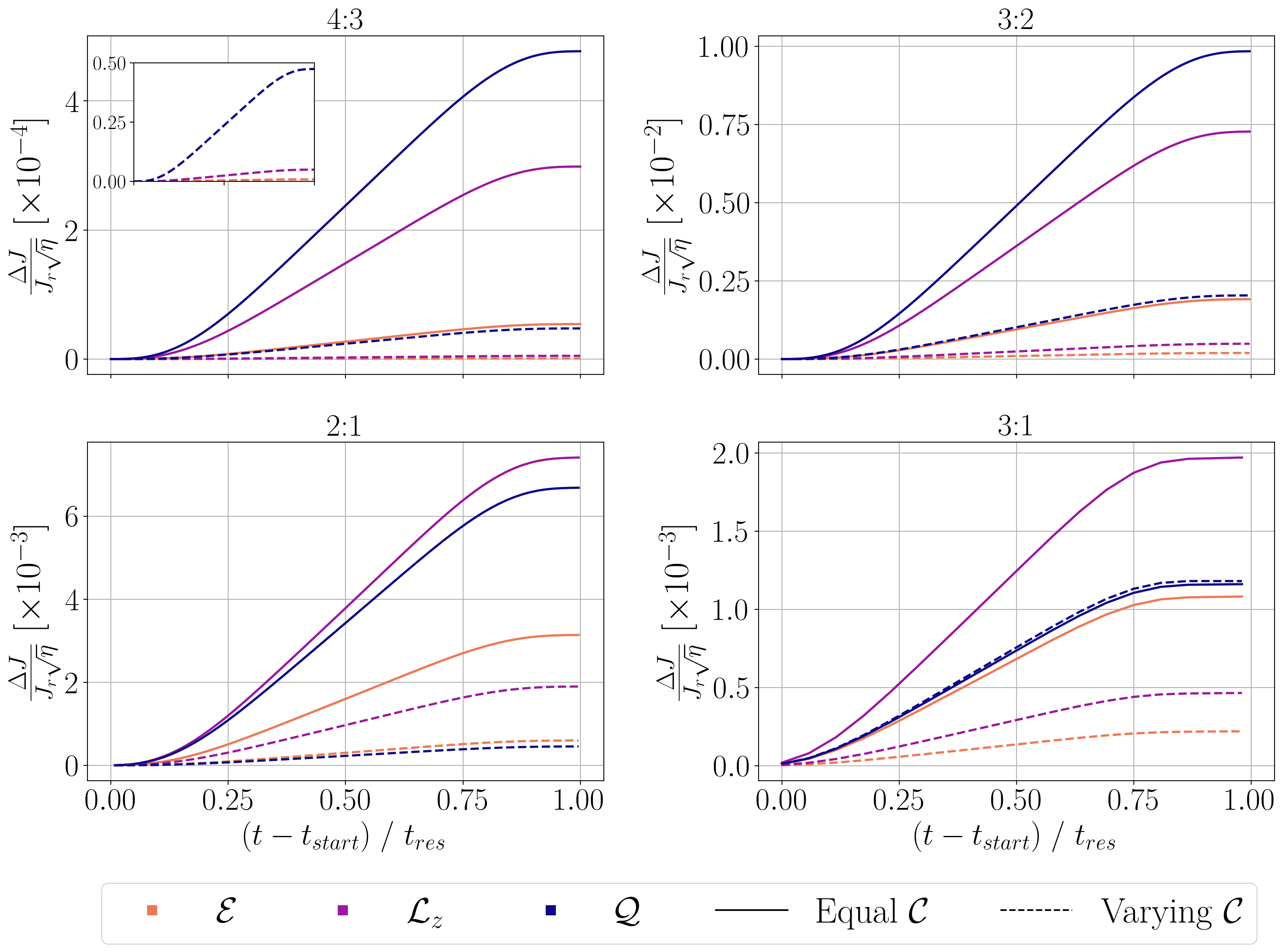} 
\caption{Evolution of the integrals of motion through single resonances ($4:3$, $3:2$, $2:1$ and $3:1$) for case (i) of Table~\ref{Table_1}, with $a = 0.9$, $\eta = 10^{-5}$ and $M = 10^6 M_{\odot}$. The solid lines depict the scenario where, for each resonance, the resonant coefficients $\mathcal{C}$ are equal, i.e., $|\mathcal{C}|_{4:3} = 0.0006$,$ |\mathcal{C}|_{3:2} = 0.01$, $|\mathcal{C}|_{2:1} = 0.007$, $|\mathcal{C}|_{3:1} = 0.003$. The dashed lines indicate the scenario where $\mathcal{C}$ are different for each integrals of motion, i.e., $|\mathcal{C}|_{4:3} = (0.00001, 0.00001, 0.00006)$, $|\mathcal{C}|_{3:2} = (0.00102, 0.00067, 0.00208)$, $|\mathcal{C}|_{2:1} = (0.00131, 0.00179, 0.00046)$, $|\mathcal{C}|_{3:1} = (0.00059, 0.00070, 0.00310)$. The values for these coefficients were computed in Ref.~\cite{FlanaganHughes} and are also shown in Table~\ref{Table_1}. The inset in the top left panel zooms in on the \textit{varying} $\mathcal{C}$ scenario for the $4:3$ resonance. As expected, using different values for these coefficients leads to smaller flux enhancements, which should therefore manifest in smaller mismatches and orbital dephasings.}
\label{kick_single_case_i}
\end{figure*}

Let us start by computing the relative changes in ($\mathcal E, \mathcal L_z, \mathcal Q$) from the start to the end of the investigated resonances, for the orbital configurations of Table~\ref{Table_1}. Figure~\ref{kick_single_case_i} shows these changes for case (i), i.e., the system with the lowest eccentricity and inclination. We find that:
\begin{itemize}
    \item \textit{Equal} $\mathcal{C}$: the ``kicks'' are higher for $\mathcal{Q}$ in the case of the $3:2$ and $4:3$ resonances, while $\mathcal{L}_z$ shows the highest flux changes for the $2:1$ and $3:1$ resonances. If we consider the integral of motion with the greatest ``kicks,'' the $3:2$ is, as expected, the strongest resonance, followed by the $2:1$, $3:1$, and $4:3$.
    \item \textit{Varying} $\mathcal{C}$: the ``kicks'' are higher for $\mathcal{Q}$ in the case of the $3:2$, $3:1$ and $4:3$ resonances, while $\mathcal{L}_z$ shows the highest flux changes for the $2:1$ resonance. We observe that the integrals of motion with the greatest ``kicks'' are the $3:2$, $2:1$ and $3:1$ resonances, which are comparable, while the $4:3$ is the weakest by nearly two orders of magnitude. 
\end{itemize}

The results for the other orbital configurations, i.e., cases (ii)-(iv) of Table~\ref{Table_1}, are shown in App. \ref{App::other_cases}: we notice that they do not vary significantly and the overall pattern is similar to the case (i) orbit.

The resonance coefficients, which are free parameters in the ERM, determine, together with the resonance duration, the impact of the resonances on the orbital evolution. This analysis suggests that one should also take into account low-order resonances in addition to the $3:2$, at least for the orbits considered in this work, as their cumulative effect may not be negligible, especially for $2:1$ and $3:1$. The results are even more prominent for the (more realistic) \textit{varying} $\mathcal{C}$ case, where the flux changes of the $3:2$, $2:1$ and $3:1$ resonances are comparable. 

\begin{figure*}[t]
    \includegraphics[scale=0.073]{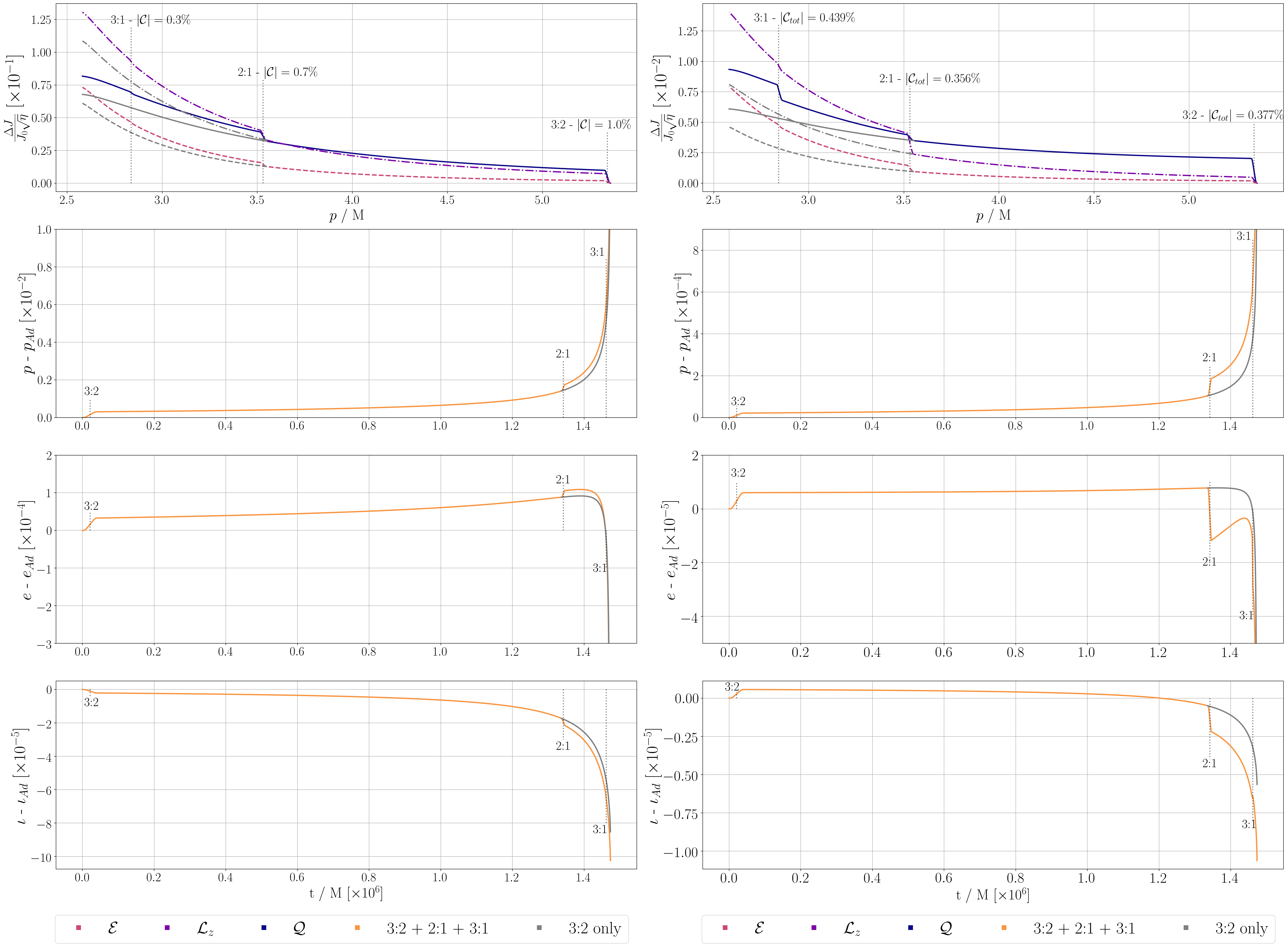}
    \caption{\emph{Left:} Evolution of the integrals of motion through multiple resonances ($3:2$ + $2:1$ + $3:1$) for the case (i) orbit of Table~\ref{Table_1}, with $a = 0.9$, $\eta = 10^{-5}$ and $M = 10^6 M_{\odot}$ and equal resonance coefficients, i.e., $|\mathcal{C}|_{3:2} = 0.01$, $|\mathcal{C}|_{2:1} = 0.007$, $|\mathcal{C}|_{3:1} = 0.003$. The solid, dashed and dot-dashed gray lines indicate the evolution where we activate only the $3:2$ resonance. The dotted gray lines show the location of the $3:2$, $2:1$ and $3:1$ resonances. \emph{Right:} Same as left with varying coefficients, i.e., $|\mathcal{C}|_{3:2} = (0.00102, 0.00067, 0.00208)$, $|\mathcal{C}|_{2:1} = (0.00131, 0.00179, 0.00046)$, $|\mathcal{C}|_{3:1} = (0.00059, 0.00070, 0.00310)$.}
    \label{kick_multi_case_i}
\end{figure*}

We further investigate the effects of multiple resonant-crossings in a single inspiral by starting the trajectory close to the $3:2$ resonance, and evolving the systems of Table~\ref{Table_1} through the $3:2$, $2:1$ and $3:1$ resonances. We terminate the evolution when the secondary reaches the separatrix. We then compute the relative changes in ($\mathcal E, \mathcal L_z, \mathcal Q$) and in ($p/M, e, \iota$), and compare them to the evolution where we activate only the $3:2$ resonance and not the low-order ones. Figure~\ref{kick_multi_case_i} shows the results for the inspiral of case (i) orbit of Table~\ref{Table_1}. We find that:
\begin{itemize}
    \item \textit{Equal} $\mathcal{C}$: The flux changes are imprinted upon the orbital elements. The coefficients being negative (positive jumps in the integrals of motion), leads to ``kicks'' in the orbital elements that act in the opposite way with respect to the adiabatic evolution \footnote{During the inspiral, both $p/M$ and $e$ decrease with time, while $\iota$ increases. The eccentricity shows an increase at the end of the inspiral, when the system is close to the separatrix.}. For example, an orbital element is decreasing during the inspiral the ``kick'' is positive, or vice versa. When the system crosses the $2:1$ resonance, the evolution clearly deviates from the one where we activate \emph{only} the $3:2$ resonance. However, the ``kicks'' in ($p/M$, $e$, $\iota$) are lower than the ones caused by the $3:2$ resonance. The $3:1$ resonance, instead, does not affect the evolution significantly, as it lasts for much shorter times, being located closer to the primary Kerr BH. Also, the coefficients are lower than the ones for the $3:2$ resonance by nearly one order of magnitude.
    \item \textit{Varying} $\mathcal{C}$: the flux changes of the $3:2$, $2:1$ and $3:1$ resonances may look comparable, if we consider the integral of motion with the largest kicks, or the values of $\mathcal{C}_{\textnormal{tot}}$ as provided in Table~\ref{Table_1}. However, the way they translate into the orbital elements depends on the individual coefficients for ($\mathcal E, \mathcal L_z, \mathcal Q$), and their relative strength. We notice different patterns in the evolution of the orbital elements, compared to the \textit{equal} $\mathcal{C}$ case. When the system crosses the $3:2$ resonance, the inclination angle increases, while the $2:1$ resonance causes a decrease in the eccentricity. These effects appear to be related to the jumps in $\mathcal Q$ ($\mathcal L_z$) being much larger than the ones in $\mathcal L_z$ ($\mathcal Q$) in the case of the $3:2$ ($2:1$) resonance. As in the previous case, the $2:1$ resonance affects the orbital evolution significantly, but the ``kicks'' in ($p/M$, $e$, $\iota$) are now greater than the ones caused only by the $3:2$ resonance. The $3:1$ resonance does not affect the evolution significantly; however, we observe a small change in the orbital elements due to the ``kick'' in $\mathcal{Q}$.
\end{itemize}

\begin{figure*}[t]
\centering
\includegraphics[scale=0.5]{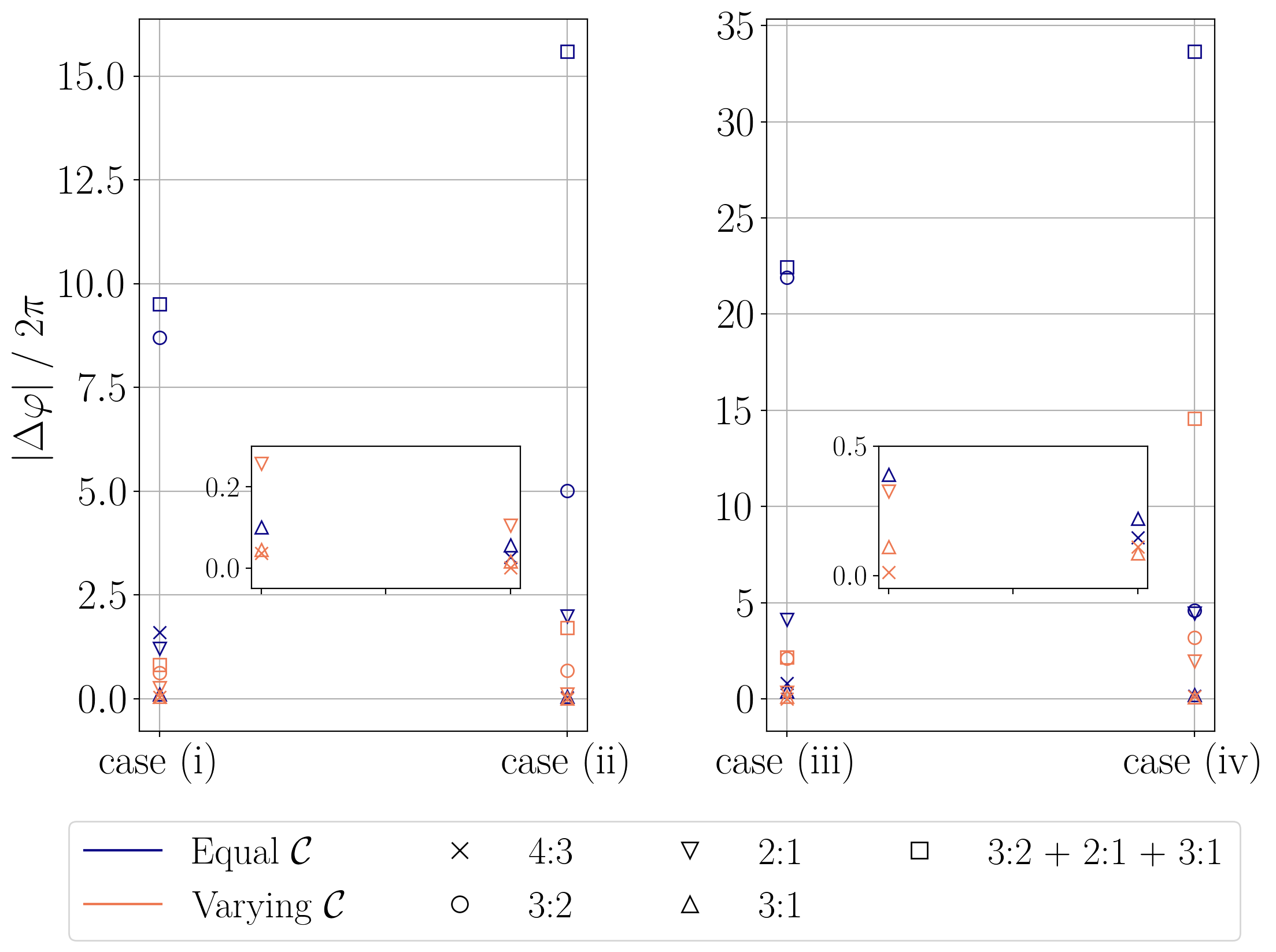} 
\caption{Orbital dephasing between ERM and NK orbits for the investigated resonances and orbital configurations of Table~\ref{Table_1}, together with the multiple resonances analysis, with $a = 0.9$, $\eta = 10^{-5}$ and $M = 10^6 M_{\odot}$. The blue markers indicate the scenario where, for each resonance, the resonance coefficients $\mathcal{C}$ are equal to the maximum value found in Table~\ref{Table_1}. The orange markers indicate the scenario where the resonance coefficients $\mathcal{C}$ are varying for each integrals of motion; the values are computed in Ref.~\cite{FlanaganHughes} and they are provided in Table~\ref{Table_1}.}
\label{dephasing}
\end{figure*}

The multiple resonant-crossings evolutions for the other orbital configurations, i.e., cases (ii)-(iv) of Table~\ref{Table_1}, are shown in App. \ref{App::other_cases}. Overall, how the flux changes in ($\mathcal E, \mathcal L_z, \mathcal Q$) translate into the orbital elements, and ultimately affect the evolution of the GW frequency, is very convoluted: it does not only depend on the values of the individual resonance coefficients ($\mathcal{C}_{\mathcal{E}}$, $\mathcal{C}_{\mathcal{L}_z}$, $\mathcal{C}_{\mathcal{Q}}$), but also on their relative strength. The results are even more prominent for the \textit{varying} $\mathcal{C}$ case. This deserves further studies, especially for the purposes of implementing data analysis for generic orbits.

\subsection{Orbital Dephasing}

In Fig.~\ref{dephasing}, we compute the orbital dephasing between ERM and NK orbits for the investigated resonances and orbital configurations of Table~\ref{Table_1}, together with the multiple resonances analysis. We focus on the $\varphi$ coordinate, where we expect the largest effects as shown in Fig.~\ref{dephasing_1}. We find that:
\begin{itemize}
    \item The dephasing caused by the $3:2$ resonance is higher with respect to the other resonances, especially when compared to the $3:1$ and $4:3$ ones, where the difference is roughly one order of magnitude.
    \item For orbits with small inclination, the dephasing caused by the multiple resonances is not significantly different from the $3:2$ case. The difference increases for the larger $\iota$ orbits, where the multiple resonance case shows the highest dephasing, both in the \textit{equal} $\mathcal{C}$ and \textit{varying} $\mathcal{C}$ scenarios.
\end{itemize}

\begin{figure*}[t]
\centering
\includegraphics[scale=0.5]{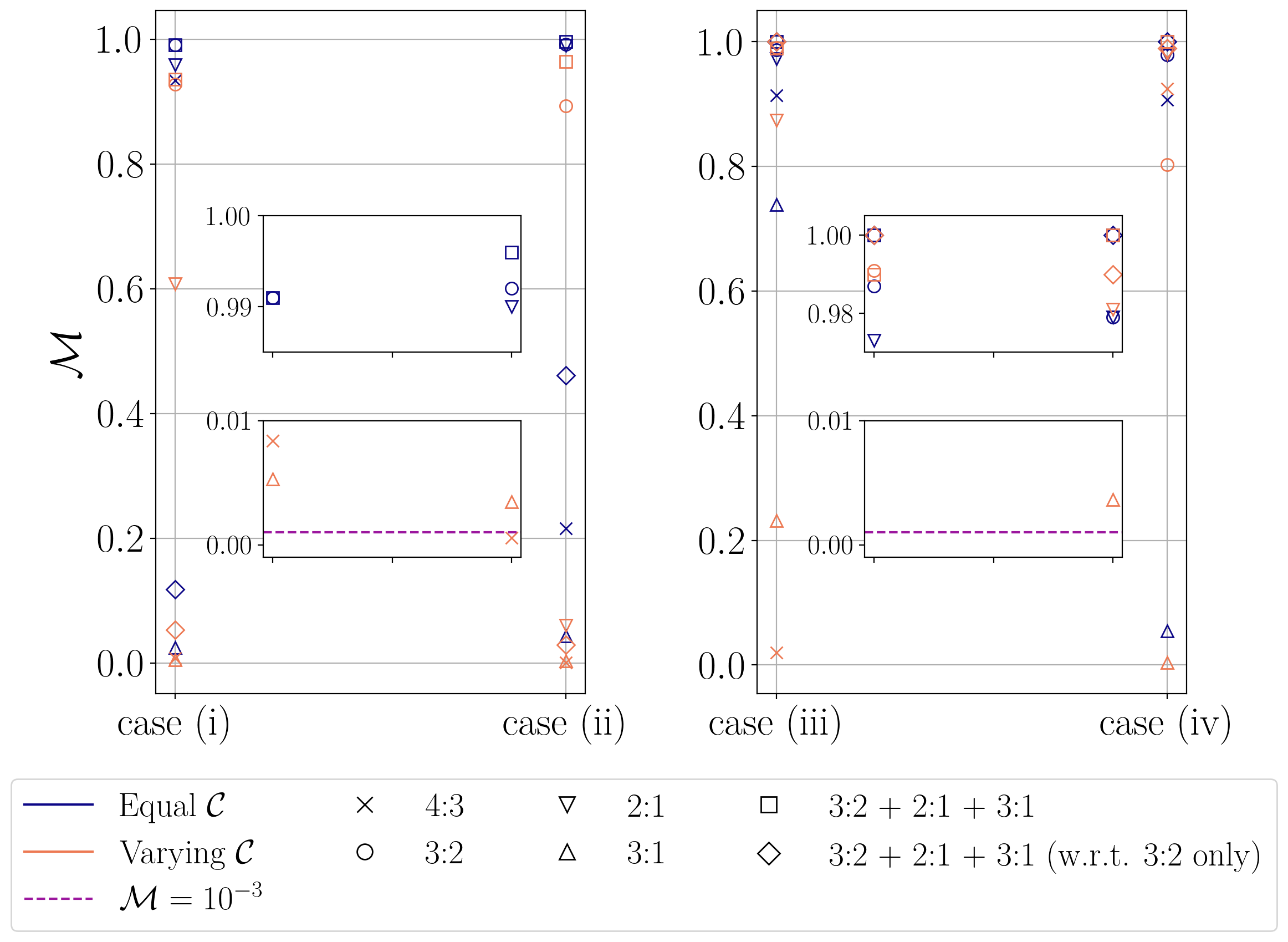} 
\caption{Mismatch between ERM and NK orbits for the investigated resonances and orbital configurations of Table~\ref{Table_1}, together with the multiple resonances analysis, with $a = 0.9$, $\eta = 10^{-5}$ and $M = 10^6 M_{\odot}$. The blue markers indicate the scenario where, for each resonance, the resonance coefficients $\mathcal{C}$ are equal to the maximum value found in Table~\ref{Table_1}. The orange markers indicate the scenario where the resonance coefficients $\mathcal{C}$ are varying for each integrals of motion; the values are computed in Ref.~\cite{FlanaganHughes} and they are provided in Table~\ref{Table_1}. For the multiple resonances case, we compute also the mismatch with respect to the ERM waveforms where we activate only the $3:2$ resonance.}
\label{GW_1}
\end{figure*}

\subsection{Waveform Mismatch}

Figure~\ref{GW_1} shows the mismatch between ERM and NK waveforms for the investigated resonances and orbital configurations of Table~\ref{Table_1}. For the multiple resonances analysis, we compute the mismatch between ERM and NK waveforms, as well as the one with respect to the ERM where we activate only the $3:2$ resonance. Overall, we find that:
\begin{itemize}
    \item For almost all the investigated resonances, we find $\mathcal{M} > 10^{-3}$; only in the case (ii), for the $4:3$ resonance with \textit{varying} $\mathcal{C}$, the mismatch is lower than the indistinguishability threshold.
    \item The mismatch is higher for the $3:2$ and $2:1$ resonances, and for the multiple resonances case (with respect to the NK waveforms), with $\mathcal{M} \approx 1$.
    \item The $3:1$ and $4:3$ resonances cause the smallest effects, except for case (iv), for the $4:3$ resonance.
    \item With respect to the ERM waveforms where we activate only the $3:2$ resonance, the mismatch caused by the multiple resonances analysis increases with the orbital eccentricity.
\end{itemize}

Based on the above results, almost all the investigated resonances are, in principle, phenomenologically likely to affect the detection of the EMRIs of Table~\ref{Table_1}. For the purposes of building a phenomenological EMRI model that is able to capture the most significant resonance effects, it seems convenient to include at least the $3:2$ and $2:1$ resonances, as the cumulative effects are not negligible, especially in the case of highly eccentric orbits, i.e., cases (iii) and (iv).

\section{Discussion and future outlook}
\label{sec::discussion}

Transient orbital resonances are paramount in EMRI modeling for space-based detectors like LISA. Even though the GSF calculations are closing in toward an endpoint of Kerr generic orbits, significant work has been performed with approximate, though highly accurate, methods to describe and account for the radiation reaction process in EMRIs. The missing conservative part of gravitational radiation has recently been modeled and augmented in the most used waveform modeling technique, namely the numerical kludge waveforms that show reasonable agreement with Teukolsky-based waveforms of generic EMRIs.

In this work, we have first compared three distinct methods of evolving the adiabatic motion of the secondary around a Kerr primary with radiation reaction, i.e., the fully-decoupled first-order equations for all degrees of freedom, and the coupled second-order differential equations for the $(r,\theta)$-motion. Second-order dynamical systems can be solved with two distinct methods. The first one uses the geodesic equations to derive the set of coupled second-order differential equations for all degrees of freedom. In the second method, the radiation reaction terms are included in the EMRI through the MiSaTaQuWa equations. We found that, compared to the NK scheme, the former method is more accurate since the error does not grow with time as in latter one. The former method for solving coupled, second-order differential equations is also more robust over large EMRI evolutions. As far as we know, this is the first time that such an analysis has been carried out in detail. Since these two second-order methods do not rely on the separability of the geodesic equations, they may be easier to be extended to primary backgrounds that are arbitrarily close to the Kerr metric, i.e. slight modifications of GR or the additions of very sparse astrophysical environments. 

Using the most accurate integration method of inspirals, i.e., the first-order NK method, we have evolved several EMRI systems with varying initial conditions, i.e., the semilatus rectum $p/M$, and configurations with small/large eccentricity $e$ and inclination angle $\iota$. We find that the orbital dephasing and mismatch of GWs emitted from such systems, for fixed resonance coefficients, between pure NK and NK+ERM evolutions, are higher than the indistinguishability threshold, therefore taking into account the most dominant resonant-crossings might be important for parameter estimation of EMRIs. 

Our results highlight the critical importance of considering varying resonance coefficients ($\mathcal{C}$) when analyzing the impact of resonances on EMRIs. We have demonstrated that under a \textit{varying} $\mathcal{C}$ scenario, the mismatch remains below the indistinguishability threshold (assuming a SNR=$20$), suggesting that constant resonance coefficients can significantly overestimate the mismatch, sometimes by an order of magnitude or more. While the mismatch for individual crossings is small, we find that its cumulative effect over multiple resonance crossings, particularly for low-order resonances such as $2:1$ and $3:1$, cannot be neglected. These findings emphasize the need to account for all resonances in orbital evolution studies, as their combined influence can substantially alter the system's detectability. Furthermore, the distinct patterns observed in orbital element evolution between \textit{equal} and \textit{varying} $\mathcal{C}$ scenarios underscore the complex interplay between resonance coefficients and their effects, which must be carefully modeled to ensure accurate predictions of EMRI dynamics.

The results of this work are in overall agreement with those presented in Ref.~\cite{Lynch:2024ohd}, which employs near-identity transformations to resolve resonant crossings continuously—switching between full averaging and a partial averaging that retains resonant phase information—to efficiently compute inspiral trajectories. In contrast, our ERM (based on Ref.~\cite{SperiGair}) simplifies the treatment by representing each resonance crossing as a windowed ``jump'' in the orbital parameters, where the fluxes are computed from Teukolsky-based calculations for resonant orbits~\cite{FlanaganHughes}. These two approaches display qualitatively comparable behavior, yet they yield different estimates for the magnitude of resonant effects on the integrals of motion and, hence, on the GW dephasing. In addition, while in Ref.~\cite{Lynch:2024ohd} they include post-adiabatic terms in their evolution—which we do not account for here—the key difference lies in the computation of trajectories and the modeling of the behavior around the resonances. One could, in principle, incorporate a resonance jump within their framework as performed here. However, the results may not necessarily directly match those from the partial average method. A detailed comparison of these methods is certainly worth pursuing in future work, as more complete self-force calculations are needed to determine whether the discrete ``jumps'' in our model can accurately reproduce the full resonance dynamics captured by the continuous approach. Ultimately, these complementary approaches underscore that no single method may capture all aspects of resonant dynamics in EMRIs; to the contrary, a hybrid strategy—such as empirically fitting resonant changes within a continuous framework—could provide an optimal balance between accuracy and computational efficiency for LISA data analysis.

In fact, our results suggest that taking into account higher-order, subdominant resonances can be paramount for accurate parameter estimation. Transient orbital resonances, such as the $2:1$, can become more effective in the orbital evolutions of the EMRI since the ``kick'' is larger than the one caused by the $3:2$ resonance when the EMRI is highly eccentric. In general, combining a cumulative number of orbital resonances is fundamental in building robust models and understanding EMRIs at their most fundamental level.

There are several prospects for this research. In addition to assessing the systematic errors in parameter estimates with this ERM we can explore more configurations and various resonance effects. This framework can be generalized to include tidal resonances, which provide a greater volume for satisfying resonant conditions. In particular, when an orbit is (nearly) circular, such as in wet EMRIs produced in active galactic nuclei~\cite{Pan:2021lyw,Pan:2021oob,Lyu:2024gnk}) the strength of orbital resonances is minimized. This is because their strength is proportional to the eccentricity for lowest-order resonances. However, there can still be tidal resonances between $\omega_\theta$ and $\omega_\phi$. It may be interesting to extend our work to a general scenario where both tidal and orbital resonances co-exist.

One can also study systems that are not integrable (e.g., due to a GR modification or an environmental effect). The problem with non-integrable systems, particularly when addressing solutions beyond GR, is that the radiation reaction might not be known. However, it is possible to examine the effects of non-integrability by employing a slow-rotation expansion of the Kerr solution~\cite{Cardenas}, or for beyond GR solutions perturbatively close to Kerr. Studying such systems is appealing because one can use the radiation reaction terms from GR, either as exact or as an approximation. One can also apply our methods to general perturbed Kerr spacetimes, the so-called ``bumpy'' BHs, by extending our methodology in the NK+ERM to include modification to the equations of motion and the PN fluxes of the integrals of motion~\cite{Pan:2023wau}. Investigating these systems is particularly promising, as we anticipate that resonances will exhibit much longer durations due to chaotic effects~\cite{Destounis:2021mqv,Destounis:2021rko,Destounis:2023khj}. These methods also can potentially reveal new insights into complex astrophysical environments~\cite{Cardoso:2021wlq,Cardoso:2022whc}.

\begin{acknowledgments}
We thank Scott Hughes, Lorenzo Speri and Lennox Keeble for their useful comments. 
E.L., K.D. and P.P. acknowledge partial support by the MUR PRIN Grant 2020KR4KN2 ``String Theory as a bridge between Gauge Theories and Quantum Gravity'' and by the MUR FARE programme (GW-NEXT, CUP:~B84I20000100001).
A.C.-A. acknowledges support from the DOE, through Los Alamos National Laboratory (LANL) Directed Research and Development, grant 20240748PRD1, as well as by the Center for Nonlinear Studies. This work is authorized for unlimited release under LA-UR-25-21190.
K.D. acknowledges financial support provided by FCT – Fundação para a Ciência e a Tecnologia, I.P., under the Scientific Employment Stimulus – Individual Call – Grant No. 2023.07417.CEECIND/CP2830/CT0008.
Computations have been performed at the Vera cluster supported by MUR and the Sapienza University of Rome, as well as using the Wake Forest University (WFU) High Performance Computing Facility, a centrally managed computational resource available to WFU researchers including faculty, staff, students, and collaborators~\cite{WakeHPC}.
\end{acknowledgments}

\subsection*{DATA AVAILABILITY} 
The supporting data for this article are openly available from the zenodo archive DOI:~\href{https://zenodo.org/record/14983924}{https://zenodo.org/record/14983924}~\cite{10.5281/zenodo.14983924}.

\appendix

\section{Adiabatic Inspirals: First-order and Second-order Methods}
\label{sec::1st_vs_2nd}

\subsection{First-order Method: The Numerical ``Kludge'' Framework}
\label{sec::NK}

The NK scheme is a fast but approximate EMRI modeling technique developed to explore EMRI phenomenology. The integrals of motion are used to decouple the geodesic equations in Kerr spacetime and derive a set of first order differential equations for the orbital motion. To simplify the treatment of the turning points in the radial and polar motion, it is useful to adopt the following parametrization of $r$ and $\theta$~\cite{Schmidt, Fujita}, such as
\begin{align}
    & r = \frac{p}{1 + e \cos \psi}, \hspace{2cm} r_p \leq r \leq r_a, \label{parameterizationR} \\
    & \cos\theta = \cos\theta_{-} \cos\chi, \hspace{1cm} \theta_{-} \leq \theta \leq \pi - \theta_{-}, \label{parameterizationTheta}
\end{align}
where $r_p$ is the periapsis, $r_a$ is the apoapsis, $p$ is the semi-latus rectum and $e$ is the eccentricity. The latter are defined by $r_p = p / (1 + e)$ and $r_a = p / (1 - e)$. The turning points in the polar motion are $\theta_{-}$ and $\pi - \theta_{-}$, where $\theta_{-}$ is the minimum value of $\theta$. We also define the inclination angle $\iota = \pi/2 - \theta_{-}$. As $\psi$ and $\chi$ evolve from $0$ to $2\pi$, $r$ and $\theta$ move through their full ranges of motion. The parameterization~(\ref{parameterizationR})-(\ref{parameterizationTheta}) maps ($\mathcal E$, $\mathcal L_z$, $\mathcal Q$) into the orbital elements ($p$, $e$, $\iota$), which are likewise conserved along a geodesic. 

With respect to the time $t$ of an observer at spatial infinity, the first-order differential equations for $\psi$, $\chi$, $\varphi$ are~\cite{Kludge, Fujita}
\begin{align}
    \begin{split}
    \dv{\psi}{t} = & \frac{M \sqrt{(1 - \mathcal E^2) \Bigl[(\frac{p}{M} - p_3) - e (\frac{p}{M} + p_3 \cos \psi)\Bigr]}}{\Bigl[\gamma + a^2 M^2 \mathcal E \cos \theta(\chi)^2\Bigr](1 - e^2)} \\
    & \times \frac{\sqrt{\Bigl[(\frac{p}{M} - p_4) + e (\frac{p}{M} - p_4\cos \psi)\Bigr]}}{\Bigl[\gamma + a^2 M^2 \mathcal E \cos \theta(\chi)^2\Bigr](1 - e^2)}, \label{momentumPsiCarter}
    \end{split} 
\end{align}
\begin{align}
    & \dv{\chi}{t} = \frac{\sqrt{\beta [z_+ - \cos \theta(\chi)^2]}}{\gamma + a^2 M^2 \mathcal E \cos \theta(\chi)^2}, \label{momentumChiCarter} \\
    & \dv{\varphi}{t} = \frac{\Phi(r, \theta)}{T(r, \theta)}, \label{momentumPhiCarter}
\end{align}
where 
\begin{equation}
\begin{split}
 T(r, \theta) = & \hspace{3pt} a M (\mathcal L_{z} - a M \mathcal E \sin^2\theta) + \frac{r^2 + a^2M^2}{\Delta(r)} \\
 & \times [\mathcal E (r^2 + a^2 M^2) - \mathcal L_{z} a M],
\end{split}
\end{equation}
\begin{equation}
\begin{split}
\Phi(r, \theta) = & \hspace{3pt} \frac{\mathcal L_{z}}{\sin^2\theta} - a M \mathcal E + \frac{a M}{\Delta(r)} [\mathcal E (r^2 + a^2M^2) \\ 
& - \mathcal L_{z} a M],
\end{split}
\end{equation}
\begin{align}
    & \gamma = \mathcal E \Biggl[ \frac{(r^2 + a^2 M^2)^2}{\Delta(r)} - a^2 M^2 \Biggr] - \frac{2 a M^2 r \mathcal L_z}{\Delta(r)},  \\
    & \beta = a^2 M^2 (1 - \mathcal E^2), \\
    & p_3 = \frac{r_3}{M} (1 - e), \\
    & p_4 = \frac{r_4}{M} (1 + e), \\
    & z_+ = \frac{\mathcal Q}{\mathcal L_z^2 \epsilon_0 \cos^2\theta_{-}}, \\
    & \epsilon_0 = \frac{a^2 M^2 (1 - \mathcal E^2)}{\mathcal L_z^2},
\end{align}
and $r_3$, $r_4$ are roots of the radial potential
\begin{equation}\label{radialpotential}
    \begin{split}
    R(r) & \equiv [\mathcal E (r^2 + a^2 M^2) - \mathcal L_{z} a M]^2 - \Delta(r) \\
    & \hspace{12pt} \times [r^2 + (\mathcal L_{z} - a M \mathcal E)^2 + \mathcal Q] \\ 
    & = (1 - \mathcal E^2) (r_a - r) (r - r_p) (r - r_3) (r - r_4),
\end{split}
\end{equation}
though do not correspond to turning points in the radial motion. 

The evolution of $J = (\mathcal E, \mathcal L_z, \mathcal Q)$ is computed using the adiabatic approximation, with the fluxes written in the form
\begin{equation}\label{ad_fluxes}
   \dv{J}{t} = f_{\textnormal{NK}}(a, M, \mu, p, e, \iota),
\end{equation}
where the functions $f_{\textnormal{NK}}$ are provided in Ref.~\cite{Gair}. Assuming linear variations, the integrals of motion are updated as
\begin{equation}\label{updateJ}
     J(t_{n+1}) = J(t_n) + \Biggl(\dv{J}{t} \Biggr) \Bigg|_{t_n} \Delta t,
\end{equation}
where $\Delta t$ is small compared to the characteristic time of adiabatic variations~\cite{Destounis:2021mqv}. 

The trajectory is computed by integrating, at each step $n$ of the inspiral, Eqs.~(\ref{momentumPsiCarter})-(\ref{momentumPhiCarter}) in the interval $[t_n, t_n +\Delta t]$ using $\mathcal E(t_n), \mathcal L_z(t_n), \mathcal Q(t_n)$, with initial positions from step $n-1$. We integrate the set of first-order differential equations with an \emph{Implicit Runge--Kutta method} (see App.~\ref{sec::methods1}.). We then compute the adiabatic fluxes and update the values of the integrals of motion using Eqs.~(\ref{ad_fluxes})-(\ref{updateJ}). We reiterate the whole procedure for step $n+1$. It has been shown that NK waveforms are remarkably faithful for $r_p \gtrsim 5M$, when compared to Teukolsky-based inspirals~\cite{Kludge}.

\subsection{Second-order Methods}
\label{sec::coupled}

If we do not utilize the integrals of motion to decouple the geodesic equation, we can evolve the system by solving second-order differential equations. In such a case, the procedure for constructing the inspiral changes and, within the adiabatic approximation, we can approach the problem in two different ways.

\subsubsection{Method 1: Direct Solution to the Geodesic Equations}

We use the geodesic equations to derive the set of coupled second-order differential equations for $t(\tau)$, $r(\tau)$, $\theta(\tau)$, $\varphi(\tau)$,
\begin{align}
    & \ddot t  = - \Gamma^{t}_{\lambda \rho} \dot x^{\lambda} \dot x^{\rho}, \label{geotTau} \\
    & \ddot r  = - \Gamma^{r}_{\lambda \rho} \dot x^{\lambda} \dot x^{\rho}, \label{georTau} \\
    & \ddot \theta  = - \Gamma^{\theta}_{\lambda \rho} \dot x^{\lambda} \dot x^{\rho}, \label{geothetaTau} \\
    & \ddot \varphi  = - \Gamma^{\varphi}_{\lambda \rho} \dot x^{\lambda} \dot x^{\rho}, \label{geophiTau} 
\end{align}
where the overdot denotes differentiation with respect to proper time $\tau$. We integrate this second-order system with a \emph{Projected Implicit Runge--Kutta method} (see App.~\ref{sec::methods2}.). We utilize
\begin{align}
    & \mathcal E = - \Bigl[g_{tt}(r, \theta) \dot t + g_{t \varphi}(r, \theta) \dot \varphi \Bigr], \label{energy} \\
    & \mathcal L_z = g_{t \varphi}(r, \theta) \dot t + g_{\varphi \varphi}(r, \theta) \dot \varphi, \label{angular} \\
    & \mathcal Q = \Bigl[g_{\theta \theta}(r, \theta) \dot \theta \Bigr]^2  + \cos^2\theta \Biggl[a^2 M^2 (1 - \mathcal E^2) + \frac{\mathcal L_{z}^2}{\sin^2\theta}\Biggr], \label{carterEq} \\
    & \dot r^2 + \frac{g_{\theta \theta}(r, \theta)}{g_{rr}(r, \theta)} \dot \theta^2 + V_\textrm{eff} = 0, \label{constraintEq}
\end{align}
as invariants to constraint the motion as close as possible to the geodesic one, by substituting, at each integration time step, the numerical solutions for $r$, $\theta$, $\dot t$, $\dot r$, $\dot \theta$, $\dot \varphi$. Here, $g_{\mu \nu}$ denotes the metric tensor components, and 
\begin{equation}
V_\textrm{eff} = \frac{1}{g_{rr}} \Biggl(1 - \frac{\mathcal E^2 g_{\varphi \varphi} +  \mathcal L_{z}^2 g_{tt} +2 g_{t \varphi} \mathcal E \mathcal L_{z}}{g_{t \varphi}^2 - g_{tt} g_{\varphi \varphi}} \Biggr),
\end{equation}
is a Newtonian-like potential that characterizes bound geodesic motion. When $V_\textrm{eff} = 0$, the resulting curve is called curve of zero velocity.

At each step $n$ of the inspiral, the trajectory results from the numerical solution of Eqs.~(\ref{geotTau})-(\ref{geophiTau}) in the interval $[\tau_n, \tau_n + \Delta \tau]$, where $\Delta \tau$ is the proper time interval between two subsequent steps of the inspiral. As in the NK case, $\Delta \tau$ is small compared to the characteristic time of the adiabatic variations. We compute the orbital elements using the numerical solutions for $r(\tau)$ and $\theta(\tau)$ in the interval $[\tau_n, \tau_n + \overline{\Delta \tau}]$, with $\overline{\Delta \tau} \gg \Delta \tau$. We find $\overline{\Delta \tau} = 1000M$ to be a reasonable compromise between accuracy and computational time, at least for the evolution of EMRI systems considered in Sec.~\ref{sec:method_comparison}. We then use $p(t_n)$, $e(t_n)$, $\iota (t_n)$ to calculate the adiabatic fluxes and update the values of the integrals of motion using Eqs.~(\ref{ad_fluxes})-(\ref{updateJ}). 

We are not explicitly evolving ($\mathcal E$, $\mathcal L_z$, $\mathcal Q$) in the equations of motions. To include the effect of radiation reaction, we re-initialize the four-velocity vector, $u^{\nu} = (\dot t, \dot r, \dot \theta, \dot \varphi)$, at each step $n+1$ by solving Eq.~(\ref{carterEq}) and
\begin{align}
    & \dot t = \frac{\mathcal E g_{\varphi \varphi}(r, \theta) + \mathcal L_z g_{t \varphi}(r, \theta)}{g_{t \varphi}^2(r, \theta) - g_{tt}(r, \theta) g_{\varphi \varphi}(r, \theta)}, \label{momentumT} \\
    &  \dot \varphi = - \frac{\mathcal E g_{t \varphi}(r, \theta) + \mathcal L_z g_{tt}(r, \theta)}{g_{t \varphi}^2(r, \theta) - g_{tt}(r, \theta) g_{\varphi \varphi}(r, \theta)}, \label{momentumPhi} \\
    & g_{\mu \nu} u^{\mu} u^{\nu} = -1, \label{4vel}
\end{align}
with $\mathcal E(t_{n+1}), \mathcal L_z(t_{n+1}), \mathcal Q(t_{n+1})$. We reiterate the whole procedure and integrate Eqs.~(\ref{geotTau})-(\ref{geophiTau}) in the interval $[\tau_{n+1}, \tau_{n+1} + \overline{\Delta \tau}]$, using the re-initialized velocities and the initial positions from step $n$.

\subsubsection{Method 2: Solving the MiSaTaQuWa Equations}

The radiation reaction is included by replacing the geodesic equations with the MiSaTaQuWa equations~\cite{Biao,MiSaTa,QuWa}
\begin{equation}\label{geodesicGSF} 
    \ddot x^{\nu} + \Gamma^{\nu}_{\lambda \rho} \dot x^{\lambda} \dot x^{\rho} = \mathcal{F}^{\nu}.
\end{equation}
We compute the derivative with respect to the proper time of Eqs.~(\ref{energy})-(\ref{constraintEq}) as
\begin{align}
& \dot{\mathcal{E}} = \mathcal{E}' \dot t  = - \Bigl(g_{tt} \mathcal{F}^{t} + g_{t \varphi} \mathcal{F}^{\varphi}\Bigr), \\
& \dot{\mathcal{L}_z} = \mathcal{L}_z' \dot t  = g_{t \varphi} \mathcal{F}^{t} + g_{\varphi \varphi} \mathcal{F}^{\varphi}, \\
\begin{split}
    \dot{\mathcal{Q}} = \mathcal{Q}' \dot t  = & \: 2 g_{\theta \theta} \dot \theta \Bigl(2 r \dot r \dot \theta + g_{\theta \theta} \ddot \theta \Bigr) - a^2 M^2 \dot \theta \sin (2 \theta) \\
    & \times \Bigl(2 \dot \theta^2 g_{\theta \theta} + (1 - \mathcal{E}^2)\Bigr) - 2 \cos^2{\theta} a^2 M^2 \mathcal{E} \dot{\mathcal{E}} \\
    & - 2 \cot{\theta} \mathcal{L}_z \Bigl(\cot^2{\theta} \mathcal{L}_z - \cot{\theta} \dot{\mathcal{L}_z} + \dot \theta \mathcal{L}_z \Bigl), \\
    \end{split} \label{carterDeriv} \\
& g_{\mu \nu} u^{\mu} \mathcal{F}^{\nu} = 0.
\end{align}
We solve this system for $\mathcal{F}^{\nu}$ by replacing the derivatives ($\mathcal{E}'$, $\mathcal{L}_z'$, $\mathcal{Q}'$) with the adiabatic fluxes, $f_{\textnormal{NK}}$, provided in Ref.~\cite{Gair}, and substituting $\ddot \theta = - \Gamma^{\theta}_{\lambda \rho} \dot x^{\lambda} \dot x^{\rho} + \mathcal{F}^{\theta}$ in Eq.~(\ref{carterDeriv}). At each step $n$ of the inspiral, we replace the expressions for $\mathcal{F}^{\nu}|_n$ in Eq.~(\ref{geodesicGSF}), and integrate the coupled set of second-order differential equations in the interval $[\tau_n, \tau_n + \Delta \tau]$, using initial positions and velocities from step $n-1$. 

The motion described by Eq.~(\ref{geodesicGSF}) is not geodesic, thus we cannot use the invariants to monitor the numerical error and constraint the trajectory. We integrate the MiSaTaQuWa equation using a \emph{Stiffness Switching method}, which detects stiffness at runtime for given accuracy and precision goals, and switches between explicit and implicit methods. Further details on how stiffness is detected, as well as the approach used to switch from non-stiff to stiff solvers, can be found in~\cite{Mathematica}.

To update the adiabatic fluxes, hence the forcing terms, we integrate Eqs.~(\ref{geotTau})-(\ref{geophiTau}) in the interval $[\tau_{n+1}, \tau_{n+1} + \overline{\Delta \tau}]$ and compute the orbital elements using the numerical solutions for $r(\tau)$ and $\theta(\tau)$. We then reiterate the whole procedure, by replacing $\mathcal{F}^{\nu}|_{n+1}$ in Eq.~(\ref{geodesicGSF}) and integrating the coupled set of second-order differential equations in the interval $[\tau_{n+1}, \tau_{n+1} + \Delta \tau]$, with initial positions and velocities from step $n$.

\subsection{Integration Methods}

In this section, we describe the numerical methods used to integrate the first-order and second-order sets of differential equations, namely the \emph{Implicit Runge--Kutta} and the \emph{Projected Implicit Runge--Kutta} algorithms. The numerical integration codes are implemented in Mathematica~\cite{Mathematica}.

\subsubsection{Implicit Runge--Kutta}
\label{sec::methods1}

The Runge--Kutta (RK) method is commonly used to numerically solve a system of differential equations written in the form
\begin{equation}
    \partial_t \Vec{y}(t) = \Vec{h}(t, \Vec{y}(t)).
\end{equation}
Given the initial conditions $\Vec{y}(t_\textrm{init})$, it solves for $\Vec{y}(t_\textrm{fin})$ by discretizing the time into steps of length $\Delta t$, and iteratively updating $\Vec{y}(t)$ at each time step. We compute $\Vec{y}(t + \Delta t)$ as 
\begin{equation}
    \Vec{y}(t + \Delta t) = \Vec{y}(t) + \Delta t \sum_{n = 1}^s b_n \Vec{k}_{(n)},
\end{equation}
where
\begin{equation}\label{functionK}
    \Vec{k}_{(n)} = \Vec{h} \Biggl ( t + c_n \Delta t, \: \Vec{y}(t) + \Delta t \sum_{m = 1}^s a_{nm} \Vec{k}_{(m)} \Biggr ),
\end{equation}
$s$ is the number of stages of the RK method, and the coefficients $a_{nm}$, $b_n$, $c_n$ define the method itself. 

Every RK method can be represented by a Butcher’s tableau~\cite{BUTCHER1996247} as
\begin{equation}
    \begin{matrix}
    \textbf{c} \:\vline & \hspace{-0.5cm} \textbf{A} \\
    \hline    
    \:\: \hspace{0.1cm} \vline & \hspace{-0.2cm} \textbf{b}^{\textbf{T}} 
    \end{matrix} 
    \:\:\:\: = \:\:\:\:
    \begin{matrix}
    c_1 \hspace{0.2cm} \vline & a_{11} & \cdots & a_{1s} \\
    \cdots \: \vline & \cdots & \cdots & \cdots \\
    c_s \hspace{0.21cm} \vline & a_{s1} & \cdots & a_{ss} \\ \hline    
    \hspace{0.54cm} \vline & b_1 & \cdots & b_s
    \end{matrix} 
\end{equation}
The structure of $\textnormal{\textbf{A}}$ decides whether the RK method is explicit or implicit. It is explicit when $a_{nm} = 0$ for $m \geq n$, that is $\textnormal{\textbf{A}}$ is strictly lower triangular. For $m < n$, $\Vec{k}_{(n)}$ depends only on $\Vec{y}(t)$ and $\Vec{k}_{(m)}$ and are sequentially obtained from $n = 1$ to $s$ by substituting the known values. When $\textnormal{\textbf{A}}$ is a generic matrix, the RK method is implicit. At each stage $s$, one must solve for $\Vec{k}_{(n)}$ the set of non-linear equations from Eq.~(\ref{functionK}), thus increasing the computational cost~\cite{Touroux:2023rkv}. 

Explicit RK methods are generally unstable when applied to stiff systems\footnote{A stiff equation includes terms that can lead to a rapid variation in the solution. Many differential equations exhibit some form of stiffness, which restricts the step size and the effectiveness of explicit methods.}, and require unacceptably small $\Delta t$ in regions where the solution curve is smooth. This happens because their stability region is small and bounded. 

Implicit RK methods, that have large stability regions, are more suitable for approximating solutions of stiff equations~\cite{2024arXiv240416665M}. The higher computational cost can be counterbalanced by taking larger $\Delta t$ in certain regions, while keeping the numerical error finite~\cite{Mathematica, Touroux:2023rkv}. Furthermore, implicit methods reach a given order of accuracy in fewer stages with respect to explicit ones~\cite{Touroux:2023rkv}.

\begin{figure}[t]
\centering
\includegraphics[scale=0.42]{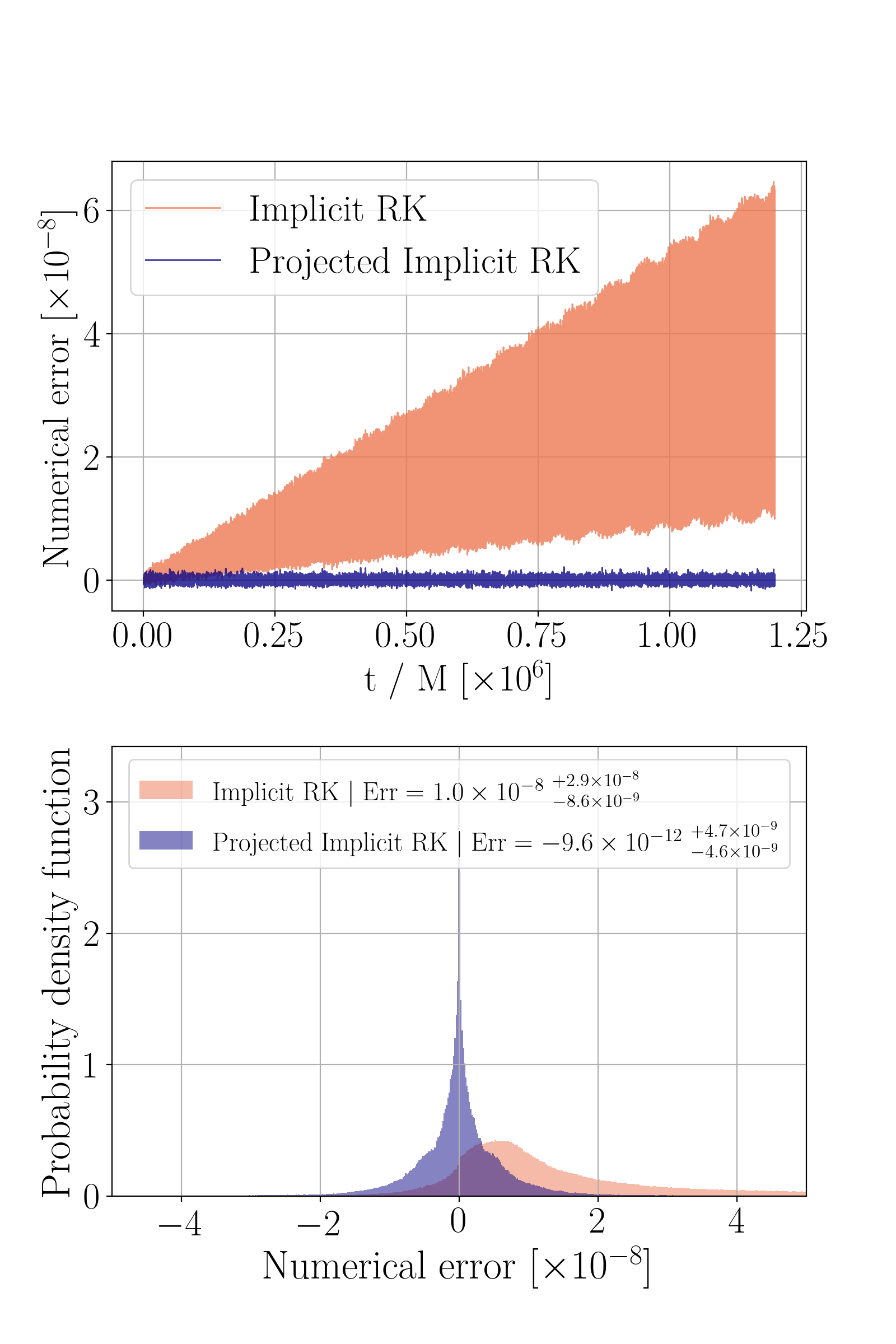} 
\caption{Comparison between Implicit RK and Projected Implicit RK methods in the second-order integration of a generic bound orbit in the Kerr spacetime, with initial parameters ($\eta$ = $10^{-5}$, $M$ = $10^6 M_{\odot}$, $a$ = $0.95$, $p/M$ = $8.91$, $e$ = $0.1$, $\iota$ = $1.3$). The total integration time is T$_{\textnormal{max}}$ = $1.2 \times 10^6 M$. We define the numerical error as the deviation from the expected value of Eq.~(\ref{constraintEq}), i.e., zero. With the Implicit RK, the numerical error grows with time and the oscillations are increasingly wider. With the Projected Implicit RK, the numerical error oscillates around zero and the oscillations are stable: this method is more robust for the required long evolution.}
\label{ImpRK_vs_ProjImpRK}
\end{figure}

\subsubsection{Projected Implicit Runge--Kutta}
\label{sec::methods2}

When a system has one or more invariants (i.e., integrals of motion), it is useful to adopt an integration method in which the numerical solution is constrained. 

The Projected Implicit RK method takes a time step using the Implicit RK, and then projects the approximate solution onto the manifold on which the exact solution evolves. The projection procedure allows the invariants of the system to be conserved along the motion~\cite{ProjRK, Mathematica}. For the purposes of this work, we use the integrals of geodesic motion as invariants to monitor the numerical error and constrain the trajectory. 

Fig.~\ref{ImpRK_vs_ProjImpRK} (top panel) shows the comparison between Implicit RK and Projected Implicit RK methods in the second-order integration of a generic bound orbit in the Kerr spacetime. With the Implicit RK, the numerical error grows with time and the oscillations are increasingly wider. With the Projected Implicit RK, the numerical error oscillates around zero and the oscillations are stable: this method is more robust for the required long evolution as it minimizes the acquired numerical error (see bottom panel of Fig.~\ref{ImpRK_vs_ProjImpRK}). 

\section{Duration of the $3 : 2$ resonance}
\label{sec::duration_3:2}

In this section, we compute the resonance duration for the EMRIs included in the parameter space of Fig.~\ref{param_space} from Eq.~(\ref{tRes}), where, in the case of the $3:2$ resonance, we set $l_* = -3$ and $m_* = 2$. To evaluate the radial and polar frequency derivatives at $t = t_0$, that is when $\omega_{\theta}/{\omega_r} = 1.5$, we evolve the EMRIs in the adiabatic approximation, and compute $\omega_r$, $\omega_{\theta}$ at each step of the inspiral using Eqs. (9)-(15)-(21) provided in Ref.~\cite{Fujita}. We then interpolate these values to obtain the functions $\omega_r(t)$, $\omega_{\theta}(t)$, and compute the derivatives $\dot \omega_{r 0}$, $\dot \omega_{\theta 0}$. 

\begin{figure}[t]
\centering
\includegraphics[scale=0.39]{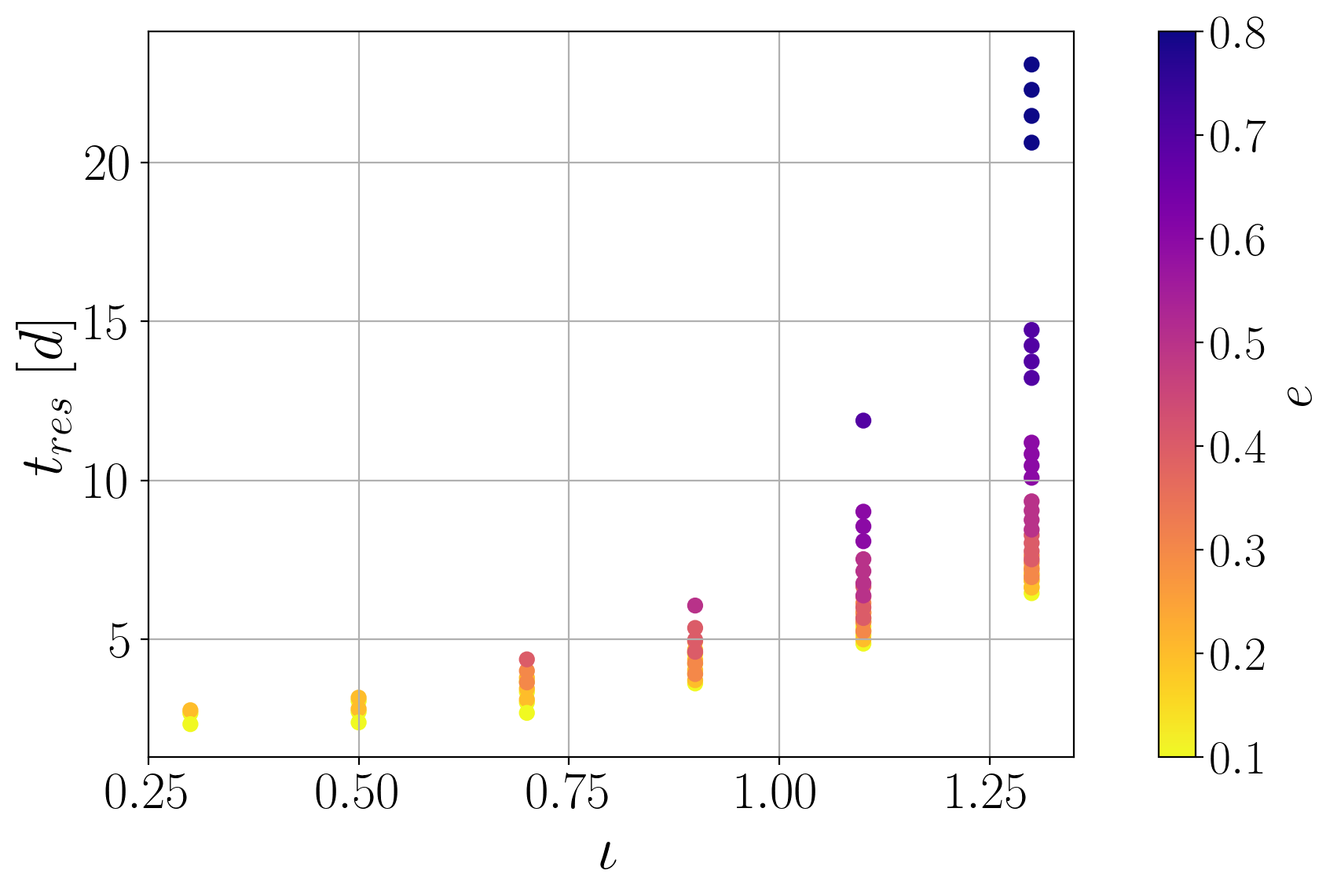}
\caption{Resonance duration, in days, as a function of the orbital inclination angle and the orbital eccentricity, in the case of the $3:2$ resonance. Using Eq.~(\ref{tRes}), we compute $t_{res}$ for the EMRIs included in the parameter space of Fig.~\ref{param_space}, with $\eta = 10^{-5}$ and $M = 10^6 M_{\odot}$. Estimates are of the order of days, with $t_{res} \in [2, 23]$ days. Systems with higher $\iota$ and $e$ show longer resonance duration. These results are consistent with App. B of Ref.~\cite{Ruangsri:2013hra}.}
\label{tResPlot2}
\end{figure}

Results are illustrated in Fig.~\ref{tResPlot2}: estimates for the resonance duration are of the order of days, with $t_{\textnormal{res}} \in [2 - 23]$ days. Systems with higher orbital inclination and eccentricity show higher resonance duration. 

We have compared the above results with the ones provided in Ref.~\cite{Ruangsri:2013hra}, where the authors use a definition of the resonance duration which differs from Eq.~(\ref{tRes}) by a factor of $\sqrt{\pi / 2}$ (see Eq. (2.17) in Ref.~\cite{Ruangsri:2013hra}). They compute, for several EMRI configurations, the number of oscillations in $\theta$ and $r$ spent near resonance as
\begin{equation}\label{Noscill}
    N_{\theta, r} = \frac{\omega_{\theta 0, \: r 0} \: \: t_{res}}{2 \pi}.
\end{equation}
These oscillations scale with the system's mass ratio as~\cite{Ruangsri:2013hra}
\begin{equation}
    N_{\theta, r} \sim \sqrt{\frac{1}{\eta}}.
\end{equation}
We have compute the resonance durations using the same definition provided in Ref.~\cite{Ruangsri:2013hra}. Then, we calculated $N_{\theta, r}$ from Eq.~(\ref{Noscill}), and rescaled the values for different mass ratios. We identify an overall agreement for the same EMRI parameter space, within a few percent. 

\section{Other Orbital Configurations}
\label{App::other_cases}

\subsection{Single resonant-crossings}
\label{sec::other_cases_single_kicks}

\begin{figure*}[t]
\centering
\includegraphics[scale=0.4]{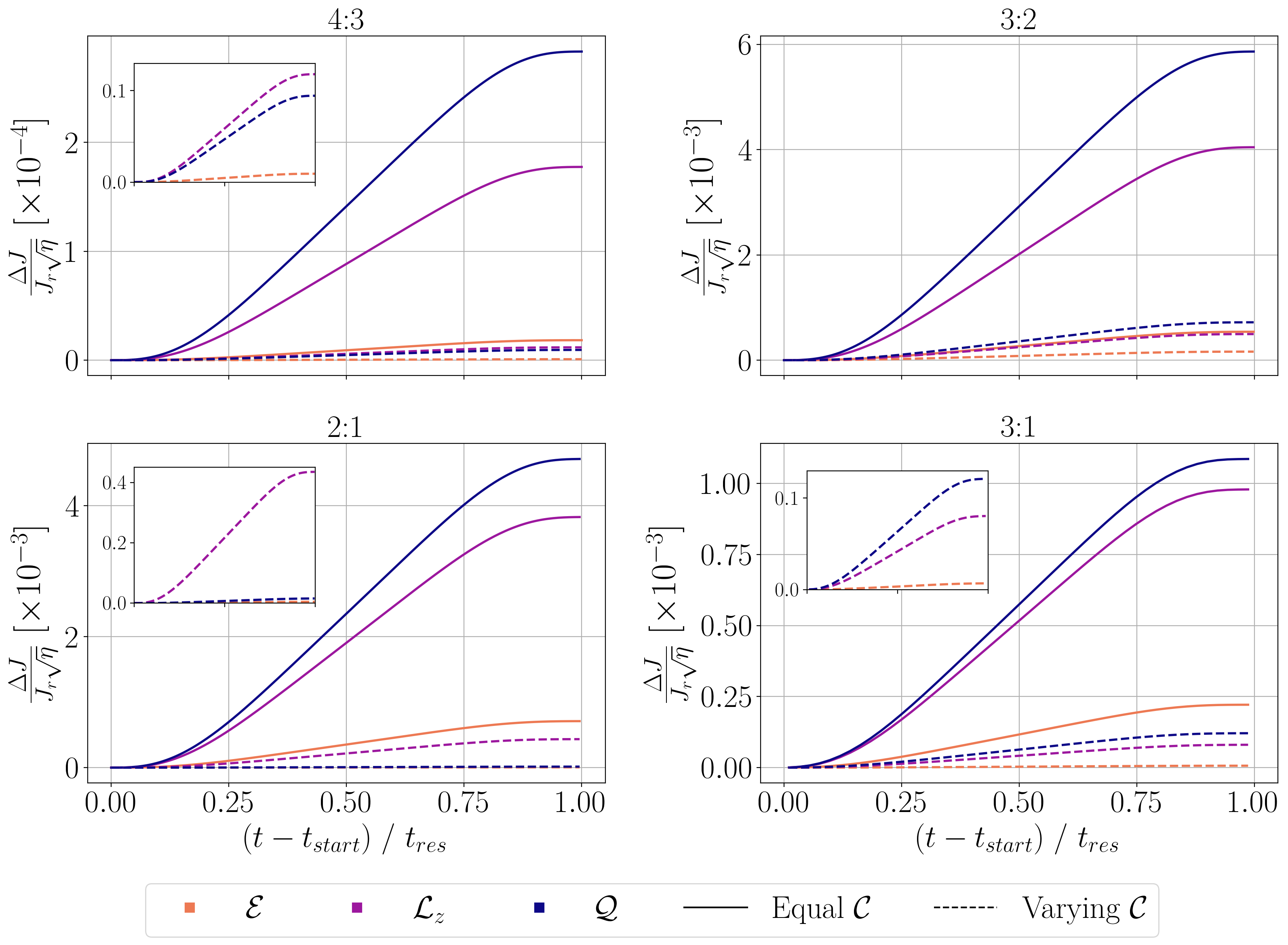} 
\caption{Evolution of the integrals of motion through single resonances ($4:3$, $3:2$, $2:1$ and $3:1$) for case (ii) of Table~\ref{Table_1}, with $a = 0.9$, $\eta = 10^{-5}$ and $M = 10^6 M_{\odot}$. The solid lines depict the scenario where, for each resonance, the resonant coefficients $\mathcal{C}$ are equal, i.e., $|\mathcal{C}|_{4:3} = 0.0006$,$ |\mathcal{C}|_{3:2} = 0.01$, $|\mathcal{C}|_{2:1} = 0.007$, $|\mathcal{C}|_{3:1} = 0.003$. The dashed lines indicate the scenario where $\mathcal{C}$ are different for each integrals of motion, i.e., $|\mathcal{C}|_{4:3} = (0.00003, 0.00004, 0.00002)$, $|\mathcal{C}|_{3:2} = (0.00303, 0.00123, 0.00123)$, $|\mathcal{C}|_{2:1} = (0.00004, 0.00080, 0.00002)$, $|\mathcal{C}|_{3:1} = (0.00008, 0.00024, 0.00033)$. The values for these coefficients were computed in Ref.~\cite{FlanaganHughes} and are also shown in Table~\ref{Table_1}. The insets zoom in on the \textit{Varying} $\mathcal{C}$ scenario for the $4:3$, $2:1$, $3:1$ resonances.}
\label{kick_single_case_ii}
\end{figure*}

\begin{figure*}[t]
\centering
\includegraphics[scale=0.4]{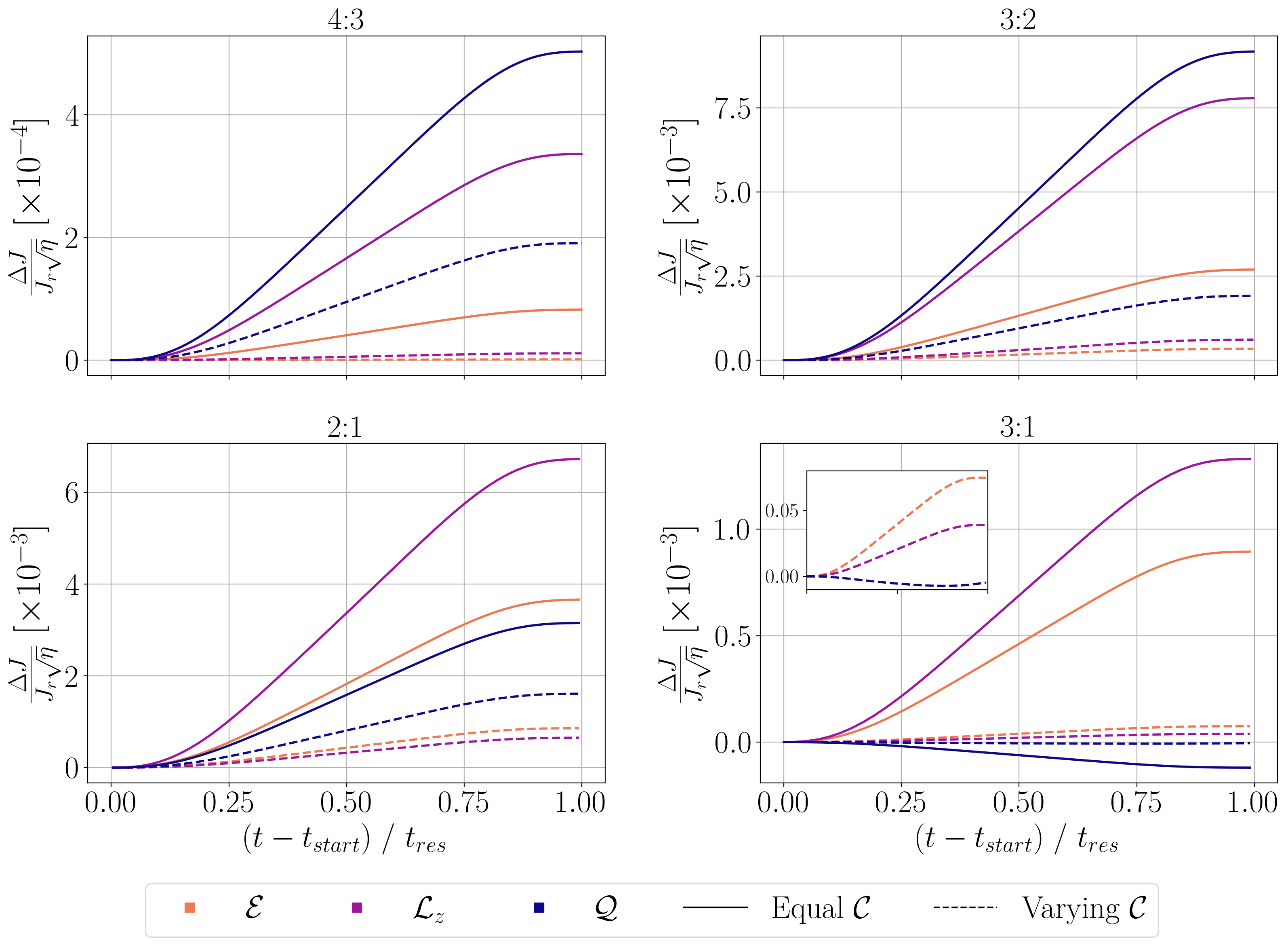} 
\caption{Evolution of the integrals of motion through single resonances ($4:3$, $3:2$, $2:1$ and $3:1$) for case (iii) of Table~\ref{Table_1}, with $a = 0.9$, $\eta = 10^{-5}$ and $M = 10^6 M_{\odot}$. The solid lines depict the scenario where, for each resonance, the resonant coefficients $\mathcal{C}$ are equal, i.e., $|\mathcal{C}|_{4:3} = 0.0006$,$ |\mathcal{C}|_{3:2} = 0.01$, $|\mathcal{C}|_{2:1} = 0.007$, $|\mathcal{C}|_{3:1} = 0.003$. The dashed lines indicate the scenario where $\mathcal{C}$ are different for each integrals of motion, i.e., $|\mathcal{C}|_{4:3} = (0.00001, 0.00002, 0.00023)$, $|\mathcal{C}|_{3:2} = (0.00127, 0.00078, 0.00210)$, $|\mathcal{C}|_{2:1} = (0.00167, 0.00067, 0.00357)$, $|\mathcal{C}|_{3:1} = (0.00026, 0.00009, 0.00035)$. The values for these coefficients were computed in Ref.~\cite{FlanaganHughes} and are also shown in Table~\ref{Table_1}. The inset in the bottom right panel zooms in on the \textit{Varying} $\mathcal{C}$ scenario for the $3:1$ resonance. For this orbital configuration, the $3:1$ resonance is located way beyond the NK limit of $r_p = 5M$, and the evolution is performed very close to the separatrix: hence, the peculiar behavior of $\mathcal{Q}$ should be checked for consistency with Teukolsky-based inspirals.}
\label{kick_single_case_iii}
\end{figure*}

\begin{figure*}[t]
\centering
\includegraphics[scale=0.4]{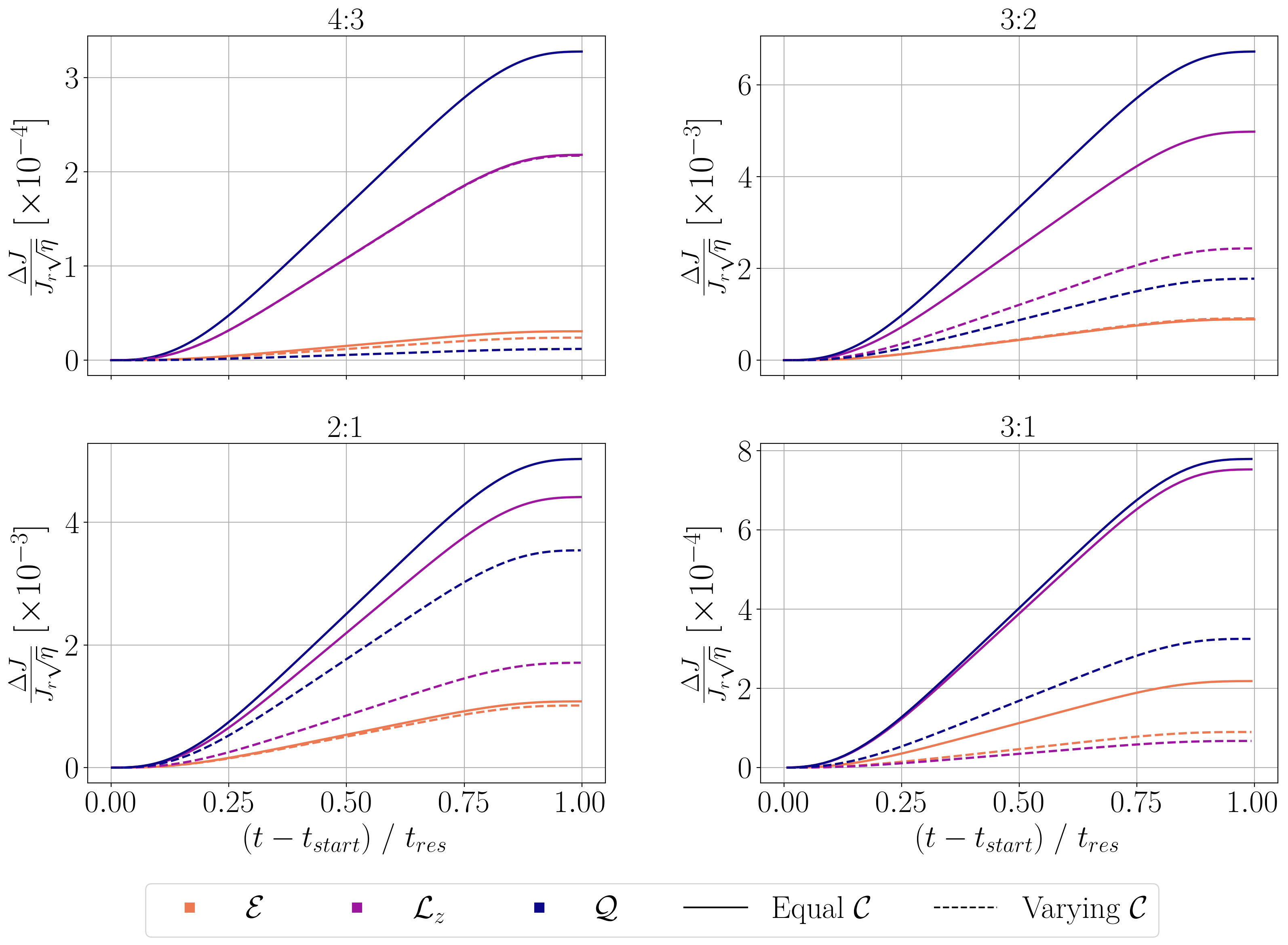} 
\caption{Evolution of the integrals of motion through single resonances ($4:3$, $3:2$, $2:1$ and $3:1$) for case (iv) of Table~\ref{Table_1}, with $a = 0.9$, $\eta = 10^{-5}$ and $M = 10^6 M_{\odot}$. The solid lines depict the scenario where, for each resonance, the resonant coefficients $\mathcal{C}$ are equal, i.e., $|\mathcal{C}|_{4:3} = 0.0006$,$ |\mathcal{C}|_{3:2} = 0.01$, $|\mathcal{C}|_{2:1} = 0.007$, $|\mathcal{C}|_{3:1} = 0.003$. The dashed lines indicate the scenario where $\mathcal{C}$ are different for each integrals of motion, i.e., $|\mathcal{C}|_{4:3} = (0.00047, 0.00060, 0.00002)$, $|\mathcal{C}|_{3:2} = (0.01030, 0.00489, 0.00261)$, $|\mathcal{C}|_{2:1} = (0.00662, 0.00270, 0.00494)$, $|\mathcal{C}|_{3:1} = (0.00125, 0.00027, 0.00126)$. The values for these coefficients were computed in Ref.~\cite{FlanaganHughes} and are also shown in Table~\ref{Table_1}.}
\label{kick_single_case_iv}
\end{figure*}

In this section we present the resonant flux augmentation of the integrals of motion from the start to the end of the investigated resonances ($4:3$, $3:2$, $2:1$, $3:1$), for the case (ii)-(iv) orbits of Table~\ref{Table_1}, which are shown in Figures~\ref{kick_single_case_ii}-\ref{kick_single_case_iii}-\ref{kick_single_case_iv} respectively. We either keep equal $\mathcal{C}$ coefficients and set them to their maximum value found in Table \ref{Table_1} (shown in bold), or varying $\mathcal{C}$ coefficients that have been calculated in Ref. \cite{FlanaganHinderer2} and are presented again in Table \ref{Table_1}.

We observe that for equal $\mathcal{C}$, the Carter constant $Q$ obtains the maximum augmentation when the EMRI is passing from all subdominant resonances, separately. The second most augmented flux is that of $\mathcal{L}_z$, while the energy $\mathcal{E}$ is the least augmented in all cases. For the varying $\mathcal{C}$ scenario, the energy $\mathcal{E}$ still has the least augmented flux, while $\mathcal{Q}$ fluxes dominate for the $3:2$ and $3:1$ resonances. Finally, $\mathcal{L}_z$ is slightly more augmented for the $4:3$ resonance-crossing while for the case of the $2:1$ resonance $\mathcal{L}_z$ is significantly more augmented than the rest of the fluxes. 

\subsection{Multiple Resonant-crossings}
\label{sec::other_cases_multi_kicks}

\begin{figure*}[t]
    \includegraphics[scale=0.073]{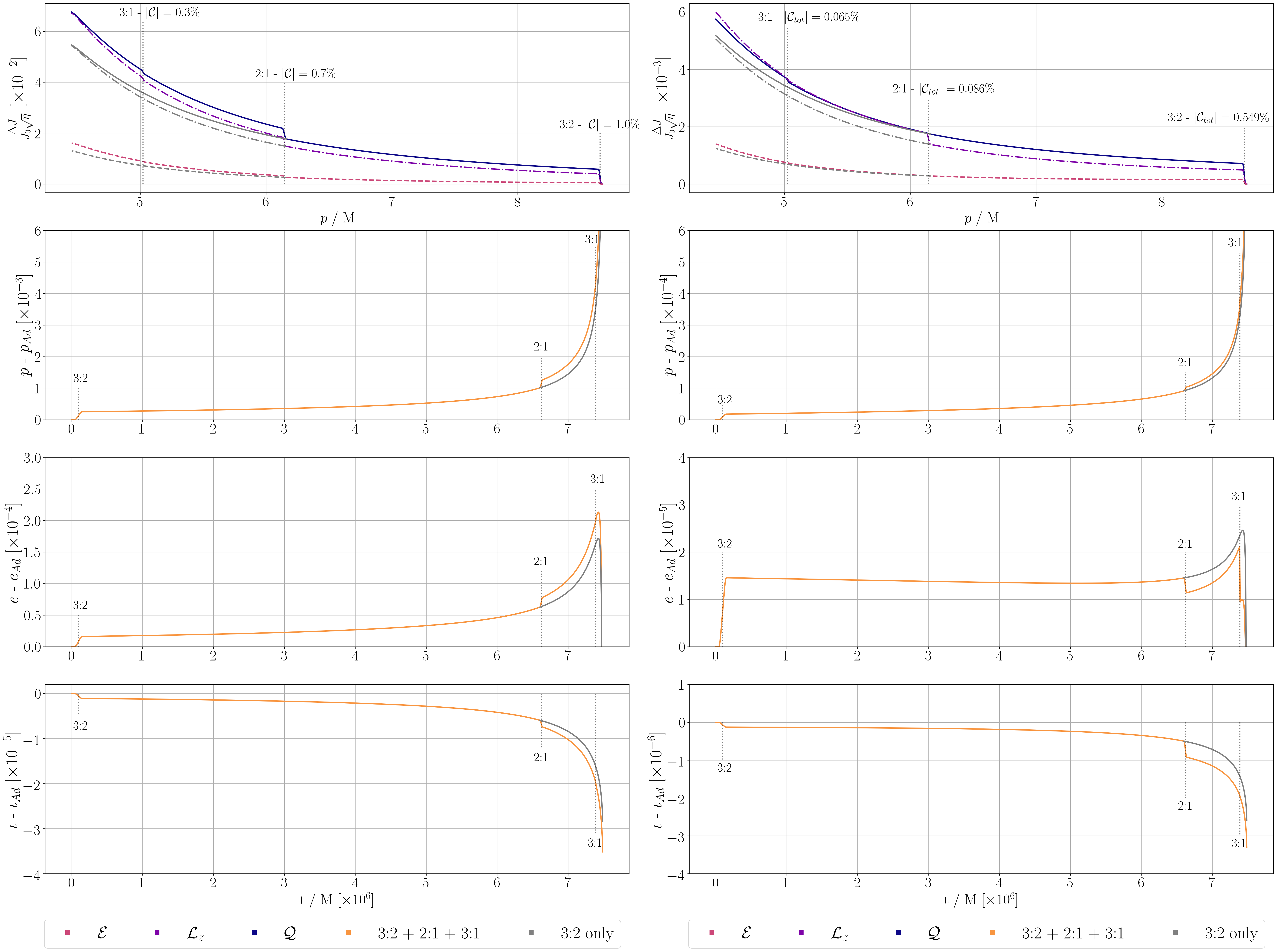} 
    \caption{\emph{Left:} Evolution of the integrals of motion through multiple resonances ($3:2$ + $2:1$ + $3:1$) for the case (ii) orbit of Table~\ref{Table_1}, with $a = 0.9$, $\eta = 10^{-5}$ and $M = 10^6 M_{\odot}$ and equal resonance coefficients, i.e., $|\mathcal{C}|_{3:2} = 0.01$, $|\mathcal{C}|_{2:1} = 0.007$, $|\mathcal{C}|_{3:1} = 0.003$. The solid, dashed and dot-dashed gray lines indicate the evolution where we activate only the $3:2$ resonance. The dotted gray lines show the location of the $3:2$, $2:1$ and $3:1$ resonances. \emph{Right:} Same as left with varying coefficients, i.e., $|\mathcal{C}|_{3:2} = (0.00303, 0.00123, 0.00123)$, $|\mathcal{C}|_{2:1} = (0.00004, 0.00080, 0.00002)$, $|\mathcal{C}|_{3:1} = (0.00008, 0.00024, 0.00033)$.}
    \label{kick_multi_case_ii}
\end{figure*}

\begin{figure*}[t]
    \includegraphics[scale=0.073]{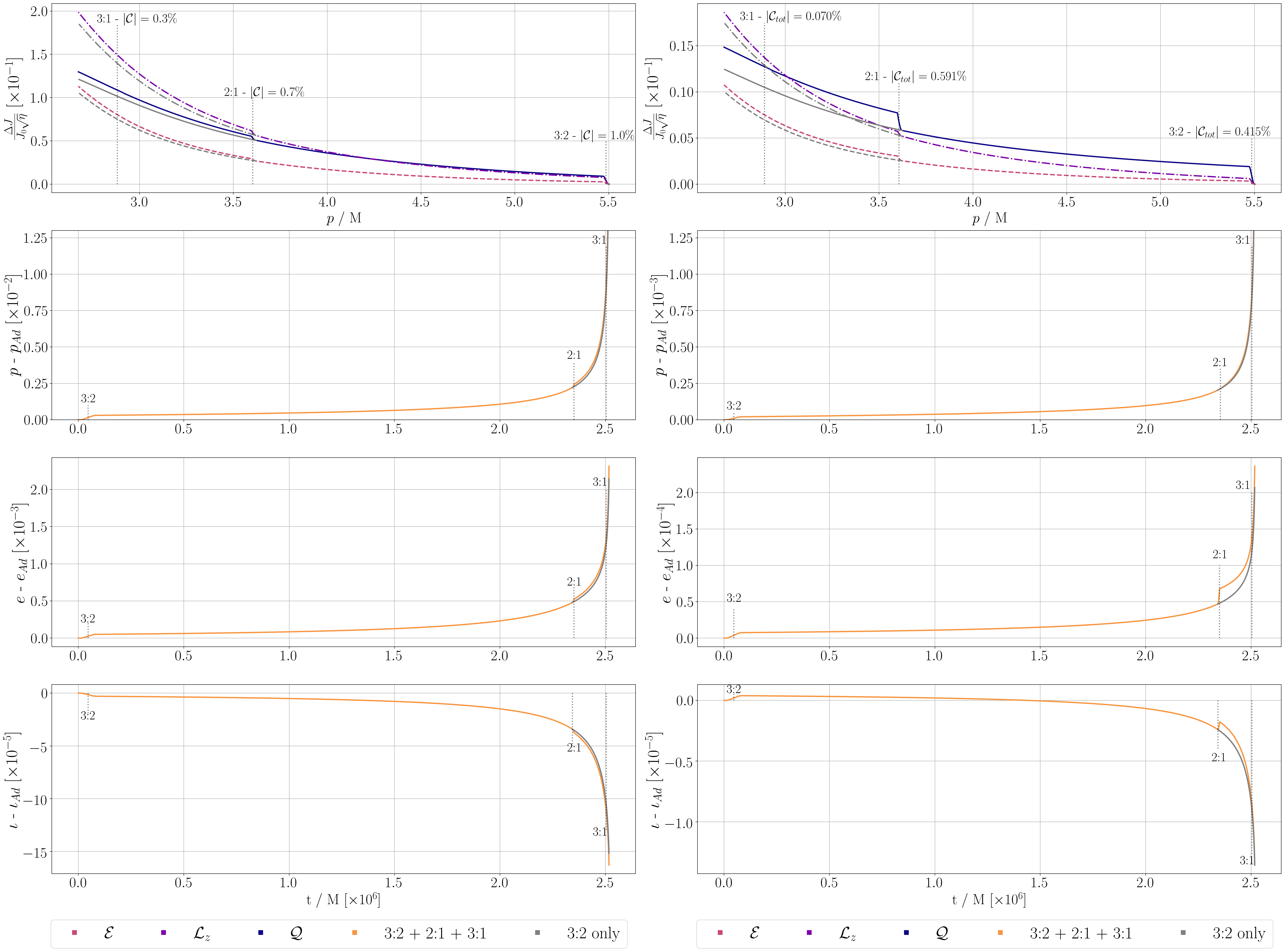}
    \caption{\emph{Left:} Evolution of the integrals of motion through multiple resonances ($3:2$ + $2:1$ + $3:1$) for the case (iii) orbit of Table~\ref{Table_1}, with $a = 0.9$, $\eta = 10^{-5}$ and $M = 10^6 M_{\odot}$ and equal resonance coefficients, i.e., $|\mathcal{C}|_{3:2} = 0.01$, $|\mathcal{C}|_{2:1} = 0.007$, $|\mathcal{C}|_{3:1} = 0.003$. The solid, dashed and dot-dashed gray lines indicate the evolution where we activate only the $3:2$ resonance. The dotted gray lines show the location of the $3:2$, $2:1$ and $3:1$ resonances. \emph{Right:} Same as left with varying coefficients, i.e., $|\mathcal{C}|_{3:2} = (0.00127, 0.00078, 0.00210)$, $|\mathcal{C}|_{2:1} = (0.00167, 0.00067, 0.00357)$, $|\mathcal{C}|_{3:1} = (0.00026, 0.00009, 0.00035)$.}
    \label{kick_multi_case_iii}
\end{figure*}

\begin{figure*}[t]
    \includegraphics[scale=0.073]{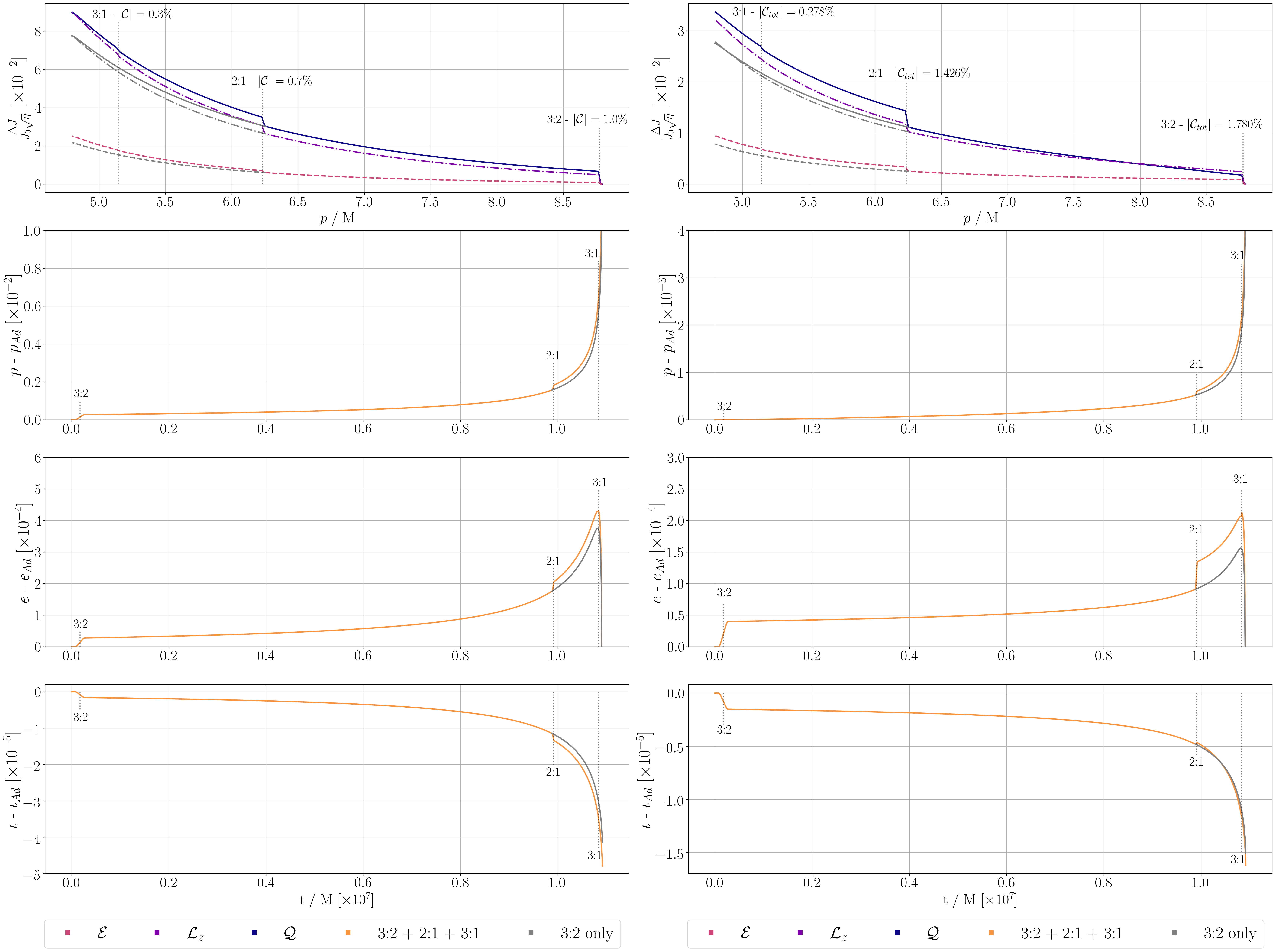}
    \caption{\emph{Left:} Evolution of the integrals of motion through multiple resonances ($3:2$ + $2:1$ + $3:1$) for the case (iv) orbit of Table~\ref{Table_1}, with $a = 0.9$, $\eta = 10^{-5}$ and $M = 10^6 M_{\odot}$ and equal resonance coefficients, i.e., $|\mathcal{C}|_{3:2} = 0.01$, $|\mathcal{C}|_{2:1} = 0.007$, $|\mathcal{C}|_{3:1} = 0.003$. The solid, dashed and dot-dashed gray lines indicate the evolution where we activate only the $3:2$ resonance. The dotted gray lines show the location of the $3:2$, $2:1$ and $3:1$ resonances. \emph{Right:} Same as left with varying coefficients, i.e., $|\mathcal{C}|_{3:2} = (0.01030, 0.00489, 0.00261)$, $|\mathcal{C}|_{2:1} = (0.00662, 0.00270, 0.00494)$, $|\mathcal{C}|_{3:1} = (0.00125, 0.00027, 0.00126)$.}
    \label{kick_multi_case_iv}
\end{figure*}

In this section we present the effects of multiple resonant-crossings in a single inspiral for the case (ii)-(iv) orbits of Table~\ref{Table_1}, which are shown in Figures~\ref{kick_multi_case_ii}-\ref{kick_multi_case_iii}-\ref{kick_multi_case_iv} respectively. We start the trajectory close to the $3:2$ resonance, and evolve the systems through the $3:2$, $2:1$ and $3:1$ resonances. We terminate the evolution when the secondary reaches the separatrix. We then compute the relative changes in ($\mathcal E, \mathcal L_z, \mathcal Q$) and in ($p/M, e, \iota$), and compare them to the evolution where we activate only the $3:2$ resonance and not the low-order ones. 

Overall, how the flux changes in ($\mathcal E, \mathcal L_z, \mathcal Q$) translate into the orbital elements, and ultimately affect the evolution of the GW frequency, is very convoluted: it does not only depend on the values of the individual resonance coefficients ($\mathcal{C}_{\mathcal{E}}$, $\mathcal{C}_{\mathcal{L}_z}$, $\mathcal{C}_{\mathcal{Q}}$), but also on their relative strength. The results are even more prominent for the \textit{varying} $\mathcal{C}$ case. This deserves further studies, especially for the purposes of implementing data analysis for generic orbits.

\bibliography{Cumulative_effect_resonances_EMRI}

\begin{thebibliography}{86}%
\makeatletter
\providecommand \@ifxundefined [1]{%
 \@ifx{#1\undefined}
}%
\providecommand \@ifnum [1]{%
 \ifnum #1\expandafter \@firstoftwo
 \else \expandafter \@secondoftwo
 \fi
}%
\providecommand \@ifx [1]{%
 \ifx #1\expandafter \@firstoftwo
 \else \expandafter \@secondoftwo
 \fi
}%
\providecommand \natexlab [1]{#1}%
\providecommand \enquote  [1]{``#1''}%
\providecommand \bibnamefont  [1]{#1}%
\providecommand \bibfnamefont [1]{#1}%
\providecommand \citenamefont [1]{#1}%
\providecommand \href@noop [0]{\@secondoftwo}%
\providecommand \href [0]{\begingroup \@sanitize@url \@href}%
\providecommand \@href[1]{\@@startlink{#1}\@@href}%
\providecommand \@@href[1]{\endgroup#1\@@endlink}%
\providecommand \@sanitize@url [0]{\catcode `\\12\catcode `\$12\catcode
  `\&12\catcode `\#12\catcode `\^12\catcode `\_12\catcode `\%12\relax}%
\providecommand \@@startlink[1]{}%
\providecommand \@@endlink[0]{}%
\providecommand \url  [0]{\begingroup\@sanitize@url \@url }%
\providecommand \@url [1]{\endgroup\@href {#1}{\urlprefix }}%
\providecommand \urlprefix  [0]{URL }%
\providecommand \Eprint [0]{\href }%
\providecommand \doibase [0]{http://dx.doi.org/}%
\providecommand \selectlanguage [0]{\@gobble}%
\providecommand \bibinfo  [0]{\@secondoftwo}%
\providecommand \bibfield  [0]{\@secondoftwo}%
\providecommand \translation [1]{[#1]}%
\providecommand \BibitemOpen [0]{}%
\providecommand \bibitemStop [0]{}%
\providecommand \bibitemNoStop [0]{.\EOS\space}%
\providecommand \EOS [0]{\spacefactor3000\relax}%
\providecommand \BibitemShut  [1]{\csname bibitem#1\endcsname}%
\let\auto@bib@innerbib\@empty
\bibitem [{\citenamefont {Abbott}\ \emph {et~al.}(2016)\citenamefont {Abbott}
  \emph {et~al.}}]{LIGO}%
  \BibitemOpen
  \bibfield  {author} {\bibinfo {author} {\bibfnamefont {B.~P.}\ \bibnamefont
  {Abbott}} \emph {et~al.} (\bibinfo {collaboration} {LIGO Scientific
  Collaboration and Virgo Collaboration}),\ }\href {\doibase
  10.1103/PhysRevLett.116.061102} {\bibfield  {journal} {\bibinfo  {journal}
  {Phys. Rev. Lett.}\ }\textbf {\bibinfo {volume} {116}},\ \bibinfo {pages}
  {061102} (\bibinfo {year} {2016})}\BibitemShut {NoStop}%
\bibitem [{\citenamefont {Davis}\ \emph {et~al.}(2021)\citenamefont {Davis}
  \emph {et~al.}}]{LIGO:2021ppb}%
  \BibitemOpen
  \bibfield  {author} {\bibinfo {author} {\bibfnamefont {D.}~\bibnamefont
  {Davis}} \emph {et~al.} (\bibinfo {collaboration} {LIGO}),\ }\href {\doibase
  10.1088/1361-6382/abfd85} {\bibfield  {journal} {\bibinfo  {journal} {Class.
  Quant. Grav.}\ }\textbf {\bibinfo {volume} {38}},\ \bibinfo {pages} {135014}
  (\bibinfo {year} {2021})},\ \Eprint {http://arxiv.org/abs/2101.11673}
  {arXiv:2101.11673 [astro-ph.IM]} \BibitemShut {NoStop}%
\bibitem [{\citenamefont {Abbott}\ \emph {et~al.}(2023)\citenamefont {Abbott}
  \emph {et~al.}}]{KAGRA:2021vkt}%
  \BibitemOpen
  \bibfield  {author} {\bibinfo {author} {\bibfnamefont {R.}~\bibnamefont
  {Abbott}} \emph {et~al.} (\bibinfo {collaboration} {KAGRA, VIRGO, LIGO
  Scientific}),\ }\href {\doibase 10.1103/PhysRevX.13.041039} {\bibfield
  {journal} {\bibinfo  {journal} {Phys. Rev. X}\ }\textbf {\bibinfo {volume}
  {13}},\ \bibinfo {pages} {041039} (\bibinfo {year} {2023})},\ \Eprint
  {http://arxiv.org/abs/2111.03606} {arXiv:2111.03606 [gr-qc]} \BibitemShut
  {NoStop}%
\bibitem [{\citenamefont {Einstein}(1916)}]{Einstein:1916cc}%
  \BibitemOpen
  \bibfield  {author} {\bibinfo {author} {\bibfnamefont {A.}~\bibnamefont
  {Einstein}},\ }\href@noop {} {\bibfield  {journal} {\bibinfo  {journal}
  {Sitzungsber. Preuss. Akad. Wiss. Berlin (Math. Phys. )}\ }\textbf {\bibinfo
  {volume} {1916}},\ \bibinfo {pages} {688} (\bibinfo {year}
  {1916})}\BibitemShut {NoStop}%
\bibitem [{\citenamefont {Barack}\ \emph {et~al.}(2019)\citenamefont {Barack}
  \emph {et~al.}}]{Barack:2018yly}%
  \BibitemOpen
  \bibfield  {author} {\bibinfo {author} {\bibfnamefont {L.}~\bibnamefont
  {Barack}} \emph {et~al.},\ }\href {\doibase 10.1088/1361-6382/ab0587}
  {\bibfield  {journal} {\bibinfo  {journal} {Class. Quant. Grav.}\ }\textbf
  {\bibinfo {volume} {36}},\ \bibinfo {pages} {143001} (\bibinfo {year}
  {2019})},\ \Eprint {http://arxiv.org/abs/1806.05195} {arXiv:1806.05195
  [gr-qc]} \BibitemShut {NoStop}%
\bibitem [{\citenamefont {Akiyama}\ \emph {et~al.}(2019)\citenamefont {Akiyama}
  \emph {et~al.}}]{EventHorizonTelescope:2019dse}%
  \BibitemOpen
  \bibfield  {author} {\bibinfo {author} {\bibfnamefont {K.}~\bibnamefont
  {Akiyama}} \emph {et~al.} (\bibinfo {collaboration} {Event Horizon
  Telescope}),\ }\href {\doibase 10.3847/2041-8213/ab0ec7} {\bibfield
  {journal} {\bibinfo  {journal} {Astrophys. J. Lett.}\ }\textbf {\bibinfo
  {volume} {875}},\ \bibinfo {pages} {L1} (\bibinfo {year} {2019})},\ \Eprint
  {http://arxiv.org/abs/1906.11238} {arXiv:1906.11238 [astro-ph.GA]}
  \BibitemShut {NoStop}%
\bibitem [{\citenamefont {Akiyama}\ \emph {et~al.}(2022)\citenamefont {Akiyama}
  \emph {et~al.}}]{EventHorizonTelescope:2022wkp}%
  \BibitemOpen
  \bibfield  {author} {\bibinfo {author} {\bibfnamefont {K.}~\bibnamefont
  {Akiyama}} \emph {et~al.} (\bibinfo {collaboration} {Event Horizon
  Telescope}),\ }\href {\doibase 10.3847/2041-8213/ac6674} {\bibfield
  {journal} {\bibinfo  {journal} {Astrophys. J. Lett.}\ }\textbf {\bibinfo
  {volume} {930}},\ \bibinfo {pages} {L12} (\bibinfo {year} {2022})},\ \Eprint
  {http://arxiv.org/abs/2311.08680} {arXiv:2311.08680 [astro-ph.HE]}
  \BibitemShut {NoStop}%
\bibitem [{\citenamefont {Bambi}(2024)}]{Bambi:2024hhi}%
  \BibitemOpen
  \bibfield  {author} {\bibinfo {author} {\bibfnamefont {C.}~\bibnamefont
  {Bambi}}\ }(\bibinfo {year} {2024})\ \Eprint
  {http://arxiv.org/abs/2408.12262} {arXiv:2408.12262 [astro-ph.HE]}
  \BibitemShut {NoStop}%
\bibitem [{\citenamefont {Schutz}(1999)}]{Schutz:1999xj}%
  \BibitemOpen
  \bibfield  {author} {\bibinfo {author} {\bibfnamefont {B.~F.}\ \bibnamefont
  {Schutz}},\ }\href {\doibase 10.1088/0264-9381/16/12A/307} {\bibfield
  {journal} {\bibinfo  {journal} {Class. Quant. Grav.}\ }\textbf {\bibinfo
  {volume} {16}},\ \bibinfo {pages} {A131} (\bibinfo {year} {1999})},\ \Eprint
  {http://arxiv.org/abs/gr-qc/9911034} {arXiv:gr-qc/9911034} \BibitemShut
  {NoStop}%
\bibitem [{\citenamefont {Bishop}(2022)}]{Bishop:2021rye}%
  \BibitemOpen
  \bibfield  {author} {\bibinfo {author} {\bibfnamefont {N.~T.}\ \bibnamefont
  {Bishop}},\ }\enquote {\bibinfo {title} {{Introduction to Gravitational Wave
  Astronomy}},}\ \ (\bibinfo {year} {2022})\ \Eprint
  {http://arxiv.org/abs/2103.07675} {arXiv:2103.07675 [gr-qc]} \BibitemShut
  {NoStop}%
\bibitem [{\citenamefont {Bailes}\ \emph {et~al.}(2021)\citenamefont {Bailes}
  \emph {et~al.}}]{Bailes:2021tot}%
  \BibitemOpen
  \bibfield  {author} {\bibinfo {author} {\bibfnamefont {M.}~\bibnamefont
  {Bailes}} \emph {et~al.},\ }\href {\doibase 10.1038/s42254-021-00303-8}
  {\bibfield  {journal} {\bibinfo  {journal} {Nature Rev. Phys.}\ }\textbf
  {\bibinfo {volume} {3}},\ \bibinfo {pages} {344} (\bibinfo {year}
  {2021})}\BibitemShut {NoStop}%
\bibitem [{\citenamefont {{Amaro-Seoane}}\ \emph {et~al.}(2017)\citenamefont
  {{Amaro-Seoane}} \emph {et~al.}}]{LISA}%
  \BibitemOpen
  \bibfield  {author} {\bibinfo {author} {\bibfnamefont {P.}~\bibnamefont
  {{Amaro-Seoane}}} \emph {et~al.},\ }\href {\doibase
  10.48550/arXiv.1702.00786} {\bibfield  {journal} {\bibinfo  {journal} {arXiv
  e-prints}\ ,\ \bibinfo {eid} {arXiv:1702.00786}} (\bibinfo {year} {2017})},\
  \Eprint {http://arxiv.org/abs/1702.00786} {arXiv:1702.00786 [astro-ph.IM]}
  \BibitemShut {NoStop}%
\bibitem [{\citenamefont {Colpi}\ \emph {et~al.}(2024)\citenamefont {Colpi}
  \emph {et~al.}}]{LISA:2024hlh}%
  \BibitemOpen
  \bibfield  {author} {\bibinfo {author} {\bibfnamefont {M.}~\bibnamefont
  {Colpi}} \emph {et~al.} (\bibinfo {collaboration} {LISA}),\ }\href@noop {} {\
   (\bibinfo {year} {2024})},\ \Eprint {http://arxiv.org/abs/2402.07571}
  {arXiv:2402.07571 [astro-ph.CO]} \BibitemShut {NoStop}%
\bibitem [{\citenamefont {{C{\'a}rdenas-Avenda{\~n}o}}\ and\ \citenamefont
  {{Sopuerta}}(2024)}]{CardenasSopuertaReview}%
  \BibitemOpen
  \bibfield  {author} {\bibinfo {author} {\bibfnamefont {A.}~\bibnamefont
  {{C{\'a}rdenas-Avenda{\~n}o}}}\ and\ \bibinfo {author} {\bibfnamefont
  {C.~F.}\ \bibnamefont {{Sopuerta}}},\ }\href {\doibase
  10.48550/arXiv.2401.08085} {\bibfield  {journal} {\bibinfo  {journal} {arXiv
  e-prints}\ ,\ \bibinfo {eid} {arXiv:2401.08085}} (\bibinfo {year} {2024})},\
  \Eprint {http://arxiv.org/abs/2401.08085} {arXiv:2401.08085 [gr-qc]}
  \BibitemShut {NoStop}%
\bibitem [{\citenamefont {Hughes}\ \emph {et~al.}(2021)\citenamefont {Hughes},
  \citenamefont {Warburton}, \citenamefont {Khanna}, \citenamefont {Chua},\
  and\ \citenamefont {Katz}}]{Hughes}%
  \BibitemOpen
  \bibfield  {author} {\bibinfo {author} {\bibfnamefont {S.~A.}\ \bibnamefont
  {Hughes}}, \bibinfo {author} {\bibfnamefont {N.}~\bibnamefont {Warburton}},
  \bibinfo {author} {\bibfnamefont {G.}~\bibnamefont {Khanna}}, \bibinfo
  {author} {\bibfnamefont {A.~J.~K.}\ \bibnamefont {Chua}}, \ and\ \bibinfo
  {author} {\bibfnamefont {M.~L.}\ \bibnamefont {Katz}},\ }\href {\doibase
  10.1103/PhysRevD.103.104014} {\bibfield  {journal} {\bibinfo  {journal}
  {Phys. Rev. D}\ }\textbf {\bibinfo {volume} {103}},\ \bibinfo {pages}
  {104014} (\bibinfo {year} {2021})}\BibitemShut {NoStop}%
\bibitem [{\citenamefont {Babak}\ \emph {et~al.}(2006)\citenamefont {Babak},
  \citenamefont {Fang}, \citenamefont {Gair}, \citenamefont {Glampedakis},\
  and\ \citenamefont {Hughes}}]{Kludge}%
  \BibitemOpen
  \bibfield  {author} {\bibinfo {author} {\bibfnamefont {S.}~\bibnamefont
  {Babak}}, \bibinfo {author} {\bibfnamefont {H.}~\bibnamefont {Fang}},
  \bibinfo {author} {\bibfnamefont {J.}~\bibnamefont {Gair}}, \bibinfo {author}
  {\bibfnamefont {K.}~\bibnamefont {Glampedakis}}, \ and\ \bibinfo {author}
  {\bibfnamefont {S.}~\bibnamefont {Hughes}},\ }\href {\doibase
  10.1103/PhysRevD.75.024005} {\bibfield  {journal} {\bibinfo  {journal}
  {Physical Review D}\ }\textbf {\bibinfo {volume} {75}} (\bibinfo {year}
  {2006}),\ 10.1103/PhysRevD.75.024005}\BibitemShut {NoStop}%
\bibitem [{\citenamefont {Babak}\ \emph {et~al.}(2017)\citenamefont {Babak},
  \citenamefont {Gair}, \citenamefont {Sesana}, \citenamefont {Barausse},
  \citenamefont {Sopuerta}, \citenamefont {Berry}, \citenamefont {Berti},
  \citenamefont {Amaro-Seoane}, \citenamefont {Petiteau},\ and\ \citenamefont
  {Klein}}]{Babak:2017tow}%
  \BibitemOpen
  \bibfield  {author} {\bibinfo {author} {\bibfnamefont {S.}~\bibnamefont
  {Babak}}, \bibinfo {author} {\bibfnamefont {J.}~\bibnamefont {Gair}},
  \bibinfo {author} {\bibfnamefont {A.}~\bibnamefont {Sesana}}, \bibinfo
  {author} {\bibfnamefont {E.}~\bibnamefont {Barausse}}, \bibinfo {author}
  {\bibfnamefont {C.~F.}\ \bibnamefont {Sopuerta}}, \bibinfo {author}
  {\bibfnamefont {C.~P.~L.}\ \bibnamefont {Berry}}, \bibinfo {author}
  {\bibfnamefont {E.}~\bibnamefont {Berti}}, \bibinfo {author} {\bibfnamefont
  {P.}~\bibnamefont {Amaro-Seoane}}, \bibinfo {author} {\bibfnamefont
  {A.}~\bibnamefont {Petiteau}}, \ and\ \bibinfo {author} {\bibfnamefont
  {A.}~\bibnamefont {Klein}},\ }\href {\doibase 10.1103/PhysRevD.95.103012}
  {\bibfield  {journal} {\bibinfo  {journal} {Phys. Rev. D}\ }\textbf {\bibinfo
  {volume} {95}},\ \bibinfo {pages} {103012} (\bibinfo {year} {2017})},\
  \Eprint {http://arxiv.org/abs/1703.09722} {arXiv:1703.09722 [gr-qc]}
  \BibitemShut {NoStop}%
\bibitem [{\citenamefont {Speri}\ \emph {et~al.}(2024)\citenamefont {Speri},
  \citenamefont {Barsanti}, \citenamefont {Maselli}, \citenamefont {Sotiriou},
  \citenamefont {Warburton}, \citenamefont {van~de Meent}, \citenamefont
  {Chua}, \citenamefont {Burke},\ and\ \citenamefont {Gair}}]{Speri:2024qak}%
  \BibitemOpen
  \bibfield  {author} {\bibinfo {author} {\bibfnamefont {L.}~\bibnamefont
  {Speri}}, \bibinfo {author} {\bibfnamefont {S.}~\bibnamefont {Barsanti}},
  \bibinfo {author} {\bibfnamefont {A.}~\bibnamefont {Maselli}}, \bibinfo
  {author} {\bibfnamefont {T.~P.}\ \bibnamefont {Sotiriou}}, \bibinfo {author}
  {\bibfnamefont {N.}~\bibnamefont {Warburton}}, \bibinfo {author}
  {\bibfnamefont {M.}~\bibnamefont {van~de Meent}}, \bibinfo {author}
  {\bibfnamefont {A.~J.~K.}\ \bibnamefont {Chua}}, \bibinfo {author}
  {\bibfnamefont {O.}~\bibnamefont {Burke}}, \ and\ \bibinfo {author}
  {\bibfnamefont {J.}~\bibnamefont {Gair}},\ }\href@noop {} {\  (\bibinfo
  {year} {2024})},\ \Eprint {http://arxiv.org/abs/2406.07607} {arXiv:2406.07607
  [gr-qc]} \BibitemShut {NoStop}%
\bibitem [{\citenamefont {Hinderer}\ and\ \citenamefont
  {Flanagan}(2008)}]{FlanaganHinderer1}%
  \BibitemOpen
  \bibfield  {author} {\bibinfo {author} {\bibfnamefont {T.}~\bibnamefont
  {Hinderer}}\ and\ \bibinfo {author} {\bibfnamefont {E.~E.}\ \bibnamefont
  {Flanagan}},\ }\href {\doibase 10.1103/PhysRevD.78.064028} {\bibfield
  {journal} {\bibinfo  {journal} {Phys. Rev. D}\ }\textbf {\bibinfo {volume}
  {78}},\ \bibinfo {pages} {064028} (\bibinfo {year} {2008})}\BibitemShut
  {NoStop}%
\bibitem [{\citenamefont {Afshordi}\ \emph {et~al.}(2023)\citenamefont
  {Afshordi} \emph {et~al.}}]{LISAConsortiumWaveformWorkingGroup:2023arg}%
  \BibitemOpen
  \bibfield  {author} {\bibinfo {author} {\bibfnamefont {N.}~\bibnamefont
  {Afshordi}} \emph {et~al.} (\bibinfo {collaboration} {LISA Consortium
  Waveform Working Group}),\ }\href@noop {} {\  (\bibinfo {year} {2023})},\
  \Eprint {http://arxiv.org/abs/2311.01300} {arXiv:2311.01300 [gr-qc]}
  \BibitemShut {NoStop}%
\bibitem [{\citenamefont {Kerr}(1963)}]{Kerr}%
  \BibitemOpen
  \bibfield  {author} {\bibinfo {author} {\bibfnamefont {R.~P.}\ \bibnamefont
  {Kerr}},\ }\href {\doibase 10.1103/PhysRevLett.11.237} {\bibfield  {journal}
  {\bibinfo  {journal} {Phys. Rev. Lett.}\ }\textbf {\bibinfo {volume} {11}},\
  \bibinfo {pages} {237} (\bibinfo {year} {1963})}\BibitemShut {NoStop}%
\bibitem [{\citenamefont {Mino}\ \emph
  {et~al.}(1997{\natexlab{a}})\citenamefont {Mino}, \citenamefont {Sasaki},\
  and\ \citenamefont {Tanaka}}]{Mino:1996nk}%
  \BibitemOpen
  \bibfield  {author} {\bibinfo {author} {\bibfnamefont {Y.}~\bibnamefont
  {Mino}}, \bibinfo {author} {\bibfnamefont {M.}~\bibnamefont {Sasaki}}, \ and\
  \bibinfo {author} {\bibfnamefont {T.}~\bibnamefont {Tanaka}},\ }\href
  {\doibase 10.1103/PhysRevD.55.3457} {\bibfield  {journal} {\bibinfo
  {journal} {Phys. Rev. D}\ }\textbf {\bibinfo {volume} {55}},\ \bibinfo
  {pages} {3457} (\bibinfo {year} {1997}{\natexlab{a}})},\ \Eprint
  {http://arxiv.org/abs/gr-qc/9606018} {arXiv:gr-qc/9606018} \BibitemShut
  {NoStop}%
\bibitem [{\citenamefont {Quinn}\ and\ \citenamefont
  {Wald}(1997{\natexlab{a}})}]{Quinn:1996am}%
  \BibitemOpen
  \bibfield  {author} {\bibinfo {author} {\bibfnamefont {T.~C.}\ \bibnamefont
  {Quinn}}\ and\ \bibinfo {author} {\bibfnamefont {R.~M.}\ \bibnamefont
  {Wald}},\ }\href {\doibase 10.1103/PhysRevD.56.3381} {\bibfield  {journal}
  {\bibinfo  {journal} {Phys. Rev. D}\ }\textbf {\bibinfo {volume} {56}},\
  \bibinfo {pages} {3381} (\bibinfo {year} {1997}{\natexlab{a}})},\ \Eprint
  {http://arxiv.org/abs/gr-qc/9610053} {arXiv:gr-qc/9610053} \BibitemShut
  {NoStop}%
\bibitem [{\citenamefont {Gralla}\ and\ \citenamefont
  {Wald}(2008)}]{Gralla:2008fg}%
  \BibitemOpen
  \bibfield  {author} {\bibinfo {author} {\bibfnamefont {S.~E.}\ \bibnamefont
  {Gralla}}\ and\ \bibinfo {author} {\bibfnamefont {R.~M.}\ \bibnamefont
  {Wald}},\ }\href {\doibase 10.1088/0264-9381/25/20/205009} {\bibfield
  {journal} {\bibinfo  {journal} {Class. Quant. Grav.}\ }\textbf {\bibinfo
  {volume} {25}},\ \bibinfo {pages} {205009} (\bibinfo {year} {2008})},\
  \bibinfo {note} {[Erratum: Class.Quant.Grav. 28, 159501 (2011)]},\ \Eprint
  {http://arxiv.org/abs/0806.3293} {arXiv:0806.3293 [gr-qc]} \BibitemShut
  {NoStop}%
\bibitem [{\citenamefont {Yang}\ and\ \citenamefont
  {Casals}(2017)}]{Yang:2017aht}%
  \BibitemOpen
  \bibfield  {author} {\bibinfo {author} {\bibfnamefont {H.}~\bibnamefont
  {Yang}}\ and\ \bibinfo {author} {\bibfnamefont {M.}~\bibnamefont {Casals}},\
  }\href {\doibase 10.1103/PhysRevD.96.083015} {\bibfield  {journal} {\bibinfo
  {journal} {Phys. Rev. D}\ }\textbf {\bibinfo {volume} {96}},\ \bibinfo
  {pages} {083015} (\bibinfo {year} {2017})},\ \Eprint
  {http://arxiv.org/abs/1704.02022} {arXiv:1704.02022 [gr-qc]} \BibitemShut
  {NoStop}%
\bibitem [{\citenamefont {Bonga}\ \emph {et~al.}(2019)\citenamefont {Bonga},
  \citenamefont {Yang},\ and\ \citenamefont {Hughes}}]{Bonga:2019ycj}%
  \BibitemOpen
  \bibfield  {author} {\bibinfo {author} {\bibfnamefont {B.}~\bibnamefont
  {Bonga}}, \bibinfo {author} {\bibfnamefont {H.}~\bibnamefont {Yang}}, \ and\
  \bibinfo {author} {\bibfnamefont {S.~A.}\ \bibnamefont {Hughes}},\ }\href
  {\doibase 10.1103/PhysRevLett.123.101103} {\bibfield  {journal} {\bibinfo
  {journal} {Phys. Rev. Lett.}\ }\textbf {\bibinfo {volume} {123}},\ \bibinfo
  {pages} {101103} (\bibinfo {year} {2019})},\ \Eprint
  {http://arxiv.org/abs/1905.00030} {arXiv:1905.00030 [gr-qc]} \BibitemShut
  {NoStop}%
\bibitem [{\citenamefont {Gupta}\ \emph {et~al.}(2022)\citenamefont {Gupta},
  \citenamefont {Speri}, \citenamefont {Bonga}, \citenamefont {Chua},\ and\
  \citenamefont {Tanaka}}]{Gupta:2022fbe}%
  \BibitemOpen
  \bibfield  {author} {\bibinfo {author} {\bibfnamefont {P.}~\bibnamefont
  {Gupta}}, \bibinfo {author} {\bibfnamefont {L.}~\bibnamefont {Speri}},
  \bibinfo {author} {\bibfnamefont {B.}~\bibnamefont {Bonga}}, \bibinfo
  {author} {\bibfnamefont {A.~J.~K.}\ \bibnamefont {Chua}}, \ and\ \bibinfo
  {author} {\bibfnamefont {T.}~\bibnamefont {Tanaka}},\ }\href {\doibase
  10.1103/PhysRevD.106.104001} {\bibfield  {journal} {\bibinfo  {journal}
  {Phys. Rev. D}\ }\textbf {\bibinfo {volume} {106}},\ \bibinfo {pages}
  {104001} (\bibinfo {year} {2022})},\ \Eprint
  {http://arxiv.org/abs/2205.04808} {arXiv:2205.04808 [gr-qc]} \BibitemShut
  {NoStop}%
\bibitem [{\citenamefont {Bronicki}\ \emph {et~al.}(2023)\citenamefont
  {Bronicki}, \citenamefont {C\'ardenas-Avenda\~no},\ and\ \citenamefont
  {Stein}}]{Bronicki:2022eqa}%
  \BibitemOpen
  \bibfield  {author} {\bibinfo {author} {\bibfnamefont {D.}~\bibnamefont
  {Bronicki}}, \bibinfo {author} {\bibfnamefont {A.}~\bibnamefont
  {C\'ardenas-Avenda\~no}}, \ and\ \bibinfo {author} {\bibfnamefont {L.~C.}\
  \bibnamefont {Stein}},\ }\href {\doibase 10.1088/1361-6382/acfcfe} {\bibfield
   {journal} {\bibinfo  {journal} {Class. Quant. Grav.}\ }\textbf {\bibinfo
  {volume} {40}},\ \bibinfo {pages} {215015} (\bibinfo {year} {2023})},\
  \Eprint {http://arxiv.org/abs/2203.08841} {arXiv:2203.08841 [gr-qc]}
  \BibitemShut {NoStop}%
\bibitem [{\citenamefont {Nasipak}\ and\ \citenamefont
  {Evans}(2021)}]{Nasipak:2021qfu}%
  \BibitemOpen
  \bibfield  {author} {\bibinfo {author} {\bibfnamefont {Z.}~\bibnamefont
  {Nasipak}}\ and\ \bibinfo {author} {\bibfnamefont {C.~R.}\ \bibnamefont
  {Evans}},\ }\href {\doibase 10.1103/PhysRevD.104.084011} {\bibfield
  {journal} {\bibinfo  {journal} {Phys. Rev. D}\ }\textbf {\bibinfo {volume}
  {104}},\ \bibinfo {pages} {084011} (\bibinfo {year} {2021})},\ \Eprint
  {http://arxiv.org/abs/2105.15188} {arXiv:2105.15188 [gr-qc]} \BibitemShut
  {NoStop}%
\bibitem [{\citenamefont {Nasipak}(2022)}]{Nasipak:2022xjh}%
  \BibitemOpen
  \bibfield  {author} {\bibinfo {author} {\bibfnamefont {Z.}~\bibnamefont
  {Nasipak}},\ }\href {\doibase 10.1103/PhysRevD.106.064042} {\bibfield
  {journal} {\bibinfo  {journal} {Phys. Rev. D}\ }\textbf {\bibinfo {volume}
  {106}},\ \bibinfo {pages} {064042} (\bibinfo {year} {2022})},\ \Eprint
  {http://arxiv.org/abs/2207.02224} {arXiv:2207.02224 [gr-qc]} \BibitemShut
  {NoStop}%
\bibitem [{\citenamefont {Speri}\ and\ \citenamefont {Gair}(2021)}]{SperiGair}%
  \BibitemOpen
  \bibfield  {author} {\bibinfo {author} {\bibfnamefont {L.}~\bibnamefont
  {Speri}}\ and\ \bibinfo {author} {\bibfnamefont {J.~R.}\ \bibnamefont
  {Gair}},\ }\href {\doibase 10.1103/PhysRevD.103.124032} {\bibfield  {journal}
  {\bibinfo  {journal} {Phys. Rev. D}\ }\textbf {\bibinfo {volume} {103}},\
  \bibinfo {pages} {124032} (\bibinfo {year} {2021})}\BibitemShut {NoStop}%
\bibitem [{\citenamefont {{Contopoulos}}(2002)}]{Contopoulos}%
  \BibitemOpen
  \bibfield  {author} {\bibinfo {author} {\bibfnamefont {G.}~\bibnamefont
  {{Contopoulos}}},\ }\href@noop {} {\emph {\bibinfo {title} {{Order and chaos
  in dynamical astronomy}}}}\ (\bibinfo {year} {2002})\BibitemShut {NoStop}%
\bibitem [{\citenamefont {Contopoulos}\ \emph {et~al.}(2011)\citenamefont
  {Contopoulos}, \citenamefont {Lukes-Gerakopoulos},\ and\ \citenamefont
  {Apostolatos}}]{Contopoulos:2011dz}%
  \BibitemOpen
  \bibfield  {author} {\bibinfo {author} {\bibfnamefont {G.}~\bibnamefont
  {Contopoulos}}, \bibinfo {author} {\bibfnamefont {G.}~\bibnamefont
  {Lukes-Gerakopoulos}}, \ and\ \bibinfo {author} {\bibfnamefont {T.~A.}\
  \bibnamefont {Apostolatos}},\ }\href {\doibase 10.1142/S0218127411029768}
  {\bibfield  {journal} {\bibinfo  {journal} {Int. J. Bifurc. Chaos}\ }\textbf
  {\bibinfo {volume} {21}},\ \bibinfo {pages} {2261} (\bibinfo {year}
  {2011})},\ \Eprint {http://arxiv.org/abs/1108.5057} {arXiv:1108.5057 [gr-qc]}
  \BibitemShut {NoStop}%
\bibitem [{\citenamefont {Lukes-Gerakopoulos}\ \emph
  {et~al.}(2010)\citenamefont {Lukes-Gerakopoulos}, \citenamefont
  {Apostolatos},\ and\ \citenamefont {Contopoulos}}]{Gerakopoulos}%
  \BibitemOpen
  \bibfield  {author} {\bibinfo {author} {\bibfnamefont {G.}~\bibnamefont
  {Lukes-Gerakopoulos}}, \bibinfo {author} {\bibfnamefont {T.}~\bibnamefont
  {Apostolatos}}, \ and\ \bibinfo {author} {\bibfnamefont {G.}~\bibnamefont
  {Contopoulos}},\ }\href {\doibase 10.1103/PhysRevD.81.124005} {\bibfield
  {journal} {\bibinfo  {journal} {Phys. Rev. D}\ }\textbf {\bibinfo {volume}
  {81}} (\bibinfo {year} {2010}),\ 10.1103/PhysRevD.81.124005}\BibitemShut
  {NoStop}%
\bibitem [{\citenamefont {Destounis}\ \emph {et~al.}(2020)\citenamefont
  {Destounis}, \citenamefont {Suvorov},\ and\ \citenamefont
  {Kokkotas}}]{Destounis}%
  \BibitemOpen
  \bibfield  {author} {\bibinfo {author} {\bibfnamefont {K.}~\bibnamefont
  {Destounis}}, \bibinfo {author} {\bibfnamefont {A.~G.}\ \bibnamefont
  {Suvorov}}, \ and\ \bibinfo {author} {\bibfnamefont {K.~D.}\ \bibnamefont
  {Kokkotas}},\ }\href {\doibase 10.1103/PhysRevD.102.064041} {\bibfield
  {journal} {\bibinfo  {journal} {Phys. Rev. D}\ }\textbf {\bibinfo {volume}
  {102}},\ \bibinfo {pages} {064041} (\bibinfo {year} {2020})}\BibitemShut
  {NoStop}%
\bibitem [{\citenamefont {Lukes-Gerakopoulos}\ and\ \citenamefont
  {Witzany}(2021)}]{Lukes-Gerakopoulos:2021ybx}%
  \BibitemOpen
  \bibfield  {author} {\bibinfo {author} {\bibfnamefont {G.}~\bibnamefont
  {Lukes-Gerakopoulos}}\ and\ \bibinfo {author} {\bibfnamefont
  {V.}~\bibnamefont {Witzany}},\ }\href {\doibase
  10.1007/978-981-15-4702-7_42-1} {\  (\bibinfo {year} {2021}),\
  10.1007/978-981-15-4702-7_42-1},\ \Eprint {http://arxiv.org/abs/2103.06724}
  {arXiv:2103.06724 [gr-qc]} \BibitemShut {NoStop}%
\bibitem [{\citenamefont {Brink}\ \emph
  {et~al.}(2015{\natexlab{a}})\citenamefont {Brink}, \citenamefont {Geyer},\
  and\ \citenamefont {Hinderer}}]{Brink:2013nna}%
  \BibitemOpen
  \bibfield  {author} {\bibinfo {author} {\bibfnamefont {J.}~\bibnamefont
  {Brink}}, \bibinfo {author} {\bibfnamefont {M.}~\bibnamefont {Geyer}}, \ and\
  \bibinfo {author} {\bibfnamefont {T.}~\bibnamefont {Hinderer}},\ }\href
  {\doibase 10.1103/PhysRevLett.114.081102} {\bibfield  {journal} {\bibinfo
  {journal} {Phys. Rev. Lett.}\ }\textbf {\bibinfo {volume} {114}},\ \bibinfo
  {pages} {081102} (\bibinfo {year} {2015}{\natexlab{a}})},\ \Eprint
  {http://arxiv.org/abs/1304.0330} {arXiv:1304.0330 [gr-qc]} \BibitemShut
  {NoStop}%
\bibitem [{\citenamefont {Brink}\ \emph
  {et~al.}(2015{\natexlab{b}})\citenamefont {Brink}, \citenamefont {Geyer},\
  and\ \citenamefont {Hinderer}}]{Brink:2015roa}%
  \BibitemOpen
  \bibfield  {author} {\bibinfo {author} {\bibfnamefont {J.}~\bibnamefont
  {Brink}}, \bibinfo {author} {\bibfnamefont {M.}~\bibnamefont {Geyer}}, \ and\
  \bibinfo {author} {\bibfnamefont {T.}~\bibnamefont {Hinderer}},\ }\href
  {\doibase 10.1103/PhysRevD.91.083001} {\bibfield  {journal} {\bibinfo
  {journal} {Phys. Rev. D}\ }\textbf {\bibinfo {volume} {91}},\ \bibinfo
  {pages} {083001} (\bibinfo {year} {2015}{\natexlab{b}})},\ \Eprint
  {http://arxiv.org/abs/1501.07728} {arXiv:1501.07728 [gr-qc]} \BibitemShut
  {NoStop}%
\bibitem [{\citenamefont {Warburton}\ \emph {et~al.}(2012)\citenamefont
  {Warburton}, \citenamefont {Akcay}, \citenamefont {Barack}, \citenamefont
  {Gair},\ and\ \citenamefont {Sago}}]{Warburton:2011fk}%
  \BibitemOpen
  \bibfield  {author} {\bibinfo {author} {\bibfnamefont {N.}~\bibnamefont
  {Warburton}}, \bibinfo {author} {\bibfnamefont {S.}~\bibnamefont {Akcay}},
  \bibinfo {author} {\bibfnamefont {L.}~\bibnamefont {Barack}}, \bibinfo
  {author} {\bibfnamefont {J.~R.}\ \bibnamefont {Gair}}, \ and\ \bibinfo
  {author} {\bibfnamefont {N.}~\bibnamefont {Sago}},\ }\href {\doibase
  10.1103/PhysRevD.85.061501} {\bibfield  {journal} {\bibinfo  {journal} {Phys.
  Rev. D}\ }\textbf {\bibinfo {volume} {85}},\ \bibinfo {pages} {061501}
  (\bibinfo {year} {2012})},\ \Eprint {http://arxiv.org/abs/1111.6908}
  {arXiv:1111.6908 [gr-qc]} \BibitemShut {NoStop}%
\bibitem [{\citenamefont {Ruangsri}\ and\ \citenamefont
  {Hughes}(2014)}]{Ruangsri:2013hra}%
  \BibitemOpen
  \bibfield  {author} {\bibinfo {author} {\bibfnamefont {U.}~\bibnamefont
  {Ruangsri}}\ and\ \bibinfo {author} {\bibfnamefont {S.~A.}\ \bibnamefont
  {Hughes}},\ }\href {\doibase 10.1103/PhysRevD.89.084036} {\bibfield
  {journal} {\bibinfo  {journal} {Phys. Rev. D}\ }\textbf {\bibinfo {volume}
  {89}},\ \bibinfo {pages} {084036} (\bibinfo {year} {2014})},\ \Eprint
  {http://arxiv.org/abs/1307.6483} {arXiv:1307.6483 [gr-qc]} \BibitemShut
  {NoStop}%
\bibitem [{\citenamefont {Osburn}\ \emph {et~al.}(2016)\citenamefont {Osburn},
  \citenamefont {Warburton},\ and\ \citenamefont {Evans}}]{Osburn:2015duj}%
  \BibitemOpen
  \bibfield  {author} {\bibinfo {author} {\bibfnamefont {T.}~\bibnamefont
  {Osburn}}, \bibinfo {author} {\bibfnamefont {N.}~\bibnamefont {Warburton}}, \
  and\ \bibinfo {author} {\bibfnamefont {C.~R.}\ \bibnamefont {Evans}},\ }\href
  {\doibase 10.1103/PhysRevD.93.064024} {\bibfield  {journal} {\bibinfo
  {journal} {Phys. Rev. D}\ }\textbf {\bibinfo {volume} {93}},\ \bibinfo
  {pages} {064024} (\bibinfo {year} {2016})},\ \Eprint
  {http://arxiv.org/abs/1511.01498} {arXiv:1511.01498 [gr-qc]} \BibitemShut
  {NoStop}%
\bibitem [{\citenamefont {Arnold}(1989)}]{Arnold:1989who}%
  \BibitemOpen
  \bibfield  {author} {\bibinfo {author} {\bibfnamefont {V.~I.}\ \bibnamefont
  {Arnold}},\ }\href {\doibase 10.1007/978-1-4757-2063-1} {\emph {\bibinfo
  {title} {{Mathematical Methods of Classical Mechanics}}}},\ Graduate Texts in
  Mathematics\ (\bibinfo  {publisher} {Springer},\ \bibinfo {year}
  {1989})\BibitemShut {NoStop}%
\bibitem [{\citenamefont {{Lichtenberg}}\ and\ \citenamefont
  {{Lieberman}}(1992)}]{Lichtenberg:1992}%
  \BibitemOpen
  \bibfield  {author} {\bibinfo {author} {\bibfnamefont {A.}~\bibnamefont
  {{Lichtenberg}}}\ and\ \bibinfo {author} {\bibfnamefont {M.}~\bibnamefont
  {{Lieberman}}},\ }\href@noop {} {\emph {\bibinfo {title} {{Regular and
  Chaotic Dynamics}}}}\ (\bibinfo {year} {1992})\BibitemShut {NoStop}%
\bibitem [{\citenamefont {van~de Meent}(2014)}]{vandeMeent:2013sza}%
  \BibitemOpen
  \bibfield  {author} {\bibinfo {author} {\bibfnamefont {M.}~\bibnamefont
  {van~de Meent}},\ }\href {\doibase 10.1103/PhysRevD.89.084033} {\bibfield
  {journal} {\bibinfo  {journal} {Phys. Rev. D}\ }\textbf {\bibinfo {volume}
  {89}},\ \bibinfo {pages} {084033} (\bibinfo {year} {2014})},\ \Eprint
  {http://arxiv.org/abs/1311.4457} {arXiv:1311.4457 [gr-qc]} \BibitemShut
  {NoStop}%
\bibitem [{\citenamefont {Destounis}\ \emph
  {et~al.}(2023{\natexlab{a}})\citenamefont {Destounis}, \citenamefont
  {Kulathingal}, \citenamefont {Kokkotas},\ and\ \citenamefont
  {Papadopoulos}}]{Destounis:2022obl}%
  \BibitemOpen
  \bibfield  {author} {\bibinfo {author} {\bibfnamefont {K.}~\bibnamefont
  {Destounis}}, \bibinfo {author} {\bibfnamefont {A.}~\bibnamefont
  {Kulathingal}}, \bibinfo {author} {\bibfnamefont {K.~D.}\ \bibnamefont
  {Kokkotas}}, \ and\ \bibinfo {author} {\bibfnamefont {G.~O.}\ \bibnamefont
  {Papadopoulos}},\ }\href {\doibase 10.1103/PhysRevD.107.084027} {\bibfield
  {journal} {\bibinfo  {journal} {Phys. Rev. D}\ }\textbf {\bibinfo {volume}
  {107}},\ \bibinfo {pages} {084027} (\bibinfo {year} {2023}{\natexlab{a}})},\
  \Eprint {http://arxiv.org/abs/2210.09357} {arXiv:2210.09357 [gr-qc]}
  \BibitemShut {NoStop}%
\bibitem [{\citenamefont {Schmidt}(2002)}]{Schmidt}%
  \BibitemOpen
  \bibfield  {author} {\bibinfo {author} {\bibfnamefont {W.}~\bibnamefont
  {Schmidt}},\ }\href {\doibase 10.1088/0264-9381/19/10/314} {\bibfield
  {journal} {\bibinfo  {journal} {Classical and Quantum Gravity}\ }\textbf
  {\bibinfo {volume} {19}},\ \bibinfo {pages} {2743} (\bibinfo {year}
  {2002})}\BibitemShut {NoStop}%
\bibitem [{\citenamefont {Berry}\ \emph {et~al.}(2016)\citenamefont {Berry},
  \citenamefont {Cole}, \citenamefont {Ca\~nizares},\ and\ \citenamefont
  {Gair}}]{Berry}%
  \BibitemOpen
  \bibfield  {author} {\bibinfo {author} {\bibfnamefont {C.~P.~L.}\
  \bibnamefont {Berry}}, \bibinfo {author} {\bibfnamefont {R.~H.}\ \bibnamefont
  {Cole}}, \bibinfo {author} {\bibfnamefont {P.}~\bibnamefont {Ca\~nizares}}, \
  and\ \bibinfo {author} {\bibfnamefont {J.~R.}\ \bibnamefont {Gair}},\ }\href
  {\doibase 10.1103/PhysRevD.94.124042} {\bibfield  {journal} {\bibinfo
  {journal} {Phys. Rev. D}\ }\textbf {\bibinfo {volume} {94}},\ \bibinfo
  {pages} {124042} (\bibinfo {year} {2016})},\ \Eprint
  {http://arxiv.org/abs/1608.08951} {arXiv:1608.08951 [gr-qc]} \BibitemShut
  {NoStop}%
\bibitem [{\citenamefont {Teukolsky}(2015)}]{Teukolsky:2014vca}%
  \BibitemOpen
  \bibfield  {author} {\bibinfo {author} {\bibfnamefont {S.~A.}\ \bibnamefont
  {Teukolsky}},\ }\href {\doibase 10.1088/0264-9381/32/12/124006} {\bibfield
  {journal} {\bibinfo  {journal} {Class. Quant. Grav.}\ }\textbf {\bibinfo
  {volume} {32}},\ \bibinfo {pages} {124006} (\bibinfo {year} {2015})},\
  \Eprint {http://arxiv.org/abs/1410.2130} {arXiv:1410.2130 [gr-qc]}
  \BibitemShut {NoStop}%
\bibitem [{\citenamefont {Carter}(1968)}]{Carter}%
  \BibitemOpen
  \bibfield  {author} {\bibinfo {author} {\bibfnamefont {B.}~\bibnamefont
  {Carter}},\ }\href {\doibase 10.1103/PhysRev.174.1559} {\bibfield  {journal}
  {\bibinfo  {journal} {Phys. Rev.}\ }\textbf {\bibinfo {volume} {174}},\
  \bibinfo {pages} {1559} (\bibinfo {year} {1968})}\BibitemShut {NoStop}%
\bibitem [{\citenamefont {Destounis}\ \emph {et~al.}(2021)\citenamefont
  {Destounis}, \citenamefont {Suvorov},\ and\ \citenamefont
  {Kokkotas}}]{Destounis:2021mqv}%
  \BibitemOpen
  \bibfield  {author} {\bibinfo {author} {\bibfnamefont {K.}~\bibnamefont
  {Destounis}}, \bibinfo {author} {\bibfnamefont {A.~G.}\ \bibnamefont
  {Suvorov}}, \ and\ \bibinfo {author} {\bibfnamefont {K.~D.}\ \bibnamefont
  {Kokkotas}},\ }\href {\doibase 10.1103/PhysRevLett.126.141102} {\bibfield
  {journal} {\bibinfo  {journal} {Phys. Rev. Lett.}\ }\textbf {\bibinfo
  {volume} {126}},\ \bibinfo {pages} {141102} (\bibinfo {year} {2021})},\
  \Eprint {http://arxiv.org/abs/2103.05643} {arXiv:2103.05643 [gr-qc]}
  \BibitemShut {NoStop}%
\bibitem [{\citenamefont {Destounis}\ and\ \citenamefont
  {Kokkotas}(2021)}]{Destounis:2021rko}%
  \BibitemOpen
  \bibfield  {author} {\bibinfo {author} {\bibfnamefont {K.}~\bibnamefont
  {Destounis}}\ and\ \bibinfo {author} {\bibfnamefont {K.~D.}\ \bibnamefont
  {Kokkotas}},\ }\href {\doibase 10.1103/PhysRevD.104.064023} {\bibfield
  {journal} {\bibinfo  {journal} {Phys. Rev. D}\ }\textbf {\bibinfo {volume}
  {104}},\ \bibinfo {pages} {064023} (\bibinfo {year} {2021})},\ \Eprint
  {http://arxiv.org/abs/2108.02782} {arXiv:2108.02782 [gr-qc]} \BibitemShut
  {NoStop}%
\bibitem [{\citenamefont {Deich}\ \emph {et~al.}(2022)\citenamefont {Deich},
  \citenamefont {C\'ardenas-Avenda\~no},\ and\ \citenamefont
  {Yunes}}]{Cardenas}%
  \BibitemOpen
  \bibfield  {author} {\bibinfo {author} {\bibfnamefont {A.}~\bibnamefont
  {Deich}}, \bibinfo {author} {\bibfnamefont {A.}~\bibnamefont
  {C\'ardenas-Avenda\~no}}, \ and\ \bibinfo {author} {\bibfnamefont
  {N.}~\bibnamefont {Yunes}},\ }\href {\doibase 10.1103/PhysRevD.106.024040}
  {\bibfield  {journal} {\bibinfo  {journal} {Phys. Rev. D}\ }\textbf {\bibinfo
  {volume} {106}},\ \bibinfo {pages} {024040} (\bibinfo {year}
  {2022})}\BibitemShut {NoStop}%
\bibitem [{\citenamefont {Destounis}\ \emph
  {et~al.}(2023{\natexlab{b}})\citenamefont {Destounis}, \citenamefont
  {Angeloni}, \citenamefont {Vaglio},\ and\ \citenamefont
  {Pani}}]{Destounis:2023khj}%
  \BibitemOpen
  \bibfield  {author} {\bibinfo {author} {\bibfnamefont {K.}~\bibnamefont
  {Destounis}}, \bibinfo {author} {\bibfnamefont {F.}~\bibnamefont {Angeloni}},
  \bibinfo {author} {\bibfnamefont {M.}~\bibnamefont {Vaglio}}, \ and\ \bibinfo
  {author} {\bibfnamefont {P.}~\bibnamefont {Pani}},\ }\href {\doibase
  10.1103/PhysRevD.108.084062} {\bibfield  {journal} {\bibinfo  {journal}
  {Phys. Rev. D}\ }\textbf {\bibinfo {volume} {108}},\ \bibinfo {pages}
  {084062} (\bibinfo {year} {2023}{\natexlab{b}})},\ \Eprint
  {http://arxiv.org/abs/2305.05691} {arXiv:2305.05691 [gr-qc]} \BibitemShut
  {NoStop}%
\bibitem [{\citenamefont {Destounis}\ \emph
  {et~al.}(2023{\natexlab{c}})\citenamefont {Destounis}, \citenamefont {Huez},\
  and\ \citenamefont {Kokkotas}}]{Destounis:2023gpw}%
  \BibitemOpen
  \bibfield  {author} {\bibinfo {author} {\bibfnamefont {K.}~\bibnamefont
  {Destounis}}, \bibinfo {author} {\bibfnamefont {G.}~\bibnamefont {Huez}}, \
  and\ \bibinfo {author} {\bibfnamefont {K.~D.}\ \bibnamefont {Kokkotas}},\
  }\href {\doibase 10.1007/s10714-023-03119-2} {\bibfield  {journal} {\bibinfo
  {journal} {Gen. Rel. Grav.}\ }\textbf {\bibinfo {volume} {55}},\ \bibinfo
  {pages} {71} (\bibinfo {year} {2023}{\natexlab{c}})},\ \Eprint
  {http://arxiv.org/abs/2301.11483} {arXiv:2301.11483 [gr-qc]} \BibitemShut
  {NoStop}%
\bibitem [{\citenamefont {Destounis}\ and\ \citenamefont
  {Kokkotas}(2023)}]{Destounis:2023cim}%
  \BibitemOpen
  \bibfield  {author} {\bibinfo {author} {\bibfnamefont {K.}~\bibnamefont
  {Destounis}}\ and\ \bibinfo {author} {\bibfnamefont {K.~D.}\ \bibnamefont
  {Kokkotas}},\ }\href {\doibase 10.1007/s10714-023-03170-z} {\bibfield
  {journal} {\bibinfo  {journal} {Gen. Rel. Grav.}\ }\textbf {\bibinfo {volume}
  {55}},\ \bibinfo {pages} {123} (\bibinfo {year} {2023})},\ \Eprint
  {http://arxiv.org/abs/2305.18522} {arXiv:2305.18522 [gr-qc]} \BibitemShut
  {NoStop}%
\bibitem [{\citenamefont {Eleni}\ \emph {et~al.}(2024)\citenamefont {Eleni},
  \citenamefont {Destounis}, \citenamefont {Apostolatos},\ and\ \citenamefont
  {Kokkotas}}]{Eleni:2024fgs}%
  \BibitemOpen
  \bibfield  {author} {\bibinfo {author} {\bibfnamefont {A.}~\bibnamefont
  {Eleni}}, \bibinfo {author} {\bibfnamefont {K.}~\bibnamefont {Destounis}},
  \bibinfo {author} {\bibfnamefont {T.~A.}\ \bibnamefont {Apostolatos}}, \ and\
  \bibinfo {author} {\bibfnamefont {K.~D.}\ \bibnamefont {Kokkotas}},\ }\href
  {\doibase 10.1103/PhysRevD.110.124004} {\bibfield  {journal} {\bibinfo
  {journal} {Phys. Rev. D}\ }\textbf {\bibinfo {volume} {110}},\ \bibinfo
  {pages} {124004} (\bibinfo {year} {2024})},\ \Eprint
  {http://arxiv.org/abs/2408.02004} {arXiv:2408.02004 [gr-qc]} \BibitemShut
  {NoStop}%
\bibitem [{\citenamefont {Barack}\ and\ \citenamefont {Pound}(2019)}]{GSF}%
  \BibitemOpen
  \bibfield  {author} {\bibinfo {author} {\bibfnamefont {L.}~\bibnamefont
  {Barack}}\ and\ \bibinfo {author} {\bibfnamefont {A.}~\bibnamefont {Pound}},\
  }\href {\doibase 10.1088/1361-6633/aae552} {\bibfield  {journal} {\bibinfo
  {journal} {Rept. Prog. Phys.}\ }\textbf {\bibinfo {volume} {82}},\ \bibinfo
  {pages} {016904} (\bibinfo {year} {2019})},\ \Eprint
  {http://arxiv.org/abs/1805.10385} {arXiv:1805.10385 [gr-qc]} \BibitemShut
  {NoStop}%
\bibitem [{\citenamefont {Glampedakis}\ and\ \citenamefont
  {Babak}(2006)}]{Glampedakis:2005cf}%
  \BibitemOpen
  \bibfield  {author} {\bibinfo {author} {\bibfnamefont {K.}~\bibnamefont
  {Glampedakis}}\ and\ \bibinfo {author} {\bibfnamefont {S.}~\bibnamefont
  {Babak}},\ }\href {\doibase 10.1088/0264-9381/23/12/013} {\bibfield
  {journal} {\bibinfo  {journal} {Class. Quant. Grav.}\ }\textbf {\bibinfo
  {volume} {23}},\ \bibinfo {pages} {4167} (\bibinfo {year} {2006})},\ \Eprint
  {http://arxiv.org/abs/gr-qc/0510057} {arXiv:gr-qc/0510057} \BibitemShut
  {NoStop}%
\bibitem [{\citenamefont {Flanagan}\ \emph {et~al.}(2014)\citenamefont
  {Flanagan}, \citenamefont {Hughes},\ and\ \citenamefont
  {Ruangsri}}]{FlanaganHughes}%
  \BibitemOpen
  \bibfield  {author} {\bibinfo {author} {\bibfnamefont {E.~E.}\ \bibnamefont
  {Flanagan}}, \bibinfo {author} {\bibfnamefont {S.~A.}\ \bibnamefont
  {Hughes}}, \ and\ \bibinfo {author} {\bibfnamefont {U.}~\bibnamefont
  {Ruangsri}},\ }\href {\doibase 10.1103/PhysRevD.89.084028} {\bibfield
  {journal} {\bibinfo  {journal} {Phys. Rev. D}\ }\textbf {\bibinfo {volume}
  {89}},\ \bibinfo {pages} {084028} (\bibinfo {year} {2014})}\BibitemShut
  {NoStop}%
\bibitem [{\citenamefont {Mino}(2003)}]{Mino}%
  \BibitemOpen
  \bibfield  {author} {\bibinfo {author} {\bibfnamefont {Y.}~\bibnamefont
  {Mino}},\ }\href {\doibase 10.1103/PhysRevD.67.084027} {\bibfield  {journal}
  {\bibinfo  {journal} {Phys. Rev. D}\ }\textbf {\bibinfo {volume} {67}},\
  \bibinfo {pages} {084027} (\bibinfo {year} {2003})}\BibitemShut {NoStop}%
\bibitem [{\citenamefont {Hughes}\ \emph {et~al.}(2005)\citenamefont {Hughes},
  \citenamefont {Drasco}, \citenamefont {Flanagan},\ and\ \citenamefont
  {Franklin}}]{PhysRevLett.94.221101}%
  \BibitemOpen
  \bibfield  {author} {\bibinfo {author} {\bibfnamefont {S.~A.}\ \bibnamefont
  {Hughes}}, \bibinfo {author} {\bibfnamefont {S.}~\bibnamefont {Drasco}},
  \bibinfo {author} {\bibfnamefont {E.~E.}\ \bibnamefont {Flanagan}}, \ and\
  \bibinfo {author} {\bibfnamefont {J.}~\bibnamefont {Franklin}},\ }\href
  {\doibase 10.1103/PhysRevLett.94.221101} {\bibfield  {journal} {\bibinfo
  {journal} {Phys. Rev. Lett.}\ }\textbf {\bibinfo {volume} {94}},\ \bibinfo
  {pages} {221101} (\bibinfo {year} {2005})}\BibitemShut {NoStop}%
\bibitem [{\citenamefont {Gair}\ and\ \citenamefont
  {Glampedakis}(2005)}]{Gair}%
  \BibitemOpen
  \bibfield  {author} {\bibinfo {author} {\bibfnamefont {J.}~\bibnamefont
  {Gair}}\ and\ \bibinfo {author} {\bibfnamefont {K.}~\bibnamefont
  {Glampedakis}},\ }\href {\doibase 10.1103/PhysRevD.73.064037} {\bibfield
  {journal} {\bibinfo  {journal} {Physical Review D}\ }\textbf {\bibinfo
  {volume} {73}} (\bibinfo {year} {2005}),\
  10.1103/PhysRevD.73.064037}\BibitemShut {NoStop}%
\bibitem [{\citenamefont {Flanagan}\ and\ \citenamefont
  {Hinderer}(2012)}]{FlanaganHinderer2}%
  \BibitemOpen
  \bibfield  {author} {\bibinfo {author} {\bibfnamefont {E.~E.}\ \bibnamefont
  {Flanagan}}\ and\ \bibinfo {author} {\bibfnamefont {T.}~\bibnamefont
  {Hinderer}},\ }\href {\doibase 10.1103/PhysRevLett.109.071102} {\bibfield
  {journal} {\bibinfo  {journal} {Phys. Rev. Lett.}\ }\textbf {\bibinfo
  {volume} {109}},\ \bibinfo {pages} {071102} (\bibinfo {year}
  {2012})}\BibitemShut {NoStop}%
\bibitem [{\citenamefont {Barack}\ and\ \citenamefont
  {Cutler}(2004)}]{Barack:2003fp}%
  \BibitemOpen
  \bibfield  {author} {\bibinfo {author} {\bibfnamefont {L.}~\bibnamefont
  {Barack}}\ and\ \bibinfo {author} {\bibfnamefont {C.}~\bibnamefont
  {Cutler}},\ }\href {\doibase 10.1103/PhysRevD.69.082005} {\bibfield
  {journal} {\bibinfo  {journal} {Phys. Rev. D}\ }\textbf {\bibinfo {volume}
  {69}},\ \bibinfo {pages} {082005} (\bibinfo {year} {2004})},\ \Eprint
  {http://arxiv.org/abs/gr-qc/0310125} {arXiv:gr-qc/0310125} \BibitemShut
  {NoStop}%
\bibitem [{\citenamefont {Nitz}\ \emph {et~al.}(2024)\citenamefont {Nitz},
  \citenamefont {Harry}, \citenamefont {Brown}, \citenamefont {Biwer},
  \citenamefont {Willis}, \citenamefont {Canton}, \citenamefont {Capano},
  \citenamefont {Dent}, \citenamefont {Pekowsky}, \citenamefont {Davies},
  \citenamefont {De}, \citenamefont {Cabero}, \citenamefont {Wu}, \citenamefont
  {Williamson}, \citenamefont {Machenschalk}, \citenamefont {Macleod},
  \citenamefont {Pannarale}, \citenamefont {Kumar}, \citenamefont {Reyes},
  \citenamefont {dfinstad}, \citenamefont {Kumar}, \citenamefont {Tápai},
  \citenamefont {Singer}, \citenamefont {Kumar}, \citenamefont {veronica
  villa}, \citenamefont {maxtrevor}, \citenamefont {Gadre}, \citenamefont
  {Khan}, \citenamefont {Fairhurst},\ and\ \citenamefont {Tolley}}]{pycbc}%
  \BibitemOpen
  \bibfield  {author} {\bibinfo {author} {\bibfnamefont {A.}~\bibnamefont
  {Nitz}}, \bibinfo {author} {\bibfnamefont {I.}~\bibnamefont {Harry}},
  \bibinfo {author} {\bibfnamefont {D.}~\bibnamefont {Brown}}, \bibinfo
  {author} {\bibfnamefont {C.~M.}\ \bibnamefont {Biwer}}, \bibinfo {author}
  {\bibfnamefont {J.}~\bibnamefont {Willis}}, \bibinfo {author} {\bibfnamefont
  {T.~D.}\ \bibnamefont {Canton}}, \bibinfo {author} {\bibfnamefont
  {C.}~\bibnamefont {Capano}}, \bibinfo {author} {\bibfnamefont
  {T.}~\bibnamefont {Dent}}, \bibinfo {author} {\bibfnamefont {L.}~\bibnamefont
  {Pekowsky}}, \bibinfo {author} {\bibfnamefont {G.~S.~C.}\ \bibnamefont
  {Davies}}, \bibinfo {author} {\bibfnamefont {S.}~\bibnamefont {De}}, \bibinfo
  {author} {\bibfnamefont {M.}~\bibnamefont {Cabero}}, \bibinfo {author}
  {\bibfnamefont {S.}~\bibnamefont {Wu}}, \bibinfo {author} {\bibfnamefont
  {A.~R.}\ \bibnamefont {Williamson}}, \bibinfo {author} {\bibfnamefont
  {B.}~\bibnamefont {Machenschalk}}, \bibinfo {author} {\bibfnamefont
  {D.}~\bibnamefont {Macleod}}, \bibinfo {author} {\bibfnamefont
  {F.}~\bibnamefont {Pannarale}}, \bibinfo {author} {\bibfnamefont
  {P.}~\bibnamefont {Kumar}}, \bibinfo {author} {\bibfnamefont
  {S.}~\bibnamefont {Reyes}}, \bibinfo {author} {\bibnamefont {dfinstad}},
  \bibinfo {author} {\bibfnamefont {S.}~\bibnamefont {Kumar}}, \bibinfo
  {author} {\bibfnamefont {M.}~\bibnamefont {Tápai}}, \bibinfo {author}
  {\bibfnamefont {L.}~\bibnamefont {Singer}}, \bibinfo {author} {\bibfnamefont
  {P.}~\bibnamefont {Kumar}}, \bibinfo {author} {\bibnamefont {veronica
  villa}}, \bibinfo {author} {\bibnamefont {maxtrevor}}, \bibinfo {author}
  {\bibfnamefont {B.~U.~V.}\ \bibnamefont {Gadre}}, \bibinfo {author}
  {\bibfnamefont {S.}~\bibnamefont {Khan}}, \bibinfo {author} {\bibfnamefont
  {S.}~\bibnamefont {Fairhurst}}, \ and\ \bibinfo {author} {\bibfnamefont
  {A.}~\bibnamefont {Tolley}},\ }\href {\doibase 10.5281/zenodo.10473621}
  {\enquote {\bibinfo {title} {gwastro/pycbc: v2.3.3 release of pycbc},}\ }
  (\bibinfo {year} {2024})\BibitemShut {NoStop}%
\bibitem [{\citenamefont {Cutler}\ and\ \citenamefont
  {Flanagan}(1994)}]{Cutler:1994ys}%
  \BibitemOpen
  \bibfield  {author} {\bibinfo {author} {\bibfnamefont {C.}~\bibnamefont
  {Cutler}}\ and\ \bibinfo {author} {\bibfnamefont {E.~E.}\ \bibnamefont
  {Flanagan}},\ }\href {\doibase 10.1103/PhysRevD.49.2658} {\bibfield
  {journal} {\bibinfo  {journal} {Phys. Rev. D}\ }\textbf {\bibinfo {volume}
  {49}},\ \bibinfo {pages} {2658} (\bibinfo {year} {1994})},\ \Eprint
  {http://arxiv.org/abs/gr-qc/9402014} {arXiv:gr-qc/9402014} \BibitemShut
  {NoStop}%
\bibitem [{\citenamefont {Owen}(1996)}]{Owen:1995tm}%
  \BibitemOpen
  \bibfield  {author} {\bibinfo {author} {\bibfnamefont {B.~J.}\ \bibnamefont
  {Owen}},\ }\href {\doibase 10.1103/PhysRevD.53.6749} {\bibfield  {journal}
  {\bibinfo  {journal} {Phys. Rev. D}\ }\textbf {\bibinfo {volume} {53}},\
  \bibinfo {pages} {6749} (\bibinfo {year} {1996})},\ \Eprint
  {http://arxiv.org/abs/gr-qc/9511032} {arXiv:gr-qc/9511032} \BibitemShut
  {NoStop}%
\bibitem [{\citenamefont {Moore}\ \emph {et~al.}(2015)\citenamefont {Moore},
  \citenamefont {Cole},\ and\ \citenamefont {Berry}}]{Moore:2014lga}%
  \BibitemOpen
  \bibfield  {author} {\bibinfo {author} {\bibfnamefont {C.~J.}\ \bibnamefont
  {Moore}}, \bibinfo {author} {\bibfnamefont {R.~H.}\ \bibnamefont {Cole}}, \
  and\ \bibinfo {author} {\bibfnamefont {C.~P.~L.}\ \bibnamefont {Berry}},\
  }\href {\doibase 10.1088/0264-9381/32/1/015014} {\bibfield  {journal}
  {\bibinfo  {journal} {Class. Quant. Grav.}\ }\textbf {\bibinfo {volume}
  {32}},\ \bibinfo {pages} {015014} (\bibinfo {year} {2015})},\ \Eprint
  {http://arxiv.org/abs/1408.0740} {arXiv:1408.0740 [gr-qc]} \BibitemShut
  {NoStop}%
\bibitem [{\citenamefont {Lynch}\ \emph {et~al.}(2024)\citenamefont {Lynch},
  \citenamefont {Witzany}, \citenamefont {van~de Meent},\ and\ \citenamefont
  {Warburton}}]{Lynch:2024ohd}%
  \BibitemOpen
  \bibfield  {author} {\bibinfo {author} {\bibfnamefont {P.}~\bibnamefont
  {Lynch}}, \bibinfo {author} {\bibfnamefont {V.}~\bibnamefont {Witzany}},
  \bibinfo {author} {\bibfnamefont {M.}~\bibnamefont {van~de Meent}}, \ and\
  \bibinfo {author} {\bibfnamefont {N.}~\bibnamefont {Warburton}},\ }\href
  {\doibase 10.1088/1361-6382/ad7dc9} {\bibfield  {journal} {\bibinfo
  {journal} {Class. Quant. Grav.}\ }\textbf {\bibinfo {volume} {41}},\ \bibinfo
  {pages} {225002} (\bibinfo {year} {2024})},\ \Eprint
  {http://arxiv.org/abs/2405.21072} {arXiv:2405.21072 [gr-qc]} \BibitemShut
  {NoStop}%
\bibitem [{\citenamefont {Pan}\ \emph {et~al.}(2022)\citenamefont {Pan},
  \citenamefont {Lyu},\ and\ \citenamefont {Yang}}]{Pan:2021lyw}%
  \BibitemOpen
  \bibfield  {author} {\bibinfo {author} {\bibfnamefont {Z.}~\bibnamefont
  {Pan}}, \bibinfo {author} {\bibfnamefont {Z.}~\bibnamefont {Lyu}}, \ and\
  \bibinfo {author} {\bibfnamefont {H.}~\bibnamefont {Yang}},\ }\href {\doibase
  10.1103/PhysRevD.105.083005} {\bibfield  {journal} {\bibinfo  {journal}
  {Phys. Rev. D}\ }\textbf {\bibinfo {volume} {105}},\ \bibinfo {pages}
  {083005} (\bibinfo {year} {2022})},\ \Eprint
  {http://arxiv.org/abs/2112.10237} {arXiv:2112.10237 [astro-ph.HE]}
  \BibitemShut {NoStop}%
\bibitem [{\citenamefont {Pan}\ \emph {et~al.}(2021)\citenamefont {Pan},
  \citenamefont {Lyu},\ and\ \citenamefont {Yang}}]{Pan:2021oob}%
  \BibitemOpen
  \bibfield  {author} {\bibinfo {author} {\bibfnamefont {Z.}~\bibnamefont
  {Pan}}, \bibinfo {author} {\bibfnamefont {Z.}~\bibnamefont {Lyu}}, \ and\
  \bibinfo {author} {\bibfnamefont {H.}~\bibnamefont {Yang}},\ }\href {\doibase
  10.1103/PhysRevD.104.063007} {\bibfield  {journal} {\bibinfo  {journal}
  {Phys. Rev. D}\ }\textbf {\bibinfo {volume} {104}},\ \bibinfo {pages}
  {063007} (\bibinfo {year} {2021})},\ \Eprint
  {http://arxiv.org/abs/2104.01208} {arXiv:2104.01208 [astro-ph.HE]}
  \BibitemShut {NoStop}%
\bibitem [{\citenamefont {Lyu}\ \emph {et~al.}(2024)\citenamefont {Lyu},
  \citenamefont {Pan}, \citenamefont {Mao}, \citenamefont {Jiang},\ and\
  \citenamefont {Yang}}]{Lyu:2024gnk}%
  \BibitemOpen
  \bibfield  {author} {\bibinfo {author} {\bibfnamefont {Z.}~\bibnamefont
  {Lyu}}, \bibinfo {author} {\bibfnamefont {Z.}~\bibnamefont {Pan}}, \bibinfo
  {author} {\bibfnamefont {J.}~\bibnamefont {Mao}}, \bibinfo {author}
  {\bibfnamefont {N.}~\bibnamefont {Jiang}}, \ and\ \bibinfo {author}
  {\bibfnamefont {H.}~\bibnamefont {Yang}},\ }\href@noop {} {\  (\bibinfo
  {year} {2024})},\ \Eprint {http://arxiv.org/abs/2501.03252} {arXiv:2501.03252
  [astro-ph.HE]} \BibitemShut {NoStop}%
\bibitem [{\citenamefont {Pan}\ \emph {et~al.}(2023)\citenamefont {Pan},
  \citenamefont {Yang}, \citenamefont {Bernard},\ and\ \citenamefont
  {Bonga}}]{Pan:2023wau}%
  \BibitemOpen
  \bibfield  {author} {\bibinfo {author} {\bibfnamefont {Z.}~\bibnamefont
  {Pan}}, \bibinfo {author} {\bibfnamefont {H.}~\bibnamefont {Yang}}, \bibinfo
  {author} {\bibfnamefont {L.}~\bibnamefont {Bernard}}, \ and\ \bibinfo
  {author} {\bibfnamefont {B.}~\bibnamefont {Bonga}},\ }\href {\doibase
  10.1103/PhysRevD.108.104026} {\bibfield  {journal} {\bibinfo  {journal}
  {Phys. Rev. D}\ }\textbf {\bibinfo {volume} {108}},\ \bibinfo {pages}
  {104026} (\bibinfo {year} {2023})},\ \Eprint
  {http://arxiv.org/abs/2306.06576} {arXiv:2306.06576 [gr-qc]} \BibitemShut
  {NoStop}%
\bibitem [{\citenamefont {Cardoso}\ \emph
  {et~al.}(2022{\natexlab{a}})\citenamefont {Cardoso}, \citenamefont
  {Destounis}, \citenamefont {Duque}, \citenamefont {Macedo},\ and\
  \citenamefont {Maselli}}]{Cardoso:2021wlq}%
  \BibitemOpen
  \bibfield  {author} {\bibinfo {author} {\bibfnamefont {V.}~\bibnamefont
  {Cardoso}}, \bibinfo {author} {\bibfnamefont {K.}~\bibnamefont {Destounis}},
  \bibinfo {author} {\bibfnamefont {F.}~\bibnamefont {Duque}}, \bibinfo
  {author} {\bibfnamefont {R.~P.}\ \bibnamefont {Macedo}}, \ and\ \bibinfo
  {author} {\bibfnamefont {A.}~\bibnamefont {Maselli}},\ }\href {\doibase
  10.1103/PhysRevD.105.L061501} {\bibfield  {journal} {\bibinfo  {journal}
  {Phys. Rev. D}\ }\textbf {\bibinfo {volume} {105}},\ \bibinfo {pages}
  {L061501} (\bibinfo {year} {2022}{\natexlab{a}})},\ \Eprint
  {http://arxiv.org/abs/2109.00005} {arXiv:2109.00005 [gr-qc]} \BibitemShut
  {NoStop}%
\bibitem [{\citenamefont {Cardoso}\ \emph
  {et~al.}(2022{\natexlab{b}})\citenamefont {Cardoso}, \citenamefont
  {Destounis}, \citenamefont {Duque}, \citenamefont {Panosso~Macedo},\ and\
  \citenamefont {Maselli}}]{Cardoso:2022whc}%
  \BibitemOpen
  \bibfield  {author} {\bibinfo {author} {\bibfnamefont {V.}~\bibnamefont
  {Cardoso}}, \bibinfo {author} {\bibfnamefont {K.}~\bibnamefont {Destounis}},
  \bibinfo {author} {\bibfnamefont {F.}~\bibnamefont {Duque}}, \bibinfo
  {author} {\bibfnamefont {R.}~\bibnamefont {Panosso~Macedo}}, \ and\ \bibinfo
  {author} {\bibfnamefont {A.}~\bibnamefont {Maselli}},\ }\href {\doibase
  10.1103/PhysRevLett.129.241103} {\bibfield  {journal} {\bibinfo  {journal}
  {Phys. Rev. Lett.}\ }\textbf {\bibinfo {volume} {129}},\ \bibinfo {pages}
  {241103} (\bibinfo {year} {2022}{\natexlab{b}})},\ \Eprint
  {http://arxiv.org/abs/2210.01133} {arXiv:2210.01133 [gr-qc]} \BibitemShut
  {NoStop}%
\bibitem [{\citenamefont
  {\rm{Information~Systems~and~Wake~Forest~University}}(2021)}]{WakeHPC}%
  \BibitemOpen
  \bibfield  {author} {\bibinfo {author} {\bibnamefont
  {\rm{Information~Systems~and~Wake~Forest~University}}},\ }\href {\doibase
  10.57682/G13Z-2362} {\enquote {\bibinfo {title} {{WFU High Performance
  Computing Facility}},}\ } (\bibinfo {year} {2021})\BibitemShut {NoStop}%
\bibitem [{\citenamefont {Levati}(2025)}]{10.5281/zenodo.14983924}%
  \BibitemOpen
  \bibfield  {author} {\bibinfo {author} {\bibfnamefont {E.}~\bibnamefont
  {Levati}},\ }\href {\doibase 10.5281/zenodo.14983924} {\enquote {\bibinfo
  {title} {When resonances kick in: the cumulative effect of orbital resonances
  in extreme-mass-ratio inspirals (concept (all versions))},}\ } (\bibinfo
  {year} {2025})\BibitemShut {NoStop}%
\bibitem [{\citenamefont {Fujita}\ and\ \citenamefont {Hikida}(2009)}]{Fujita}%
  \BibitemOpen
  \bibfield  {author} {\bibinfo {author} {\bibfnamefont {R.}~\bibnamefont
  {Fujita}}\ and\ \bibinfo {author} {\bibfnamefont {W.}~\bibnamefont
  {Hikida}},\ }\href {\doibase 10.1088/0264-9381/26/13/135002} {\bibfield
  {journal} {\bibinfo  {journal} {Class. Quant. Grav.}\ }\textbf {\bibinfo
  {volume} {26}},\ \bibinfo {pages} {135002} (\bibinfo {year} {2009})},\
  \Eprint {http://arxiv.org/abs/0906.1420} {arXiv:0906.1420 [gr-qc]}
  \BibitemShut {NoStop}%
\bibitem [{\citenamefont {Xin}\ \emph {et~al.}(2019)\citenamefont {Xin},
  \citenamefont {Han},\ and\ \citenamefont {Yang}}]{Biao}%
  \BibitemOpen
  \bibfield  {author} {\bibinfo {author} {\bibfnamefont {S.}~\bibnamefont
  {Xin}}, \bibinfo {author} {\bibfnamefont {W.-B.}\ \bibnamefont {Han}}, \ and\
  \bibinfo {author} {\bibfnamefont {S.-C.}\ \bibnamefont {Yang}},\ }\href
  {\doibase 10.1103/PhysRevD.100.084055} {\bibfield  {journal} {\bibinfo
  {journal} {Phys. Rev. D}\ }\textbf {\bibinfo {volume} {100}},\ \bibinfo
  {pages} {084055} (\bibinfo {year} {2019})},\ \Eprint
  {http://arxiv.org/abs/1812.04185} {arXiv:1812.04185 [gr-qc]} \BibitemShut
  {NoStop}%
\bibitem [{\citenamefont {Mino}\ \emph
  {et~al.}(1997{\natexlab{b}})\citenamefont {Mino}, \citenamefont {Sasaki},\
  and\ \citenamefont {Tanaka}}]{MiSaTa}%
  \BibitemOpen
  \bibfield  {author} {\bibinfo {author} {\bibfnamefont {Y.}~\bibnamefont
  {Mino}}, \bibinfo {author} {\bibfnamefont {M.}~\bibnamefont {Sasaki}}, \ and\
  \bibinfo {author} {\bibfnamefont {T.}~\bibnamefont {Tanaka}},\ }\href
  {\doibase 10.1103/PhysRevD.55.3457} {\bibfield  {journal} {\bibinfo
  {journal} {Phys. Rev. D}\ }\textbf {\bibinfo {volume} {55}},\ \bibinfo
  {pages} {3457} (\bibinfo {year} {1997}{\natexlab{b}})}\BibitemShut {NoStop}%
\bibitem [{\citenamefont {Quinn}\ and\ \citenamefont
  {Wald}(1997{\natexlab{b}})}]{QuWa}%
  \BibitemOpen
  \bibfield  {author} {\bibinfo {author} {\bibfnamefont {T.~C.}\ \bibnamefont
  {Quinn}}\ and\ \bibinfo {author} {\bibfnamefont {R.~M.}\ \bibnamefont
  {Wald}},\ }\href {\doibase 10.1103/PhysRevD.56.3381} {\bibfield  {journal}
  {\bibinfo  {journal} {Phys. Rev. D}\ }\textbf {\bibinfo {volume} {56}},\
  \bibinfo {pages} {3381} (\bibinfo {year} {1997}{\natexlab{b}})}\BibitemShut
  {NoStop}%
\bibitem [{\citenamefont {\rm{Wolfram Research{,} Inc.}}()}]{Mathematica}%
  \BibitemOpen
  \bibfield  {author} {\bibinfo {author} {\bibnamefont {\rm{Wolfram Research{,}
  Inc.}}},\ }\href {https://www.wolfram.com/mathematica} {\enquote {\bibinfo
  {title} {Mathematica, {V}ersion 14.0},}\ }\bibinfo {note} {Champaign, IL,
  2024}\BibitemShut {NoStop}%
\bibitem [{\citenamefont {Butcher}(1996)}]{BUTCHER1996247}%
  \BibitemOpen
  \bibfield  {author} {\bibinfo {author} {\bibfnamefont {J.}~\bibnamefont
  {Butcher}},\ }\href {\doibase https://doi.org/10.1016/0168-9274(95)00108-5}
  {\bibfield  {journal} {\bibinfo  {journal} {Applied Numerical Mathematics}\
  }\textbf {\bibinfo {volume} {20}},\ \bibinfo {pages} {247} (\bibinfo {year}
  {1996})}\BibitemShut {NoStop}%
\bibitem [{\citenamefont {Touroux}\ \emph {et~al.}(2024)\citenamefont
  {Touroux}, \citenamefont {Kitazawa}, \citenamefont {Murase},\ and\
  \citenamefont {Nahrgang}}]{Touroux:2023rkv}%
  \BibitemOpen
  \bibfield  {author} {\bibinfo {author} {\bibfnamefont {N.}~\bibnamefont
  {Touroux}}, \bibinfo {author} {\bibfnamefont {M.}~\bibnamefont {Kitazawa}},
  \bibinfo {author} {\bibfnamefont {K.}~\bibnamefont {Murase}}, \ and\ \bibinfo
  {author} {\bibfnamefont {M.}~\bibnamefont {Nahrgang}},\ }\href {\doibase
  10.1093/ptep/ptae058} {\bibfield  {journal} {\bibinfo  {journal} {PTEP}\
  }\textbf {\bibinfo {volume} {2024}},\ \bibinfo {pages} {063D01} (\bibinfo
  {year} {2024})},\ \Eprint {http://arxiv.org/abs/2306.12696} {arXiv:2306.12696
  [nucl-th]} \BibitemShut {NoStop}%
\bibitem [{\citenamefont {{Mizerov{\'a}}}\ and\ \citenamefont
  {{Tvrd{\'a}}}(2024)}]{2024arXiv240416665M}%
  \BibitemOpen
  \bibfield  {author} {\bibinfo {author} {\bibfnamefont {H.}~\bibnamefont
  {{Mizerov{\'a}}}}\ and\ \bibinfo {author} {\bibfnamefont {K.}~\bibnamefont
  {{Tvrd{\'a}}}},\ }\href {\doibase 10.48550/arXiv.2404.16665} {\bibfield
  {journal} {\bibinfo  {journal} {arXiv e-prints}\ ,\ \bibinfo {eid}
  {arXiv:2404.16665}} (\bibinfo {year} {2024})},\ \Eprint
  {http://arxiv.org/abs/2404.16665} {arXiv:2404.16665 [math.NA]} \BibitemShut
  {NoStop}%
\bibitem [{\citenamefont {Ascher}\ and\ \citenamefont
  {Petzold}(1991)}]{ProjRK}%
  \BibitemOpen
  \bibfield  {author} {\bibinfo {author} {\bibfnamefont {U.~M.}\ \bibnamefont
  {Ascher}}\ and\ \bibinfo {author} {\bibfnamefont {L.~R.}\ \bibnamefont
  {Petzold}},\ }\href {\doibase 10.1137/0728059} {\bibfield  {journal}
  {\bibinfo  {journal} {SIAM Journal on Numerical Analysis}\ }\textbf {\bibinfo
  {volume} {28}},\ \bibinfo {pages} {1097} (\bibinfo {year} {1991})},\ \Eprint
  {http://arxiv.org/abs/https://doi.org/10.1137/0728059}
  {https://doi.org/10.1137/0728059} \BibitemShut {NoStop}%
\end{thebibliography}%

\end{document}